\documentclass[11pt, a4paper]{article}
\usepackage{jcappub}

\usepackage{color}
\usepackage{booktabs}
\usepackage{multirow}
\usepackage{amssymb}
\usepackage[export]{adjustbox}
\usepackage{floatrow}
\newfloatcommand{capbtabbox}{table}[][3.5in]
\newfloatcommand{capbfigbox}{figure}[][2.3in]
\usepackage{blindtext}
\usepackage{amsmath}
\usepackage{booktabs}
\usepackage{aas_macros}
\usepackage{epsfig}
\usepackage{graphicx}
\usepackage{subfigure}
\usepackage{comment}

\def\lsim{\mathrel{\raise.3ex\hbox{$<$\kern-.75em\lower1ex\hbox{$\sim$}}}}
\def\gsim{\mathrel{\raise.3ex\hbox{$>$\kern-.75em\lower1ex\hbox{$\sim$}}}}

\begin{document}

\title{Gamma-rays and Neutrinos from Giant Molecular Cloud Populations in the Galactic Plane}

\author[a,]{Abhijit Roy}
\author[a,b]{Jagdish C. Joshi,}
\author[c,]{Martina Cardillo,}
\author[d,]{Prantik Sarmah}
\author[e,f]{Ritabrata Sarkar}
\author[d,]{and Sovan Chakraborty}

\affiliation[a]{Aryabhatta Research Institute of Observational Sciences (ARIES), \\ Manora peak, Beluwakhan, Uttarakhand 263001, India}

\affiliation[b]{Centre for Astro-Particle Physics (CAPP) and Department of Physics, University of Johannesburg, PO Box 524, Auckland Park 2006, South Africa}

\affiliation[c]{INAF - Istituto di Astrofisica e Planetologia Spaziali (IAPS), \\ Via del Fosso del Cavaliere 100, 00133 Roma, Italy}

\affiliation[d]{Indian Institute of Technology (IITG), Guwahati-781039, India}

\affiliation[e]{Institute of Astronomy Space and Earth Science (IASES), \\ P 177, CIT Road, Scheme 7m, Kolkata-700054, West Bengal, India}

\affiliation[f]{Gran Sasso Science Institute (GSSI), \\ Viale Francesco Crispi 7, 67100, LAquila (AQ), Italy}

\emailAdd{abhijitroy@aries.res.in (AR)}
\emailAdd{jagdish@aries.res.in (JCJ)}
\emailAdd{martina.cardillo@inaf.it (MC)}
\emailAdd{ritabrata.s@gmail.com (RS)}
\emailAdd{prantik@iitg.ac.in (PS)}
\emailAdd{sovan@iitg.ac.in (SC)}

\newcommand\mc[1]{{\bf \color{magenta} #1}}
\newcommand\rs[1]{{\bf \color{cyan} #1}}

\abstract{
The recent IceCube detection of significant neutrino flux from the inner Galactic plane has provided us valuable insights on the spectrum of cosmic rays in our Galaxy. This flux can be produced either by a population of Galactic point sources or by diffused emission from cosmic ray interactions with the interstellar medium or by a mixture of both. In this work, we compute diffused gamma-ray and neutrino fluxes produced by a population of giant molecular clouds (GMCs) in our Galaxy, assuming different parametrizations of the Galactic diffused cosmic ray distribution. In particular, we take into account two main cases: (I) constant cosmic ray luminosity in our Galaxy, and (II) space-dependent cosmic ray luminosity, based on the supernovae distribution in our Galaxy. For Case-I, we found that the neutrino flux from GMCs is a factor of $\sim 10$ below compared to $\pi^0$ and KRA$_\gamma$ best-fitted models of IceCube observations at $10^5$ GeV. Instead, for Case-II the model can explain up to $\sim 90 \%$ of the neutrino flux at that energy. Moreover, for this last scenario IceCube detector could be able to detect neutrino events from the Galactic centre regions. We then calculated gamma-ray and neutrino fluxes from individual GMCs and noticed that several current and future Cherenkov telescopes and neutrino observatories have the right sensitivities to study these objects. In particular, very neutrino-bright region such as Aquila Rift is favourable for detection by the IceCube-Gen2 observatory.}

\maketitle

\section{Introduction}

Our Galactic plane is a plausible source of neutrino emission. Indeed, in a recent paper by the IceCube collaboration \citep{2023Sci380338I}, a $4.5 \sigma$ neutrino signal from the Galactic plane has been discovered using 10 years (2011-2021) of observations. Their work provides a significant improvement compared to the previous IceCube paper, where the authors reported only a $2 \sigma$ hint of neutrino signal from our Galaxy \citep{Aartsen2019ApJ12A}. The observed gamma-ray spectrum in the inner Galactic plane exhibits a harder spectral index compared to the one derived from the observed cosmic ray (CR) spectrum on Earth \citep{nerenov2023PhRvD3044N}. However, for neutrinos the situation is still unclear and the spectral index is model dependent. The origin of this neutrino flux could be due to secondary pions, produced in the interactions of Galactic CR (GCR) sea with the target gas in the interstellar medium \citep{stecker1979ApJ_19S,Gupta2013AP75G,zhang2023ApJ3Z,2023arXiv230707451V} or to freshly accelerated CR interactions in the vicinity of their accelerators in our Galaxy \citep{Gabici2007ApJ6651G, kappes_2007ApJ70K, Gonza2009APh437G,zhang2023ApJ3Z,2023arXiv230707451V}. In both cases, the neutrino emission is accompanied by gamma-rays produced by the decay of neutral pions \citep{stecker1979ApJ_19S,kelner2006energy}.


In the past, the inner Galactic plane has also been observed in diffused MeV-GeV gamma-rays by the Fermi-LAT \citep{ackermann2012fermi} and in diffused TeV gamma-rays by the Milagro observatory \citep{abdo2008ApJ_078A}. Using these observations, phenomenological CR transport models were developed, also known as KRA$_{\gamma}$ models \citep{gaggero2015gamma}. Further, ARGO-YBJ \citep{bartoli2015Ap20B}, H.E.S.S. \citep{hess2018AnA..1H}, HAWC \citep{abdalla2021ApJ_hesshawc}, Tibet-AS$\gamma$ \citep{amenomori2021first} and LHAASO \citep{2023PhRvL.131o1001C} gamma-ray observatories have also reported the detection of TeV-PeV diffused gamma-ray spectra from Galactic plane. These observations are important to constrain the neutrino flux levels \citep{Gabici2007ApJ6651G,gabici2008_180G,ackermann2012fermi,Lunar2015PhRv_1301L,2023PhRvD.108f1305S,zhang2023ApJ3Z,2023arXiv230707451V,nerenov2023PhRvD3044N}. By including KRA$_{\gamma}$ models for the Galactic plane neutrino search analysis, \cite{2023Sci380338I} found that these models are consistent with neutrino flux detected from the Galactic plane. 
Neutrino detection from the Galactic plane is an important ingredient in our understanding of the origin of GCRs and of the nature of their sources. These observations also help us to investigate CR transport in our Galaxy and their interactions with the ambient gas, radiation and magnetic fields \citep{evoli2007JCAP003E, 2014PhRvD..89j3002N,gagg2015ApJ25G}. Using the IceCube results, \citep{2023arXiv230707451V} studied the sample of Galactic TeV gamma-ray sources and found that the upper bound of their contribution in the diffuse Galactic neutrino flux is $\sim 40 \%$ for a maximum CR proton energy of 500 TeV, which lowers to $\sim 20 \%$ for maximum CR proton energies of 5 PeV. 

The difficulty in determining the origin of neutrino flux from our Galaxy is due to the ambiguity in the possible origin of CRs producing it. From one side, the interaction of pre-existing GCRs with interstellar gas and dust elements mainly produces diffused emissions. On the other side, freshly accelerated CRs from sources interacting with dense material in their surroundings may also contribute. After the very recent results obtained by ultra-high energy (UHE) and very-high energy  (VHE) gamma-ray instruments, the most plausible sources of freshly accelerated GCRs are the following\citep{hess2018AnA..1H, ahar2019NatA61A,Abe2020PhRv02A,cao2021ultrahigh, bao2021ApJ2B,cristofari2021hunt,cardillo2023lhaaso,olmi2023pulsar,vieu2023massive,gabici2023cosmic,cao2023first,2023arXiv230617285A}; supernova remnants (SNRs), Young Massive Stellar Clusters (YMSCs), Young Massive Stars,  Microquasars and also the Pulsar Wind Nebulae (PWNe), etc. In all cases, however, the Galactic dense gas regions likely irradiated by GCRs are the most useful target to understand gamma-ray and neutrino fluxes origin \citep{2014PhRvD90h3002K}. Gamma-ray emission from GMCs has been used to probe the CR spectrum at several locations in our Galaxy \citep{peron2022A57P}. Using gamma-ray data of 8 local GMCs (distances up to a few hundred pc from the solar neighbourhood) in the Gould Belt, a power-law CR spectrum was found with an energy spectral index of $2.3$ \citep{2014AnA66A142Y}. The importance of GMC population for diffused gamma-ray and neutrino emission is also supported by the fact that individual GMCs can produce secondary gamma-rays through $p-p$ inelastic interaction process \citep{2012ApJ7564A,Yang2015AnA580A..90Y,bagh20} as expected in theory \citep{1973ApJ185L7B,1983AdSpR23W, 2001SSRv187A, 2007ApnSS.309..365G, bagh20,2023JCAP08047R}. Hence, utilizing the neutrino and gamma-ray connection, we can investigate the origin of neutrino emission in connection with CR spectrum \citep{ahlers2014PhRvDb3010A, 2014MNRAS.439.3414J,Sarmah2023MNRAS144S}. In a recent work \citep{2023JCAP08047R}, we found that Aquila Rift is a very promising candidate to explain the detected neutrino signal. This motivated us to study the role of all the other known GMCs in our Galaxy in neutrino production. 

In this paper, we have included all the clouds based on the available GMC catalogs. Then the multimessenger (gamma-rays + neutrino) connection of these GMCs is explored by assuming a constant and radially dependent GCR distribution model. The detectability of multimessenger fluxes from individual GMCs is checked with the current and upcoming gamma-ray and neutrino detectors. Further, the stacked neutrino flux is calculated from these GMCs and corroborated by the corresponding observed gamma-ray flux. In the end, we compare the spatial correlation of TeV Galactic point sources with the observed neutrino significance map of the IceCube.

The paper is organised as follows. In Section \ref{Sec:catalog}, we discuss about the GMC catalogs used for this purpose and their properties; then, we describe the spectrum of CR interacting with GMCs in Section \ref{Sec:ProtonDist}. In Section \ref{sec:modelling}, we present our estimation of gamma-ray and neutrino fluxes, and we show our results in Section \ref{Sec:results}. In Section \ref{sec:spatialCorrelation}, we analyse the implications of spatial correlations of GMCs and Galactic TeV-sources compared to the IceCube neutrino observations. We present our conclusions in Section \ref{sec:conclusions}.

\section{Details on the GMC Catalogs}
\label{Sec:catalog}

The ISM has several components: coronal gas, intercloud gas, diffused clouds, dark clouds, bok globules, molecular clouds, etc \cite{myers1978compilation}. These components are mostly comprised of hydrogen (in atomic $H_I$, molecular $H_2$ or ionised $H_{II}$ form) and of a small fraction of heavy elements. Molecular hydrogen ($H_{2}$) is found mainly in dense and cold regions of the ISM (e.g., dark clouds, bok globules, molecular clouds). This is difficult to detect in such regions because of its homonuclear diatomic nature and of the extremely cold temperature of the regions themselves ($\sim 10^\circ$ kelvin), at which the $H_{2}$ is predominantly in its ground state. Consequently, emission from carbon monoxide (CO) molecules is used to trace its presence, and the density of $H_2$ regions is calculated using a constant conversion factor $X_{CO}$. Atomic hydrogen regions, on the other hand, are largely found in the intercloud gas and diffused clouds, and they can be easily probed with the 21 cm radio emission line. The coronal gas is an intercloud gas comprising mainly fully ionized hydrogen ($H_{II}$) that becomes extremely hot (with a temperature similar to the solar corona) due to the shock waves from SNRs or stellar objects \citep{mccray1987soapM,Jen197845J}. 


\begin{figure}[!ht]
    \centering
    \subfigure{\includegraphics[height=0.48\linewidth, width=0.48\linewidth]{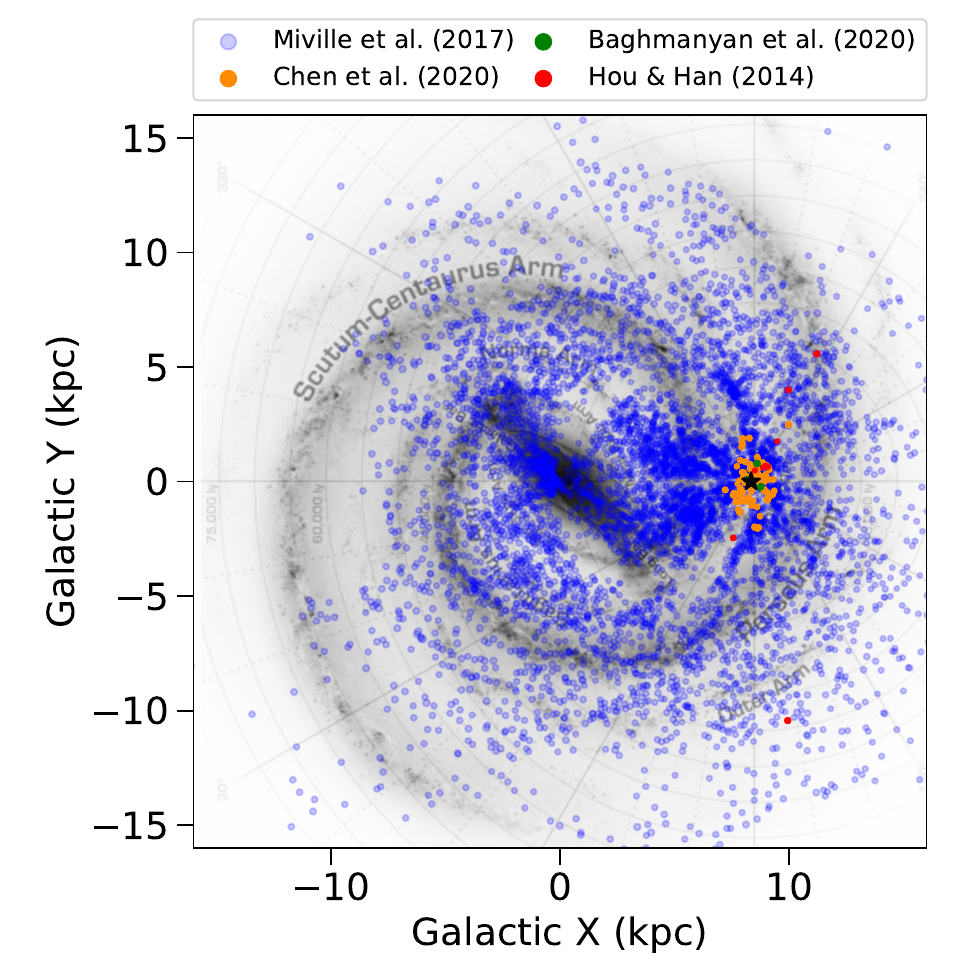}}
    \subfigure{\includegraphics[height=0.51\linewidth, width=0.51\linewidth]{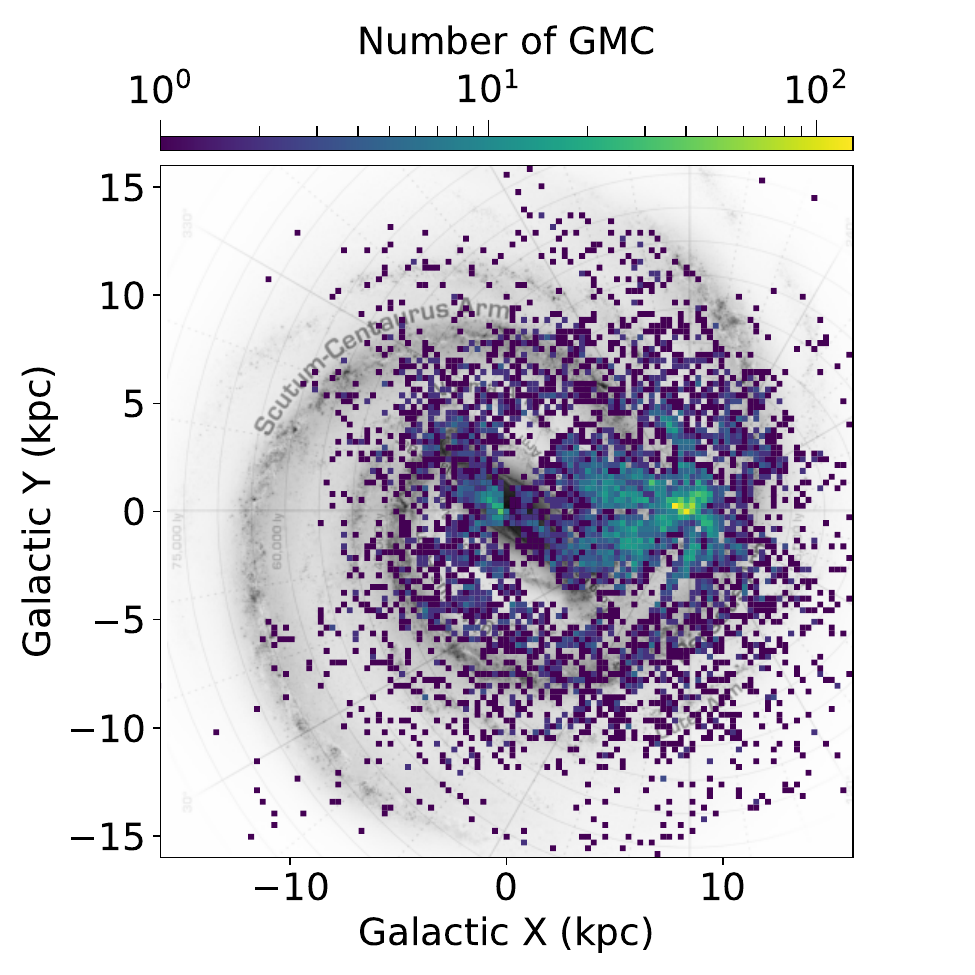}}
    \caption{{\it Left panel}: Face-on view of the MW Galaxy with the scattered distribution of GMCs on it. The marker colour here corresponds to the GMCs selected from different catalogs (\cite[blue points: Miville et al. (2017)]{mivi17}, \cite[orange points: Chen et al. (2020)]{chen20}, \cite[green points: Baghmanyan et al. (2020)]{2016ApJ...822...52R}, and \cite[red points: Hou \& Han (2014)]{hou14}) and the marker size of each GMC is proportional to $log(R)$. The Sun here is indicated by the black-coloured star marker. {\it Right panel}: The 2D-density distribution of all the GMCs considered within the combined cloud catalog.}
    \label{fig:gmc_dist_bird_view}
\end{figure}
 
The identification of isolated $H_2$ structures in the ISM is a difficult task because they are arbitrarily distributed between well-defined gravitationally bounded spherical structures and poorly structured clumpy multi-phase geometry \cite{mivi17}. Despite such difficulties, several authors have identified isolated $H_2$ structures in the Milky Way (MW) and collected their results in cloud catalogs \cite{chen20,hou14,2016ApJ...822...52R,mivi17}. Most of them are limited to certain Galactic regions of space, and the cataloged clouds can explain only a maximum of $\sim$ 40\% CO emissions data observed by \cite{2001ApJ...547..792D}. However, recently a detailed catalog of GMCs in the Galactic plane was published, restricted to a region with Galactic latitude $|b| \leq 5^\circ$ \cite{mivi17} recovering 98\% of the observed CO emissions \cite{2001ApJ...547..792D}. It is worth noticing that taking into account different detection schemes for the same CO data could lead to different cloud catalogs. However, all of them are expected to follow the same statistical features, such as similar distributions of cloud mass and similar size vs line width relations \cite{mivi17}. In this work, we have used four different cloud catalogs: 8246 clouds are taken from \cite{mivi17} within $|b| \leq 5^\circ$, while beyond that Galactic latitude, 182 clouds are taken from \cite{chen20}, 10 clouds are taken from \cite{hou14}, and 6 clouds are taken from \cite{bagh20}. A face-on view of the GMC distribution in the MW Galaxy can be seen in the left panel of Fig. \ref{fig:gmc_dist_bird_view}, where different colour indicates clouds from different catalogs. In the right panel of Fig. \ref{fig:gmc_dist_bird_view}, we show a 2D-density distribution of GMCs in the MW Galaxy considered here for the calculations. 

\begin{figure}[!ht]
    \centering
\subfigure[]{\includegraphics[width=0.32\linewidth]{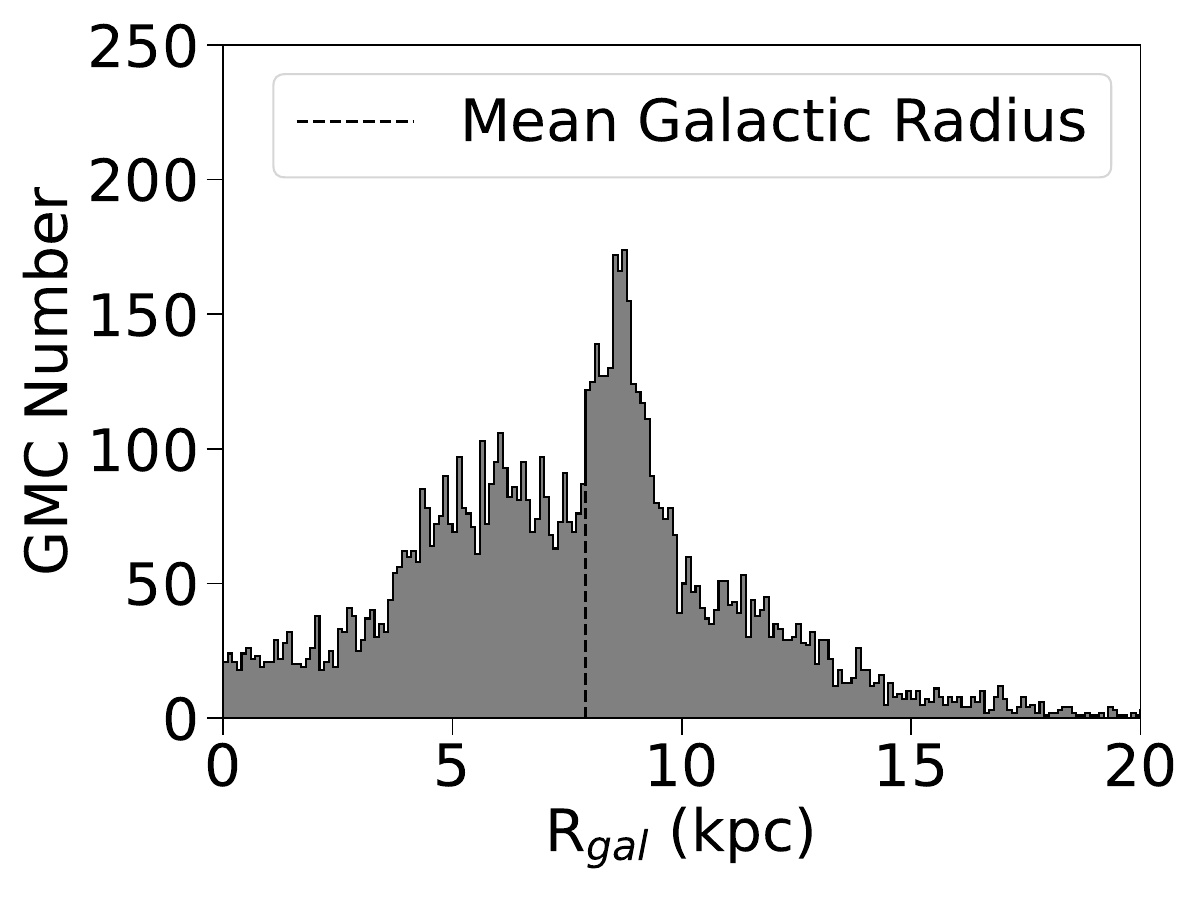}}
\subfigure[]{\includegraphics[width=0.32\linewidth]{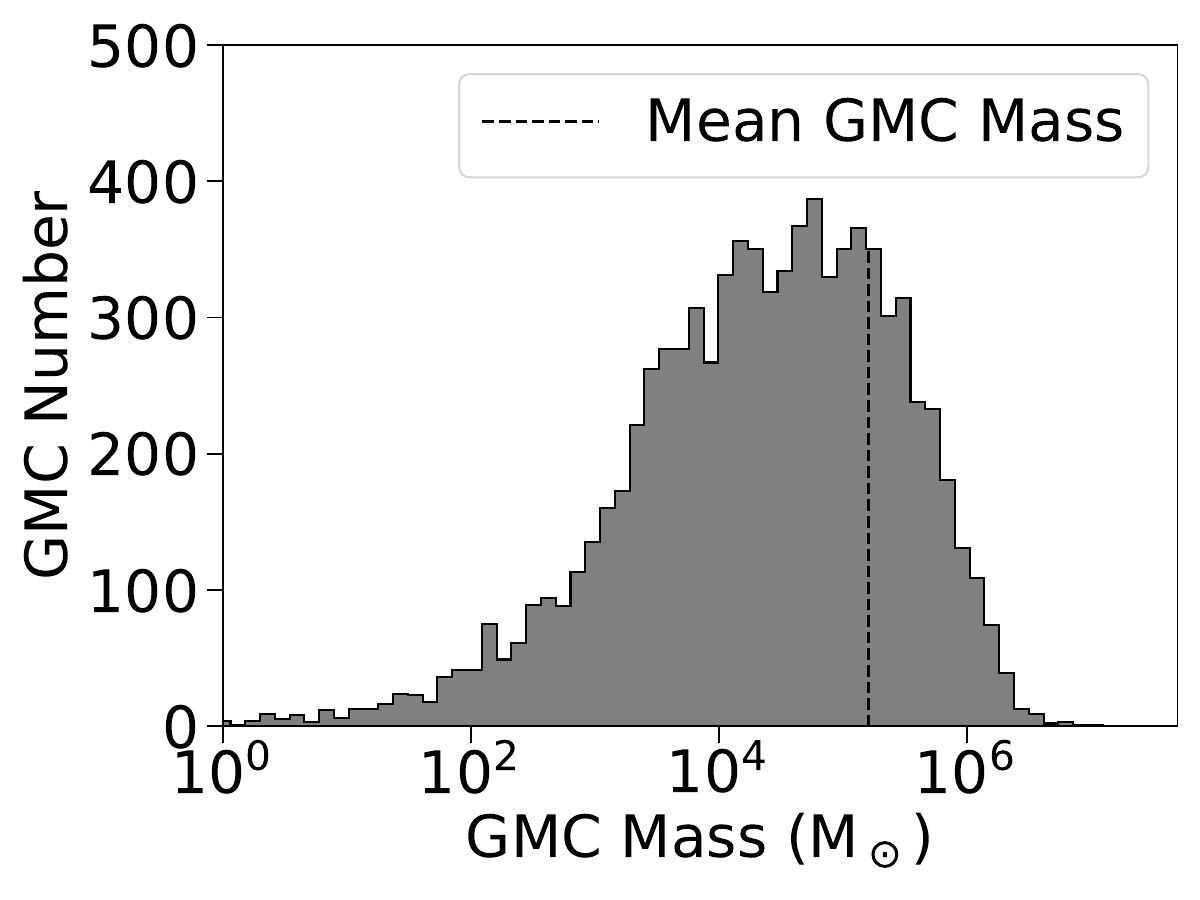}}
\subfigure[]{\includegraphics[width=0.32\linewidth]{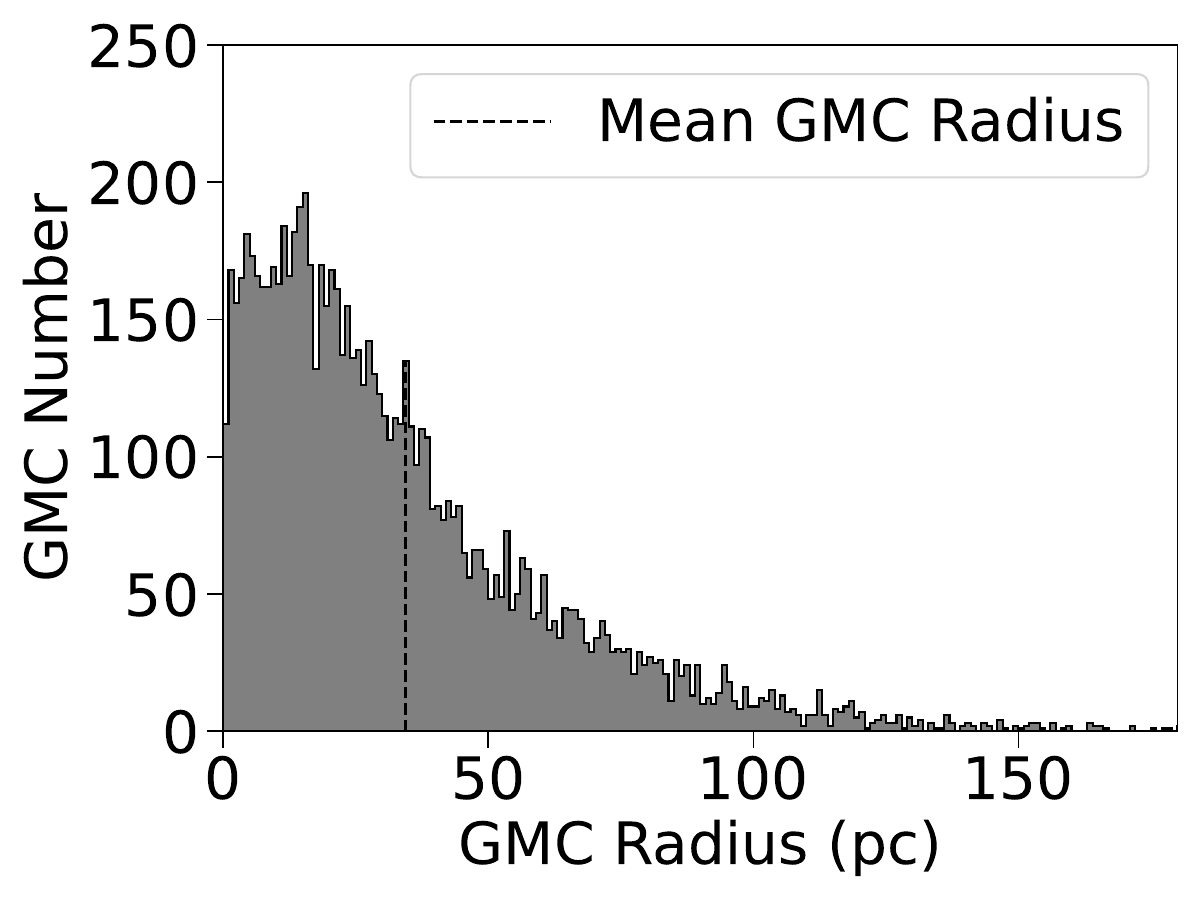}}\\
\subfigure[]{\includegraphics[width=0.32\linewidth]{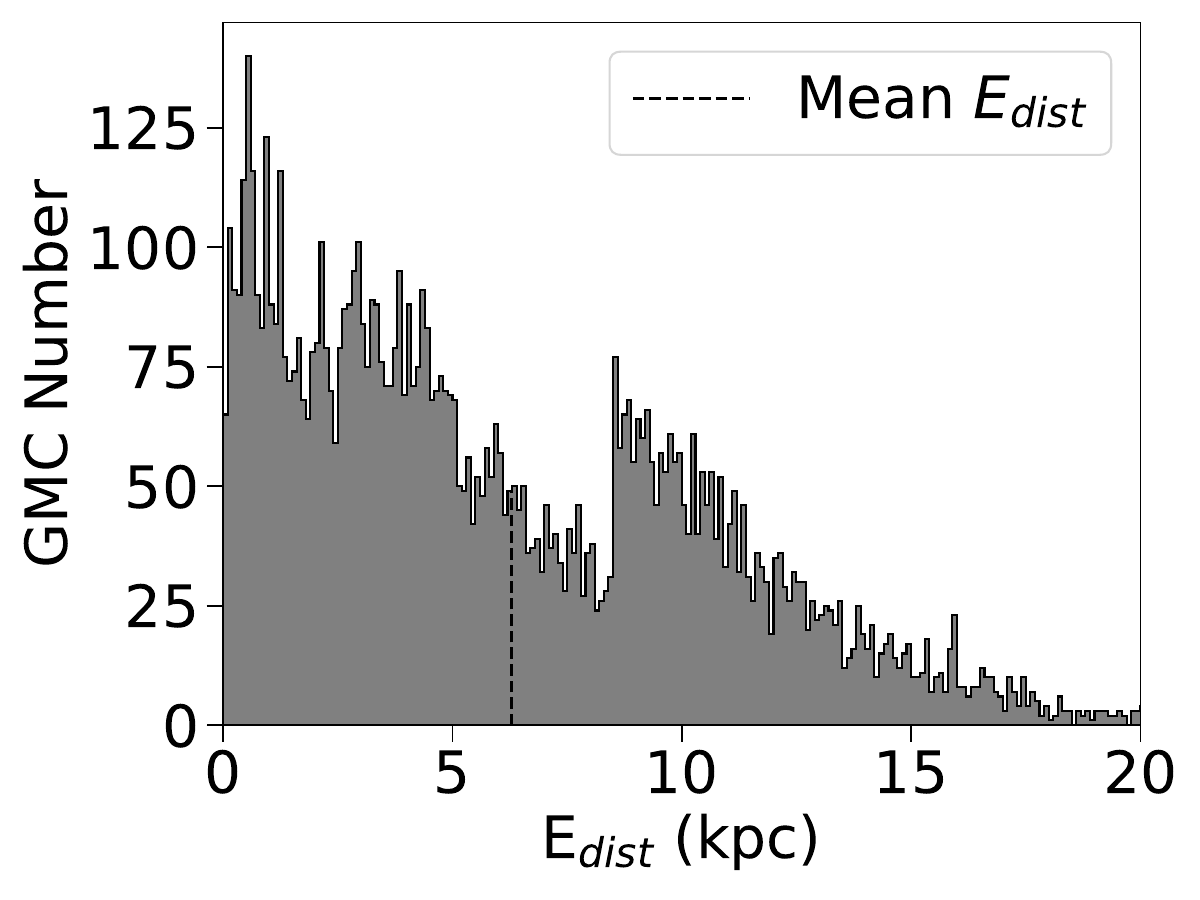}}
\subfigure[]{\includegraphics[width=0.32\linewidth]{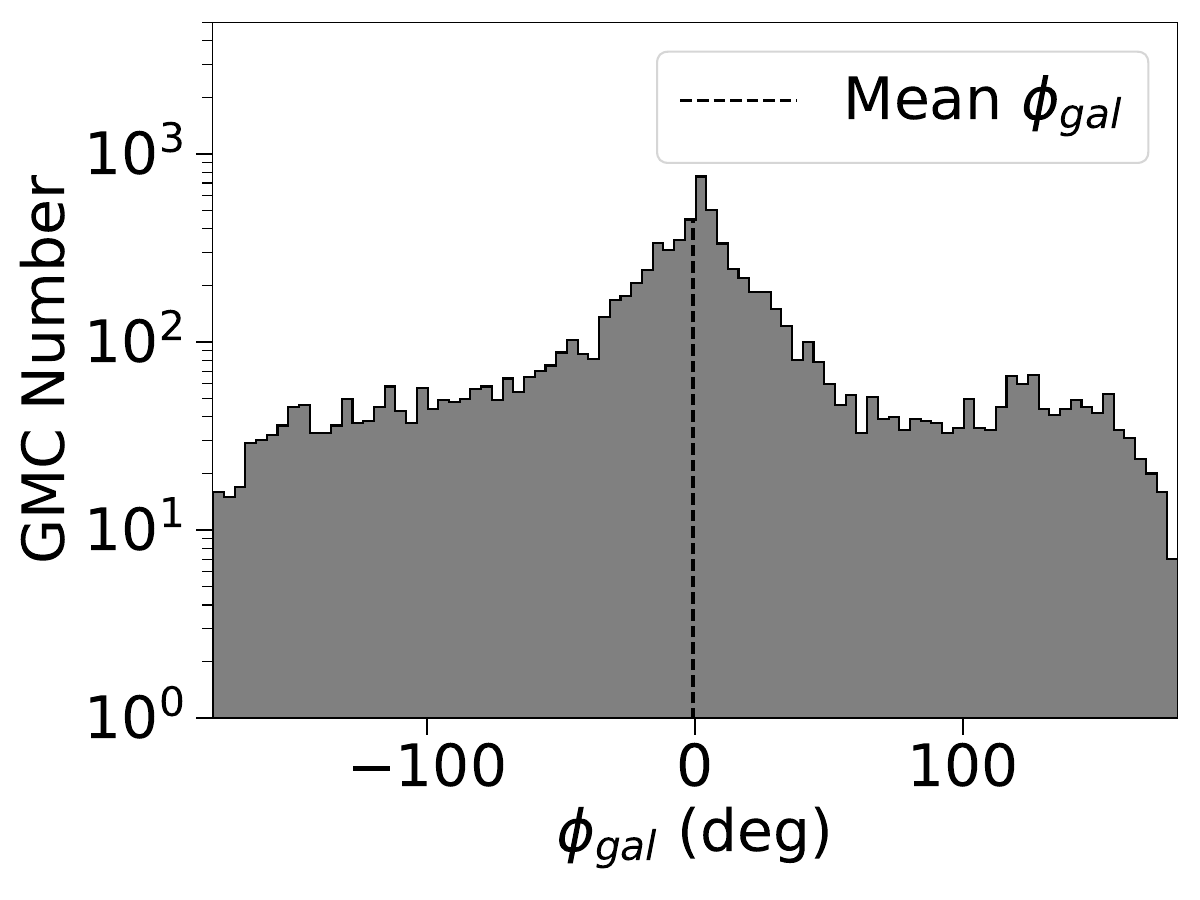}}
\subfigure[]{\includegraphics[width=0.32\linewidth]{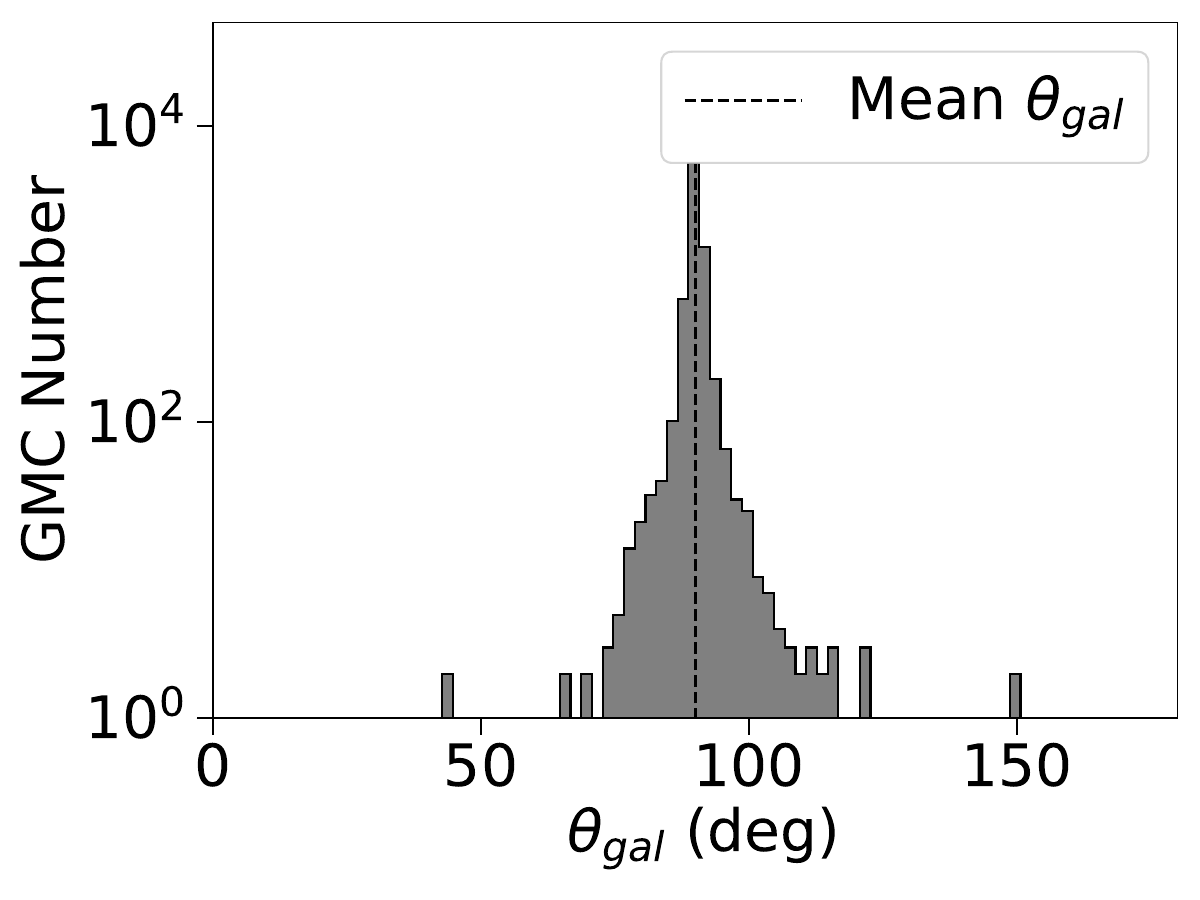}}
\caption{  Distributions of various GMC parameters in our Galaxy [a]: distance from the Galactic centre ($R_{gal}$), [b]: mass (M), [c]: radius (r), [d]: distance from the Earth ($E_{dist}$), [e]: population w.r.t. the galactocentric azimuthal angle ($\phi_{gal}$) and [f]: polar angle ($\theta_{gal}$). In these histograms the dotted vertical line indicates the mean value of each parameter.}
    \label{fig:RgalMR_dist}
\end{figure}

The total mass of the interstellar gas in the MW is about $\sim 7\times 10^9 M_{\odot}$, where about $\sim$ 20\% of it is in the form of molecular hydrogen ($H_2$) \cite{draine2010physics}. The total mass of estimated $H_2$ in the combined GMC catalog is about $\sim 1.3\times10^9$ M$_{\odot}$, within the uncertainty range of the previous estimation (1 $\pm$ 0.3)$\times10^9$ M$_{\odot}$ by \cite{heyer2015molecular}. We have calculated the average distance of the GMCs from the Galactic centre ($\bar{R}_{gal}$), their mass ($\bar{M}$), and their radius ($\bar{r}$) in the combined catalog. The values for these quantities are 7.9 kpc, $1.6 \times 10^5$ M$_{\odot}$, and 34 pc, respectively, and they are shown by the vertical dotted lines in Fig. \ref{fig:RgalMR_dist}.

We have also computed the surface mass density ($\Sigma_{H_2}^{GMC}$) of all the clouds, splitting the Galaxy in concentric rings of 1 kpc radius from the Galactic centre, dividing the total mass of GMCs in each ring by its surface area: 
 
\begin{equation}
\begin{aligned}
\Sigma_{H_2}^{GMC} &= \frac{M_{ring}^{GMC}}{\pi (r^2_{out} - r^2_{in})}
\end{aligned}
\label{equ:SMass}
\end{equation}

\begin{figure}[!ht]
    \centering
    \includegraphics[width=0.65\linewidth]{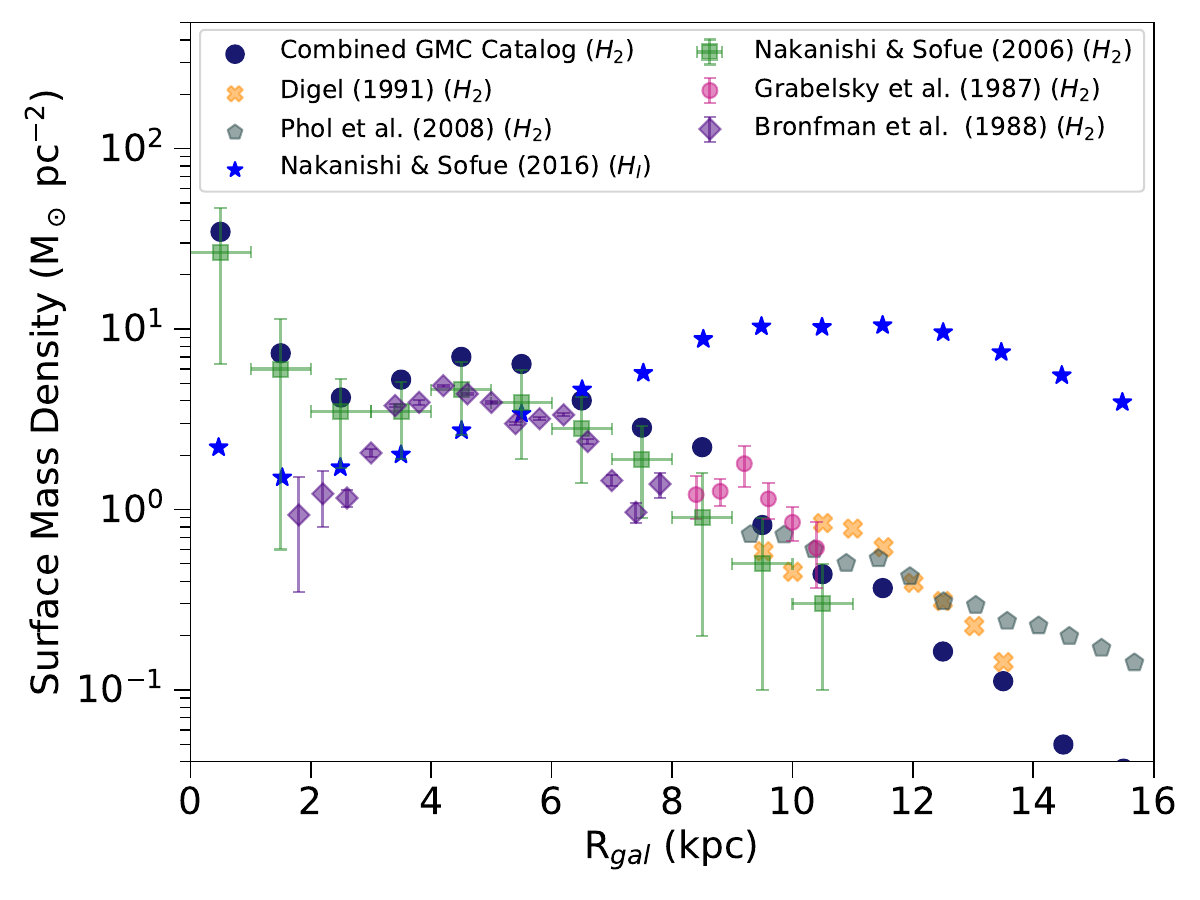}
    \caption{The distribution of $H_2$ surface mass density ($\Sigma_{H_2}^{GMC}$) along the galactocentric radius ($R_{gal}$). The blue star markers represent the variation in surface mass density of diffused atomic hydrogen gas in the MW Galaxy. The data points are extracted from Fig. 9 in \cite{mivi17} and  Fig. 3.5 in \cite{menchiari2023probing}, and corresponds to \cite{digel91} in orange colour, \cite{pohl2008three} in gray colour,  \cite{nakanishi2016three} in blue colour, \cite{nakanishi2006three} in green colour, \cite{grabelsky1987molecular} in pink colour and \cite{bronfman1988co} in violet colour points respectively.}
    \label{fig:surface_dens_Rgal}
\end{figure}

It's important to note that we are looking at the projected area of these rings (projection of a three-dimensional GMC onto a 2D plane). The computed values of the surface mass density are compared with the reference data as shown in Fig. \ref{fig:surface_dens_Rgal}. The highest $\Sigma_{H_2}^{GMC}$ of the GMCs is in the Galactic centre region, decreasing gradually thereafter, in agreement with the reported values in the literature \citep{nakanishi2006three,bronfman1988co,grabelsky1987molecular,digel91,pohl2008three}. In contrast, the surface mass density of atomic hydrogen (star points in Fig. \ref{fig:surface_dens_Rgal}) increases with galactocentric radius, dominating after $\sim8$ kpc. These values are taken from \citep{nakanishi2016three} and are provided for reference. Hence, $H_2$ in the GMCs is expected to contribute to multimessenger emission towards the Galactic centre more than the diffused $H_I$. However, \citep{2014PhRvD89j3003T} found that the role of $H_I$ could be important for neutrino production in the halo of the Galaxy  \citep{2014PhRvD89j3003T}. 
Other authors \citep{ackermann2012fermi,gagg2015ApJ25G,schwefer2023ApJ6S,fang2023milky} have estimated the neutrino flux using diffused $H_I$ and $H_2$ gas distributions in the MW. In this work, we have taken a step forward, computing gamma-ray and neutrino fluxes from the individual GMCs in the MW and also showing their total contribution towards the multimessenger emission.

\section{Spectral Distribution of Primary GCRs in MW Galaxy}
\label{Sec:ProtonDist}

\begin{figure}[!ht]
    \centering
    \includegraphics[width=0.65\linewidth]{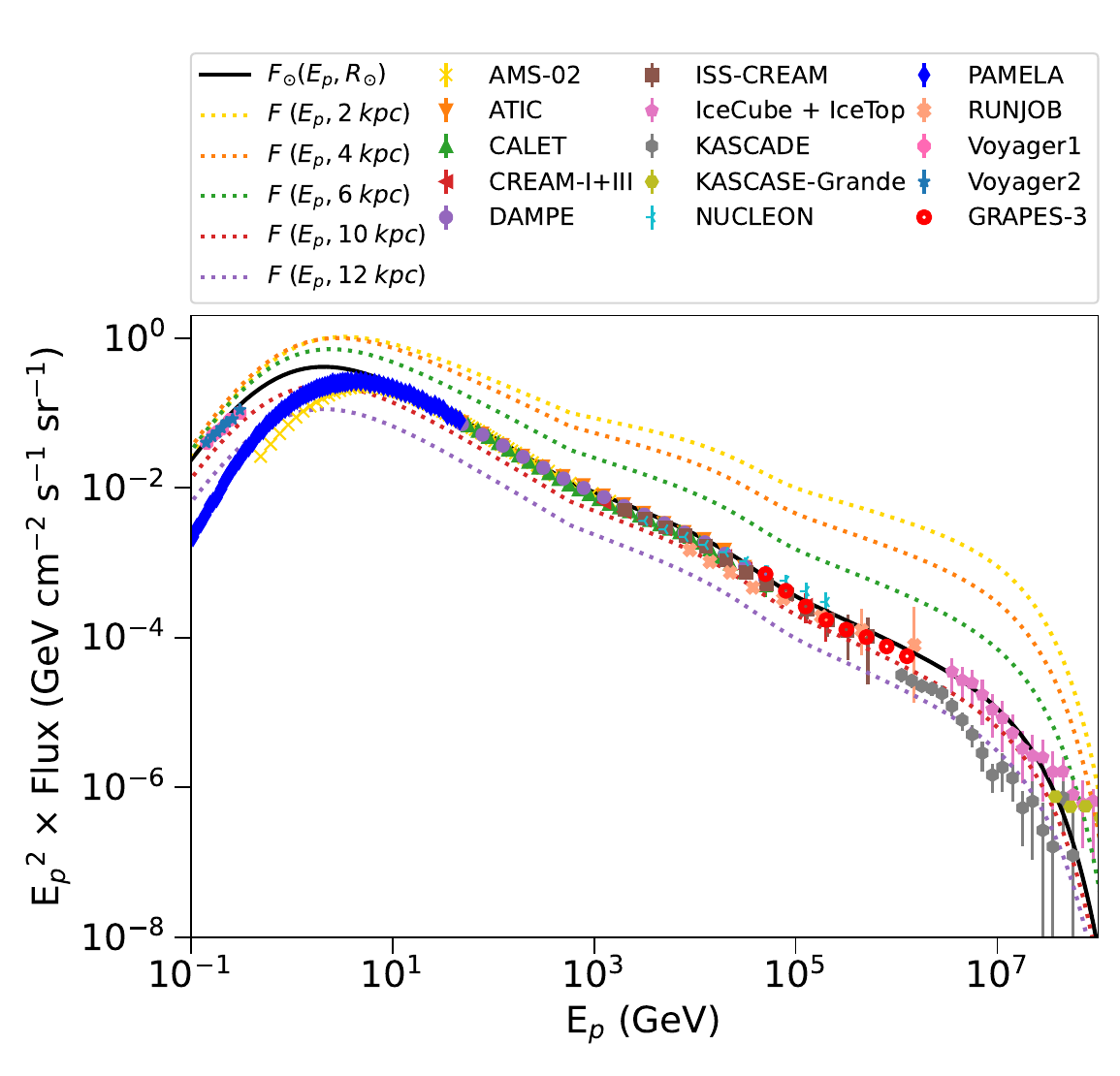}
    \caption{The spectral distribution of GCR protons, from hundreds of MeV to above PeVs, observed by different detectors (AMS-02 (blue points) \citep{2021PhR...894....1A}, ATIC (orange points) \citep{2009BRASP..73..564P}, CALET (green points) \citep{2022PhRvL.129j1102A}, CREAM-I+III (red points) \citep{2017ApJ...839....5Y}, DAMPE (purple points) \citep{2019SciA....5.3793A}, ISS-CREAM (brown points) \citep{2022ApJ...940..107C}, IceCube + IceTop(pink points) \citep{2019PhRvD.100h2002A}, KASCADE (gray points) \citep{2005APh....24....1A}, KASCADE-Grande (olive points) \citep{2017ICRC...35..316A}, NUCLEON (cyan points) \citep{2019AdSpR..64.2546G}, PAMELA (blue points) \citep{2013ApJ...765...91A}, RUNJOB (light salmon points) \citep{2005ApJ...628L..41D}, Voyager1 (hot pink points) \citep{2013Sci...341..150S}, Voyager2 (gold points) \citep{2019NatAs...3.1013S}, GRAPES-3 (red circle) \citep{2024PhRvL.132e1002V}). The black line represents the combined spectral form described in the Eq \ref{equ:LIS}.}
    \label{fig:proton_sed}
\end{figure}

The distribution of GCRs is not very well known throughout the Galaxy; however, their spectrum is expected to be harder in the inner regions of the Galactic plane \citep{2023Sci380338I}. These particles randomly traverse the Galactic volume and interact with ISM, radiation, and magnetic fields \citep{strong2007ARNPS_285S}. These interactions modify the GCR spectral distribution observed near the Earth, ranging over a broader energy range. The spectral distribution of GCR proton observed near the Earth can be approximated by Eq. \ref{equ:LIS} (Fig. \ref{fig:proton_sed}, black solid line), which is a superimposition of two spectral distributions; one E$_p \leq 550$ GeV \cite{vos2015new}, and the other for E$_p > 550$ GeV \cite{2023PhRvL.131o1001C}. Here we consider the extreme GCR proton flux distribution from \cite{2023PhRvL.131o1001C} because the post-LHC hadronic interaction models favoured the extreme data points, including the newly published GCR proton distribution by GRAPES-3 experiment \cite{2024PhRvL.132e1002V}. In fact, the pre-LHC interaction models exhibit a significant difference ranging from 20\% to 50\% in muon production when compared to the post-LHC model \cite{PhysRevD.102.063002}.

\begin{equation}
\begin{aligned}
F_{\odot} (E_p, R_{\odot}) &= N_0 \frac{E_p^{1.12}}{\beta^2} \left( \frac{E_p + E^\prime}{1 + E^\prime}\right)^{-3.93},\;\;E_p \leq 550\;{ \rm GeV}
\\
&= A_1 \left(\frac{E_p}{\rm GeV}\right)^{-B_1} exp\left(-\frac{E_p}{C_1}\right) + A_2 \left(\frac{E_p}{\rm GeV}\right)^{-B_2} exp\left(-\frac{E_p}{C_2}\right), \;\;E_p > 550 \; {\rm GeV}
\end{aligned}
\label{equ:LIS}
\end{equation}
where $E^{\prime} = 0.67$~GeV, $N_0 = 2.7\times10^{-1}$ proton cm$^{-2}$ sr$^{-1}$ s$^{-1}$ GeV$^{-1}$, $A_1$ $\approx 3.81 \times 10^{-2}$ proton cm$^{-2}$ sr$^{-1}$ s$^{-1}$ GeV$^{-1}$, $A_2$ $\approx 3.47 \times 10^{-1}$ proton cm$^{-2}$ sr$^{-1}$ s$^{-1}$ GeV$^{-1}$ are the normalization factors, $E_p$ is the kinetic energy (in GeV) of GCR proton, $\beta$ is particle velocity divided by speed of light, $B_1$ = 2.35, $B_2$ = 2.6 are the spectral index and $C_1$ = $25\times10^3$ GeV, $C_2$ = $15\times10^6$ GeV (i.e., 15 PeV) are the exponential cutoff energy in the power-law distribution. 

The investigations of GCRs and their secondary radiation have provided the clue that GCR distribution in the MW is not constant, as discussed by \cite{2013APh42.70E, 2014AnA66A142Y, 2016ApJS..223...26A,2018PhRvDd3003L}. These authors reported a dependence of GCR proton density on the galactocentric distance ($R_{\rm gal}$), showing that GCR proton spectral index near the Galactic centre is harder compared to the one observed in the Solar neighbourhood. Following the methodology defined in \cite{pagliaroli2016expectations, cataldo2019probing}, we also consider a position-dependent GCR distribution at different locations in the MW Galaxy. The variation of GCR spectral distribution with the $R_{\rm gal}$ can be approximated by the Eq. \ref{equ:GCRIS}:

\begin{align}
F (E_p, R_{\rm gal}) &=  F_{\odot}(E_p, R_{\odot}) \times S(R_{\rm gal}) \times H(E_p,R_{\rm gal})
\label{equ:GCRIS}
\end{align}

\begin{align}
H(E_p,R_{\rm gal}) = \left(\frac{E_p}{20\;\rm GeV}\right)^{\Delta(R_{\rm gal})}
, \;\; {\rm where} \;\;
\Delta(R_{\rm gal}) = 0.3\left( 1 - \frac{R_{\rm gal}}{R_{\odot}}\right)
\label{equ:SpecIdxRgal}
\end{align}

and $F_{\odot}(E_p, R_{\odot})$ is the observed GCR distribution near the Earth (i.e. at the solar neighbourhood, $R_{\odot} = 8.34$ kpc), $S(R_{\rm gal})$ is the observed SNR rate normalised to the rate at the solar neighbourhood, $H(E_p,R_{\rm gal})$ is the function for position-dependent spectral index variation, dependent on particle energy and galactocentric distances.

In this work, we assume that SNRs are the GCR sources \cite{evoli2017cosmic}, and we consider their distribution along $R_{\rm gal}$, parameterised by \cite{case1998new, green2015constraints}. Since we are mainly interested in the calculation of diffused emission from the Galactic plane, we neglect the dependency of both $S(R_{gal})$ and $H(E_p, R_{\rm gal})$ on the z-direction. The value of $H(E_p, R_{\rm gal})$ is normalised at 20 GeV because, above that energy, the integrated CR proton density roughly follows the GCR source (SNR) distribution $S(R_{\rm gal})$ \cite{recchia2016radial, 2016ApJS..223...26A}. The hardening in the GCR proton spectral index near the Galactic centre, as observed by Fermi-LAT above 20 GeV \cite{2016ApJS..223...26A, yang2016radial}, is approximated by $\Delta(R_{\rm gal})$, that represents the deviation in the spectral index of GCR particles along the $R_{\rm gal}$ from the observed one at $R_{\odot}$ (i.e., $\sim$ 2.7). It is also worth noting that, to our knowledge, observations at PeV energies or above remain scarce. Moreover, we recommend that readers refer to the \cite{schwefer2023ApJ6S} for a thorough discussion on the uncertainty due to CR source distribution, gas maps, cross-section, etc. Their study provides an in-depth examination of the complexities and challenges in accurately characterizing and quantifying the parameters that influence CR propagation, interaction and their subsequent emission. In our calculation of diffuse gamma-ray and neutrino fluxes from GMCs, we consider two different cases of GCR distribution: `Case I' --- we consider the observed GCR flux near the solar neighbourhood, i.e., $F_{\odot}(E_p, R_{\odot})$ at $R_{\odot}$, constant in the whole Galaxy (the black solid line in Fig. \ref{fig:proton_sed}). `Case-II' --- more realistic, takes into account a position-dependent GCR flux model based on observations.

\section{Modelling of Gamma-Ray and Neutrino Fluxes from GCR Interactions with GMCs}
\label{sec:modelling}

Gamma-ray flux, $F_{\gamma} (E_{\gamma})$,  and neutrino flux, $F_{\nu} (E_{\nu})$, emitted from the interaction of GCR with GMCs, having mass $M$ and located at a distance $d$, can be calculated as

\begin{equation}
\begin{aligned}
F_{\gamma, \nu} (E_{\gamma, \nu}) &= \frac{M}{d^2} \frac{\xi_N}{m_p} \int \frac{d \sigma (E_{\gamma,\nu}, E_p)}{dE_{\gamma, \nu}} F (E_p, R_{gal}) \;dE_p \\
&\simeq 1.25\times10^{19}\; \mathcal{A}\;\xi_N \int \frac{d \sigma (E_{\gamma,\nu}, E_p)}{dE_{\gamma, \nu}} F (E_p, R_{gal}) \;dE_p ,
\end{aligned}
\label{equ:GammaNuFlux}
\end{equation}

where $F (E_p, R_{gal})$ is the flux of GCR protons incident on the surface of GMCs from outside, $\frac{d \sigma (E_{\gamma,\nu}, E_p)}{dE_{\gamma, \nu}}$ is the differential cross-section of $\gamma$-ray and $\nu$ production from $p - p$ inelastic interactions \cite{kafexhiu2014parametrization, kelner2006energy}, $\mathcal{A} = M_5/d_{\rm kpc}^2$, $M_5$ = $M/10^5 M_\odot$ and $\xi_N$ here represents the nuclear enhancement factor (NEF), used to incorporate contributions from the heavy elements presents in both GCRs and GMCs. The neutrino multiplicity parameters are given in \cite{kelner2006energy}.

\begin{figure}[!ht]
    \centering
    \includegraphics[width=0.65\linewidth]{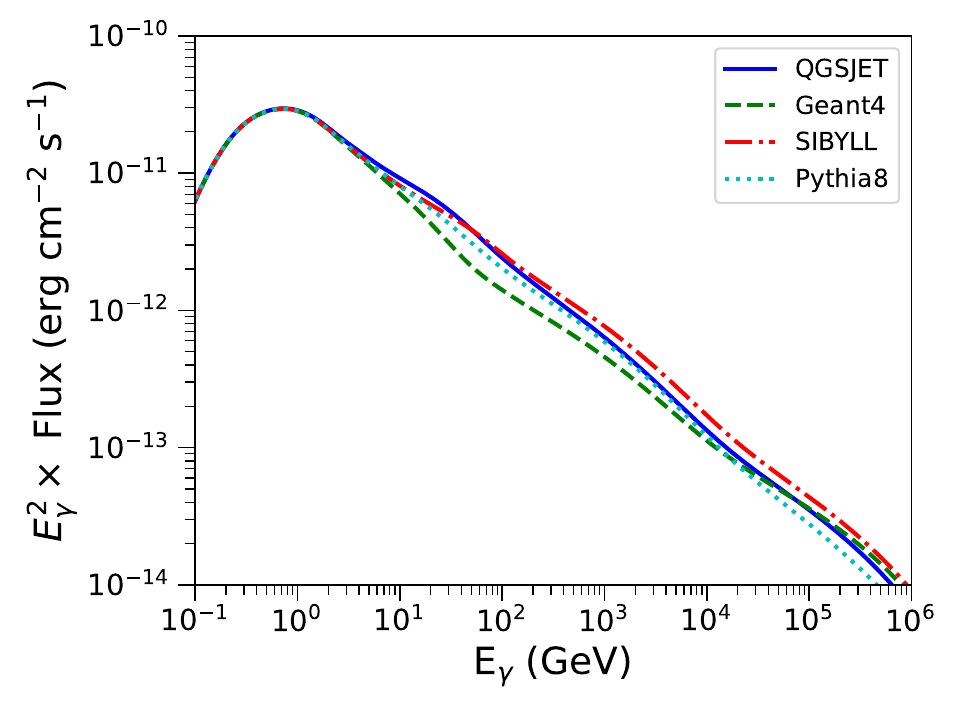}
    \caption{The gamma-ray flux resulting from the interaction of Case-I type GCR flux with a GMC having parameter $\mathcal{A} = 1$, taking into account different physical interaction models.}
    \label{fig:intModels}
\end{figure}

The possible variation of GCR flux due to different positions of GMCs in the Galaxy, $F (E_p, R_{\rm gal})$, is calculated taking into account the GCR source distribution (see Sec. \ref{Sec:ProtonDist} for details). The resulting gamma-ray flux can also vary with the interaction model of $p - p$ collision considered, as shown in Fig. \ref{fig:intModels} and described in \cite{kafexhiu2014parametrization}. The fluxes have a little different behaviour above 10 GeV. Throughout this paper, we decided to use the \textit{SIBYLL} \cite{engel1992nucleus, fletcher1994s, ahn2009cosmic} model because it has the maximum gamma-ray flux compared to the other model contributions, allowing us to optimize gamma-ray flux from GMCs. 

The value of NEF, $\xi_N$, is actually energy-dependent \cite{kafexhiu2014parametrization, kachelriess2014nuclear, mori2009nuclear, peron2022A57P,2011A&A...531A..37D}. Moreover, the NEF is also affected by the composition of CRs and ISM. As such, it fluctuates throughout the Galaxy and is not constant in energy. For example, \citep{peron2022A57P} has reported a value 1.8 at 10 GeV and that increases to 2.6 at 1 PeV. Also, discerning the effects of a change in the ISM or CR density on this factor can be challenging, as the two are intertwined and difficult to untangle \cite{2022A&A...661A..72B}. For all these difficulties, fixed value is generally used in the past \cite{kachelriess2014nuclear,mori2009nuclear, peron2022A57P,2014PhRvD..90l2007A, bagh20, aharonian2020probing}. Several authors have calculated NEF independently, showing variation in its value in a range from 1.45 to 2.1 \cite{kachelriess2014nuclear}. The reason for this variation is strictly correlated with the use of different $p - p$ interaction models, heavy nuclei abundance values, spectral shape of GCR particles, etc. In our work, we used a constant NEF, ($\xi_N$) = 2.09, based on the value reported by \cite{kachelriess2014nuclear} at a photon energy $10^3$ GeV (or $\sim 10^4$ GeV proton energy). The value of NEF may vary w.r.t. energies. Its value is lower at 10 GeV (10.05\% lower) and 100 GeV (3.35\% lower) in comparison to the NEF value, 2.09 at $10^3$ GeV \citep{kachelriess2014nuclear}. Further, at $10^6$ GeV (i.e., 1 PeV), the NEF value is higher, and its value is 19.62\% larger \citep{peron2022A57P}. This energy dependence of the NEF value would, in turn, introduce some uncertainty into our flux estimation.



\section{Results and Discussions}
\label{Sec:results}

As anticipated in Sec. \ref{Sec:ProtonDist}, we have tested two scenarios for the GCR distribution: Case-I and Case-II. In the first case, we consider a constant GCR distribution equal to the one measured in the solar neighbour, instead in the second case, we consider a radial dependent distribution. For both cases, we calculated the differential energy spectrum of gamma-rays and neutrinos for all the 8444 GMCs, following Eq. \ref{equ:GammaNuFlux}. 

\begin{figure}[!ht]
\centering
\subfigure{\includegraphics[width=0.7\linewidth]{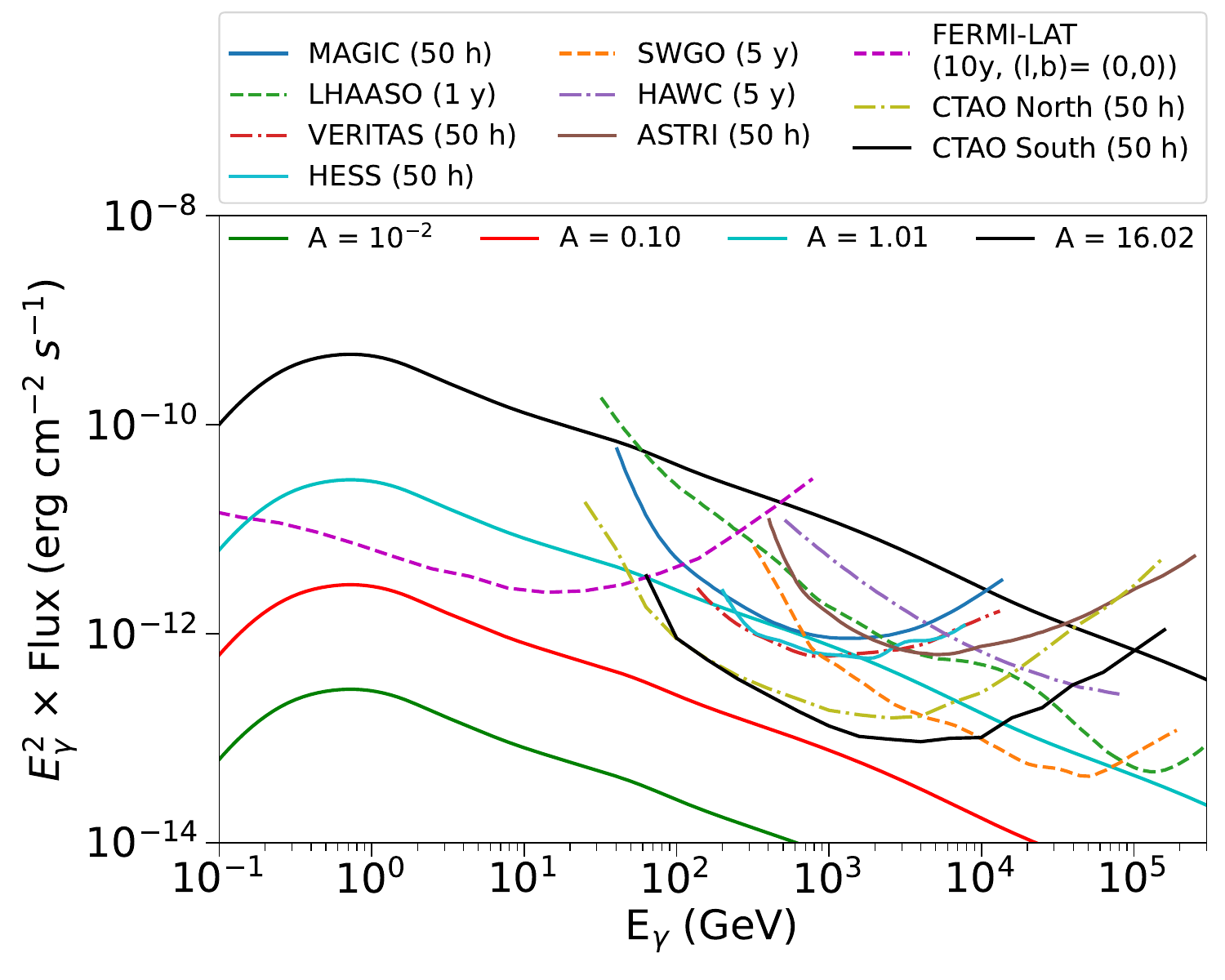}}
\subfigure{\includegraphics[width=0.7\linewidth]{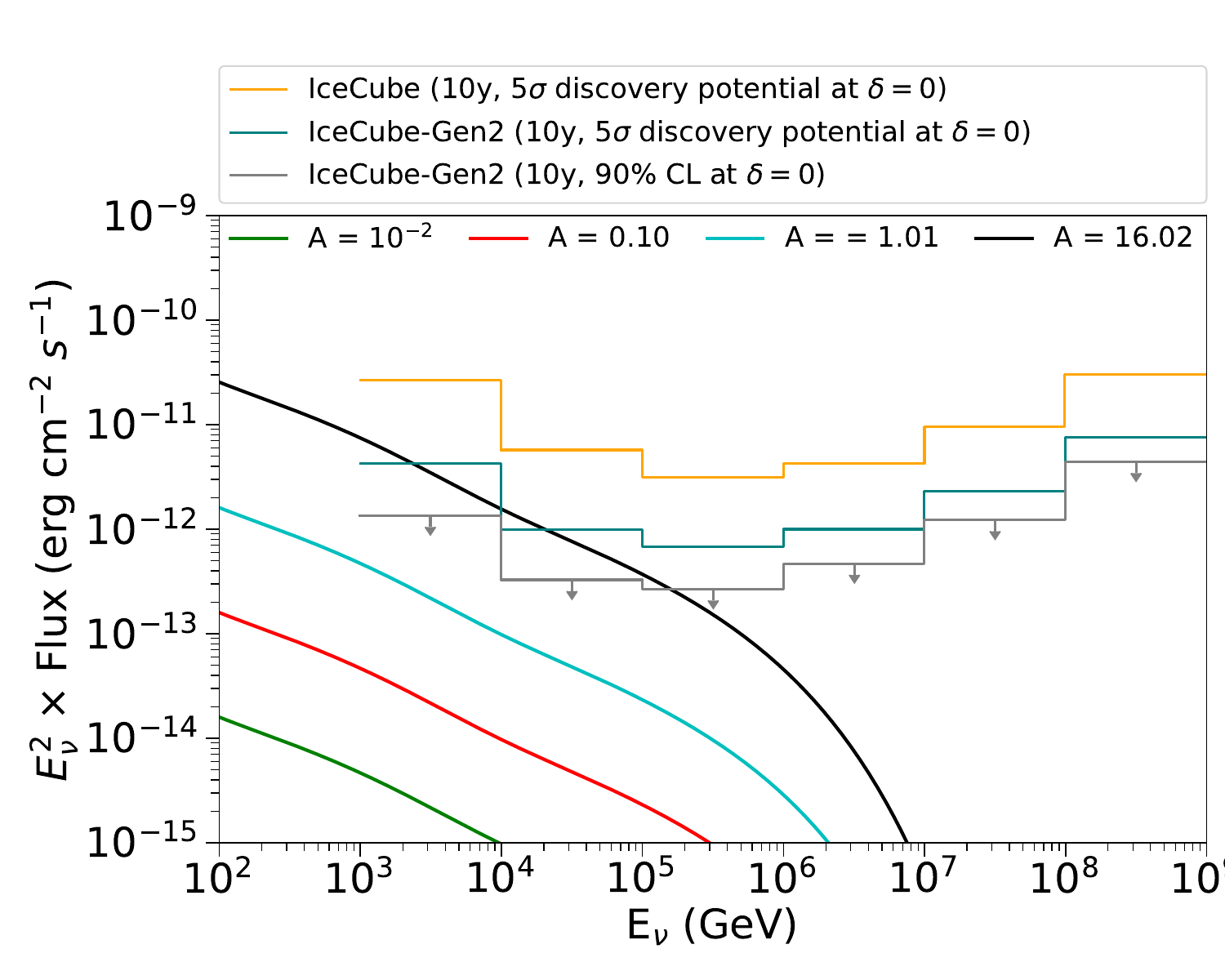}}
\caption{{\it Top panel}: Gamma-ray flux from a selected individual clouds in the combined GMC catalog. The sensitivity limits of different gamma-ray detectors are also depicted here by different coloured lines \cite{gammaSens,fermiSens}. {\it Bottom panel}: Similarly, the neutrino emission ($\nu + \bar{\nu}$: all flavor) from individual GMCs having unique $\mathcal{A}$ values are depicted here with the sensitivity limits of IceCube detectors \cite{aartsen2021icecube}. The sensitivities for IceCube-Gen2 pertain solely to the sensitivity of the optical array.}
\label{fig:fluxVSsensitivity}
\end{figure}

\begin{figure}[!ht]
    \centering
    \subfigure{\includegraphics[width=0.45\linewidth]{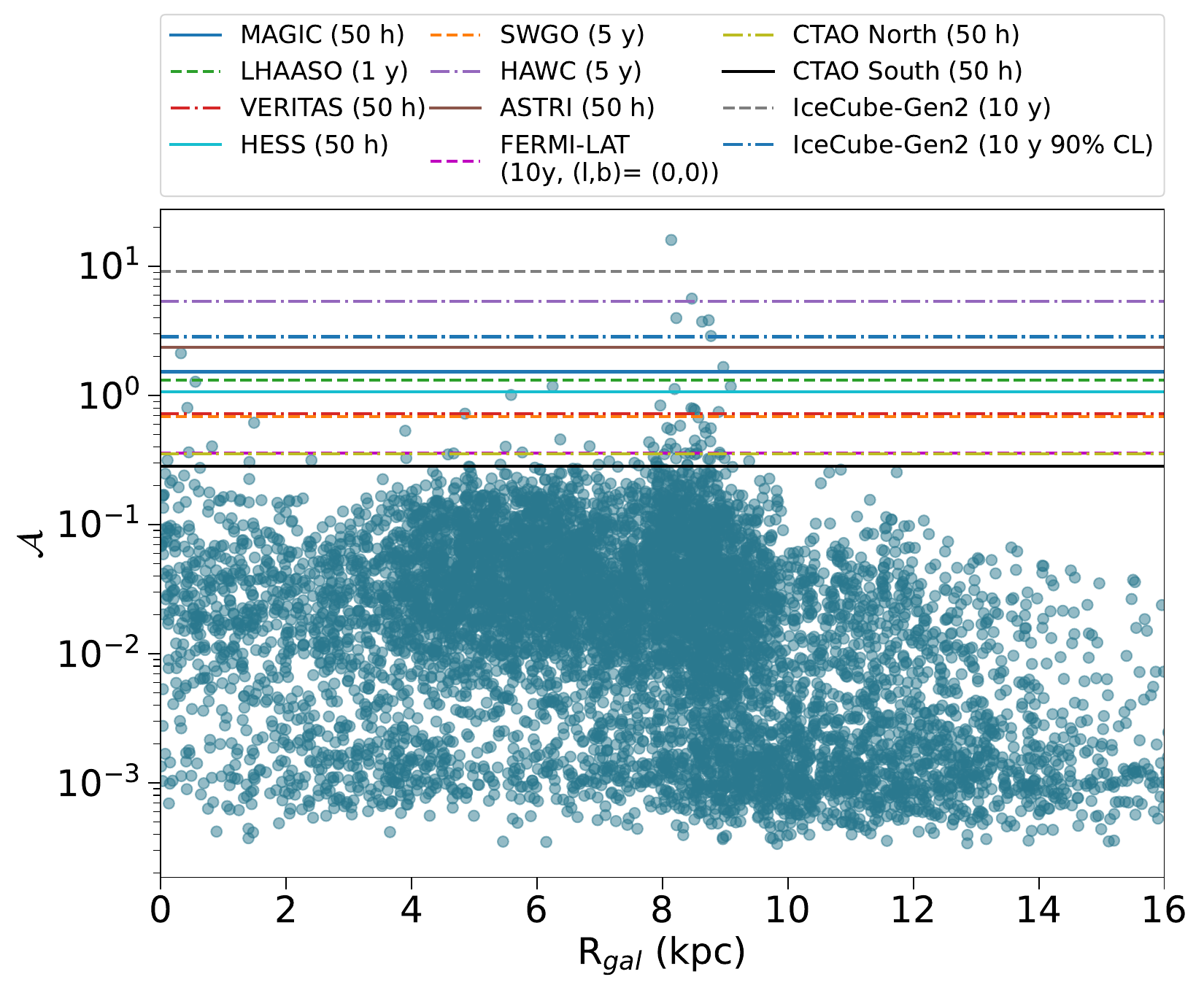}}
    \subfigure{\includegraphics[width=0.45\linewidth]{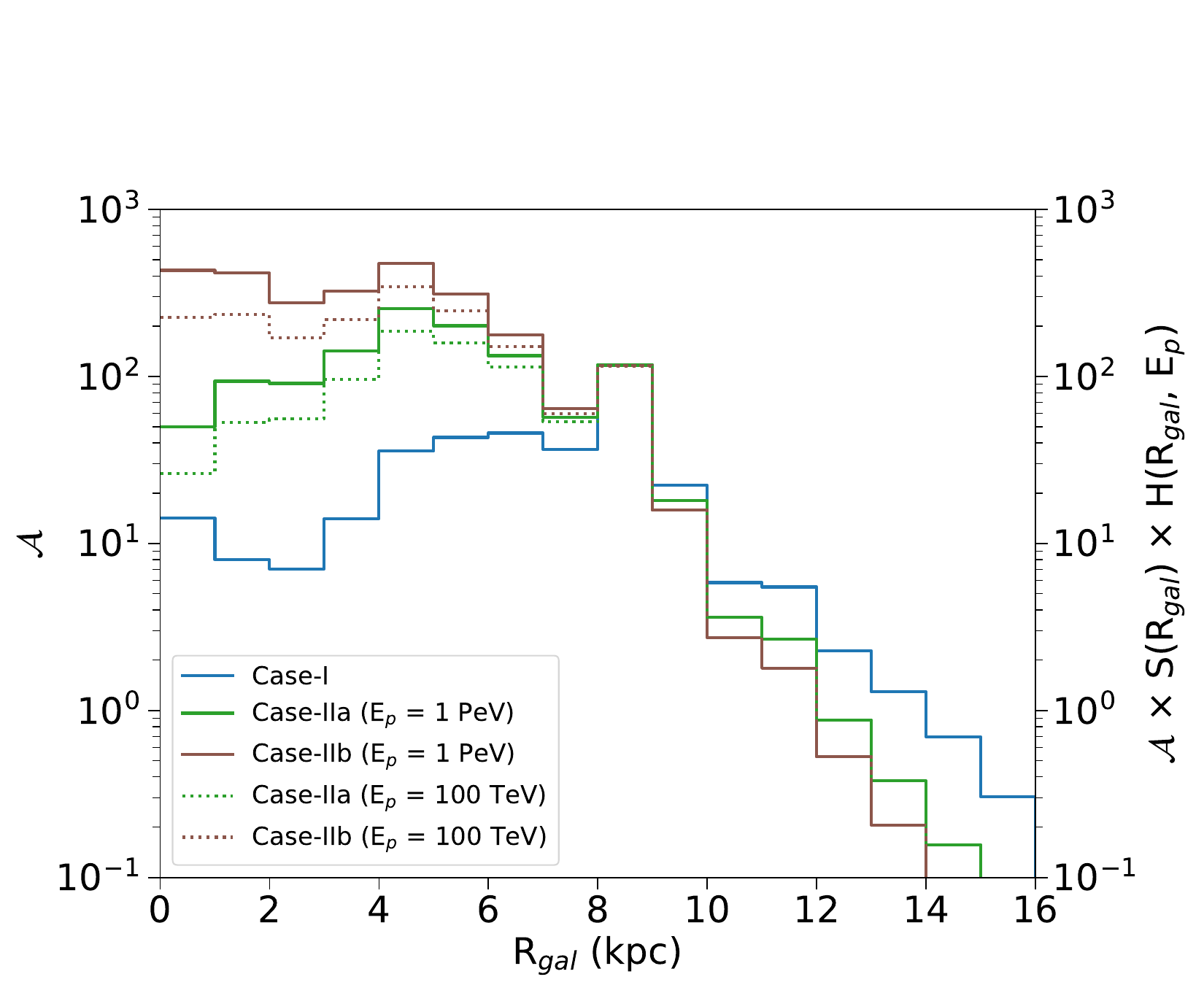}}
    \caption{{\it Left panel}: The $\mathcal{A}$ parameters distribution along the galactocentric radius and the minimum detection limit of current and future generation detectors. {\it Right panel}:  The variation of $\mathcal{A}$ and  $\mathcal{A} \times S(R_{\rm gal}) \times H(R_{\rm gal}$, $E_p$) as a histogram along the galactocentric radius with a constant bin width of 1 kpc.}
    \label{fig:MinADetectionLimit}
\end{figure}

\begin{figure}[!ht]
\centering
\subfigure{\includegraphics[width=0.7\linewidth]{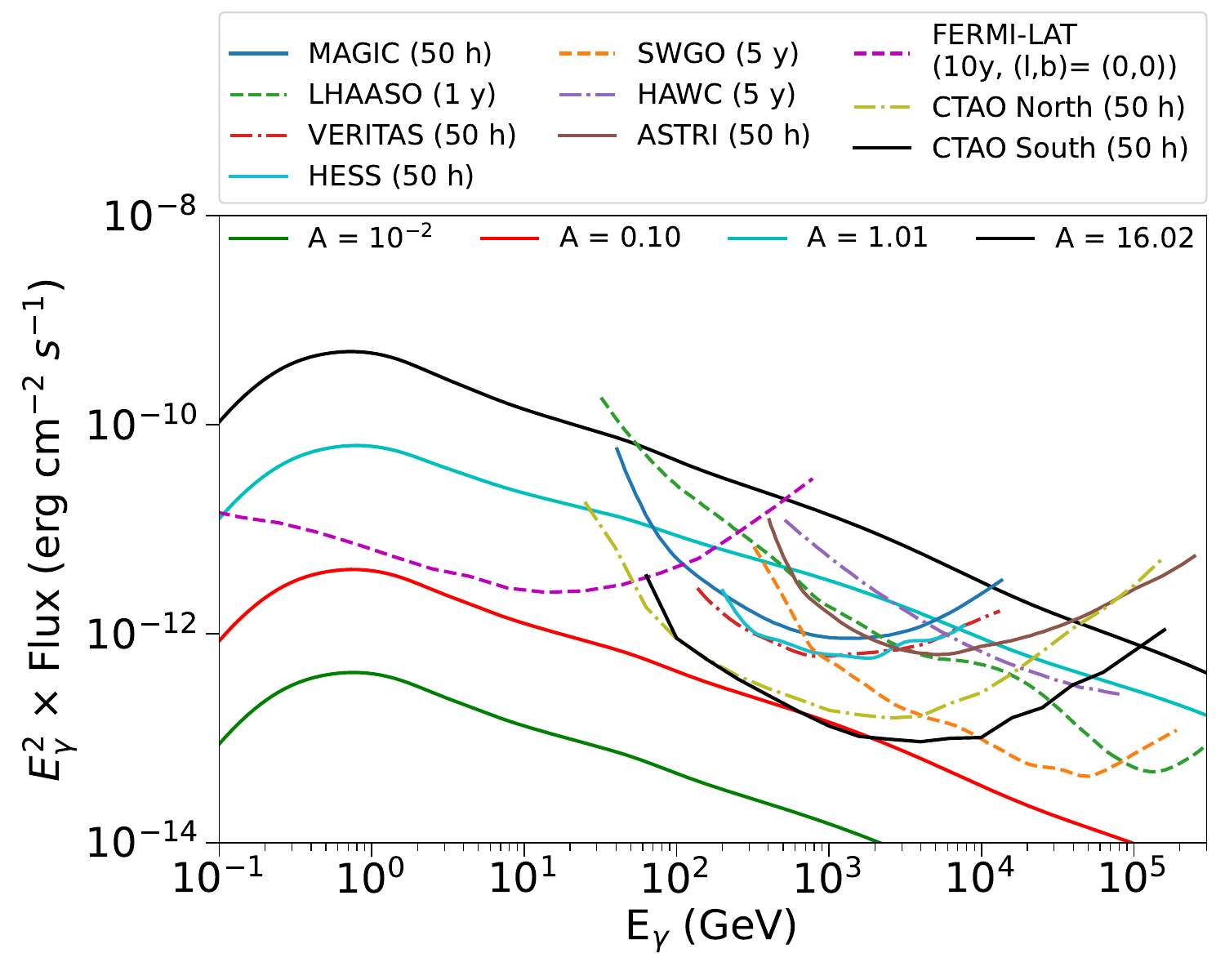}}
\subfigure{\includegraphics[width=0.7\linewidth]{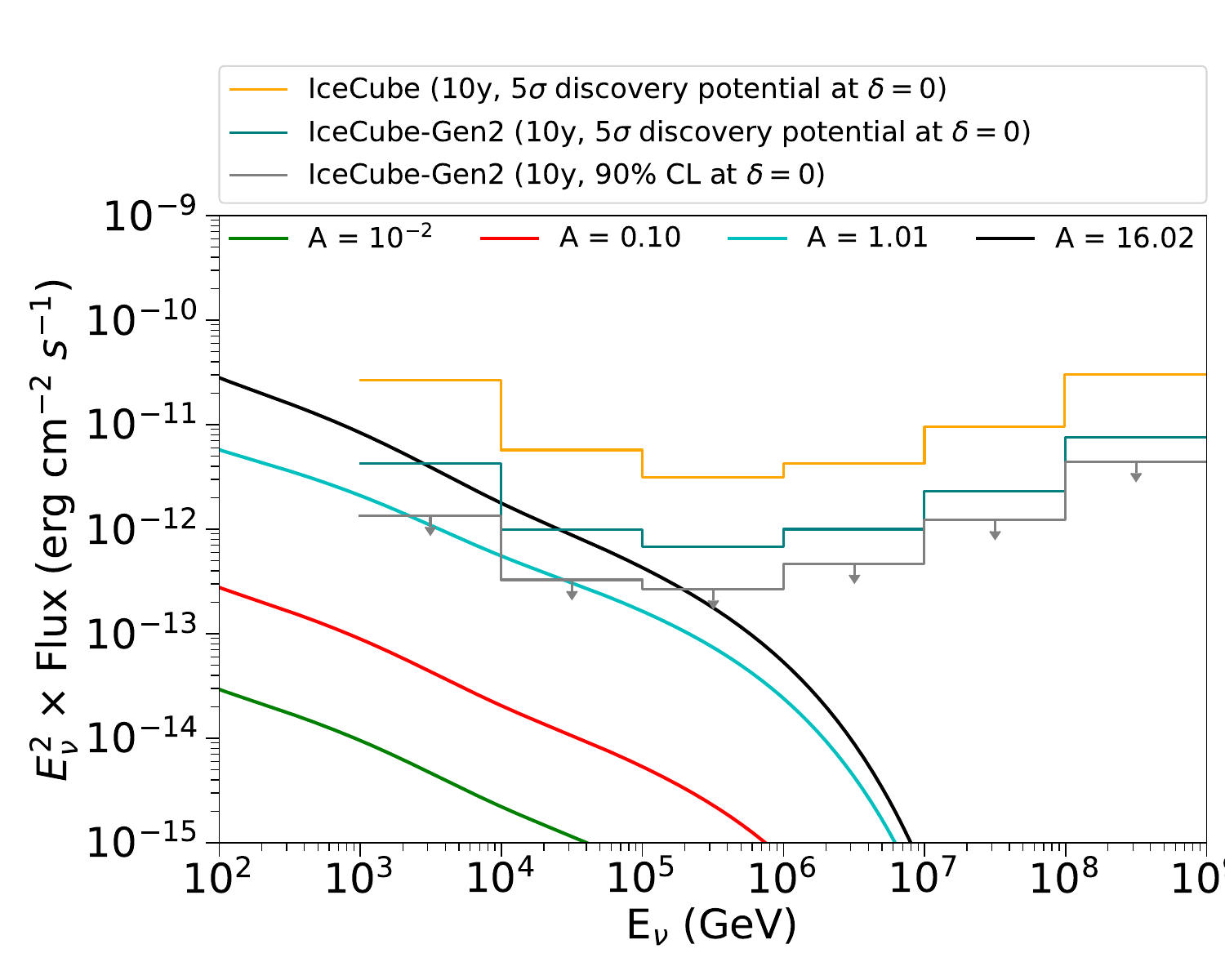}}
\caption{{\it Top panel}: The gamma-ray flux for GMCs having unique $\mathcal{A}$ parameters in the combined cloud catalog as indicated by the coloured lines. The sensitivity limit of different gamma-ray detectors is also displayed by different coloured lines \cite{gammaSens,fermiSens}. {\it Bottom panel}: The neutrino flux ($\nu + \bar{\nu}$: all flavor) from the same set of GMCs, along with the sensitivity limits of the IceCube detector is depicted here \cite{aartsen2021icecube}. The sensitivities for IceCube-Gen2 pertain solely to the sensitivity of the optical array.}
\label{fig:fluxVSsensitivityCaseII}
\end{figure}

{\bf Case-I:} 
In this scenario, CR flux is assumed to be the same in the whole Galaxy (see Sec. \ref{Sec:ProtonDist} and Fig. \ref{fig:proton_sed} for details), implying that multimessenger emission due to GCR/GMC interaction depends only on $\mathcal{A}$. Gamma-ray and neutrino fluxes resulting from hadronic interactions for each of the individual clouds are then calculated using Eq. \ref{equ:GammaNuFlux}, and the results for a few GMCs are depicted in Fig. \ref{fig:fluxVSsensitivity}. The same figure also shows the sensitivity limits of various current and future-generation gamma-ray and neutrino detectors for reference. It is evident that only a fraction of GMCs can be detected for Case-I. These clouds have a larger $\mathcal{A}$ parameter value, implying that they are either massive in size and/or located relatively close to us. However, in the interpretation of these results, it is important to keep in mind that the angular area of GMCs can vary significantly (See Fig. \ref{fig:AngArea}), and that they are located in different parts of the Galaxy. We calculated the sensitivity degradation that occurs when the source extension changes from 0.1 to 2 degrees. The available sensitivity data for CTA and KM3NeT telescopes \cite{ambrogi2018potential,aharonian2020probing} indicates a maximum decrease in sensitivity of approximately 128 and 14 times, respectively, when the source extension changes from 0.1 to 2 degrees. Hence, the sensitivity limits of detectors shown in Fig. \ref{fig:fluxVSsensitivity} and \ref{fig:fluxVSsensitivityCaseII} can vary consequently.
The parameter $\mathcal{A}$ w.r.t. the galactocentric radius $R_{\rm gal}$ for all GMCs is plotted in Fig. \ref{fig:MinADetectionLimit} (left panel). Here, the horizontal lines indicate the minimum detection limits of GMCs by several gamma-ray and neutrino detectors. The minimum value of $\mathcal{A}$ observable by a specific instrument is calculated by dividing its sensitivity limit with the corresponding gamma-ray or neutrino flux at $\mathcal{A} = 1$. Among all the GMCs considered, only a few have $\mathcal{A} > 1$ values and are located close to the Galactic centre or to the Solar system, as can be seen from Fig. \ref{fig:MinADetectionLimit} (left panel). The right panel of Fig. \ref{fig:MinADetectionLimit} shows the variation of $\mathcal{A}$ on left side of the y-axis and  $\mathcal{A} \times S(R_{\rm gal}) \times H(R_{\rm gal}$, $E_p$) on right side of the y-axis as histograms with 1 kpc binning. It reaches its maximum value between 8 - 9 kpc (blue line), close to the Sun. So, for Case-I, where the emission from GMCs is proportional to this parameter, $\mathcal{A}$ will be higher in correspondence with this region consequently resulting in a more locally generated flux. 


{\bf Case-II:}
In this scenario, multimessenger emission due to GCR/GMC interaction depends on GCR source distribution $S(R_{gal})$ and on the position-dependent spectral index variation function $H(E_p,R_{gal})$, both motivated by observations (see Sec. \ref{Sec:ProtonDist} and Eq. \ref{equ:GCRIS}, \ref{equ:SpecIdxRgal}, \ref{equ:GammaNuFlux} for details). Further, here we also study the effects of the SNR distribution based on two different catalogs. We distinguish a Case-IIa, where we used their distribution from \citep{case1998new}, and a Case-IIb, where an updated SNR distribution based on a more recent observation of SNR distribution and has a larger sample size\citep{green2015constraints} is included. The source distribution $S(R_{gal})$ reaches its maximum value at $R_{gal}$ around $\sim$ 2 -- 3 kpc and then it gradually decreases \cite{green2015constraints}. The factor $S(R_{gal}) \times H(R_{gal}, E_p)$ reaches its maximum value of about $\sim$ 53 at $R_{gal} = 1.3$ kpc for $E_p = 1$ PeV, implying that GCR flux at this location will be greater than the one around the Sun ($F_\odot (E_p, R_\odot)$) by this factor. Gamma-ray and neutrino emissions from all individual GMCs taken into account are calculated for this radially dependent GCR flux (i.e., $F (E_p, R_{\rm gal})$) variation model. The flux for a few GMCs are shown in Fig. \ref{fig:fluxVSsensitivityCaseII} (considering Case-IIb type GCR flux distribution), together with the sensitivity curves of various detectors. The fluxes are relatively harder than those obtained in Case-I, and the flux levels of the GMCs are also different and dependent on the GMC location. 

\subsection{Implication for the IceCube observations}
\label{Subsec: Icecube}

\begin{figure}[!ht]
\centering
\subfigure{\includegraphics[width=0.7\linewidth]{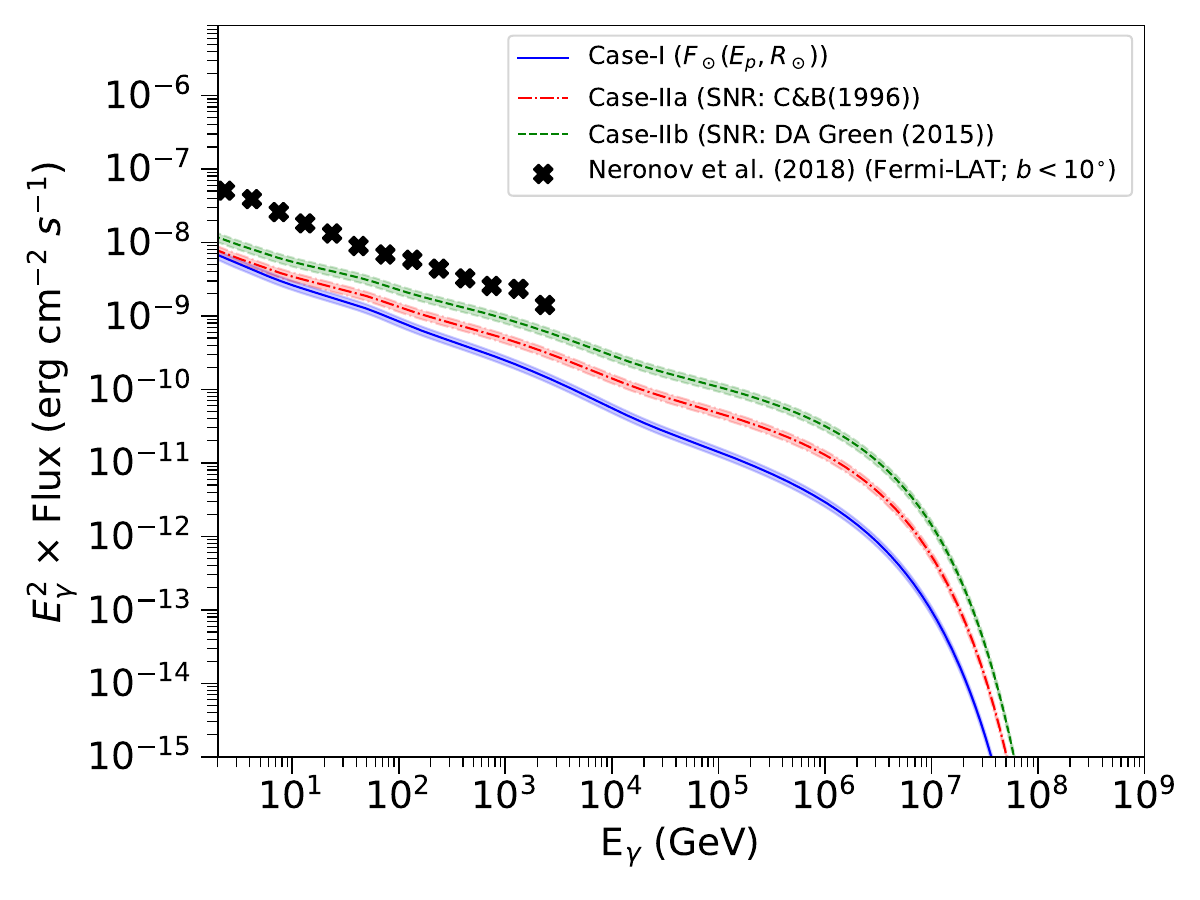}}
\subfigure{\includegraphics[width=0.7\linewidth]{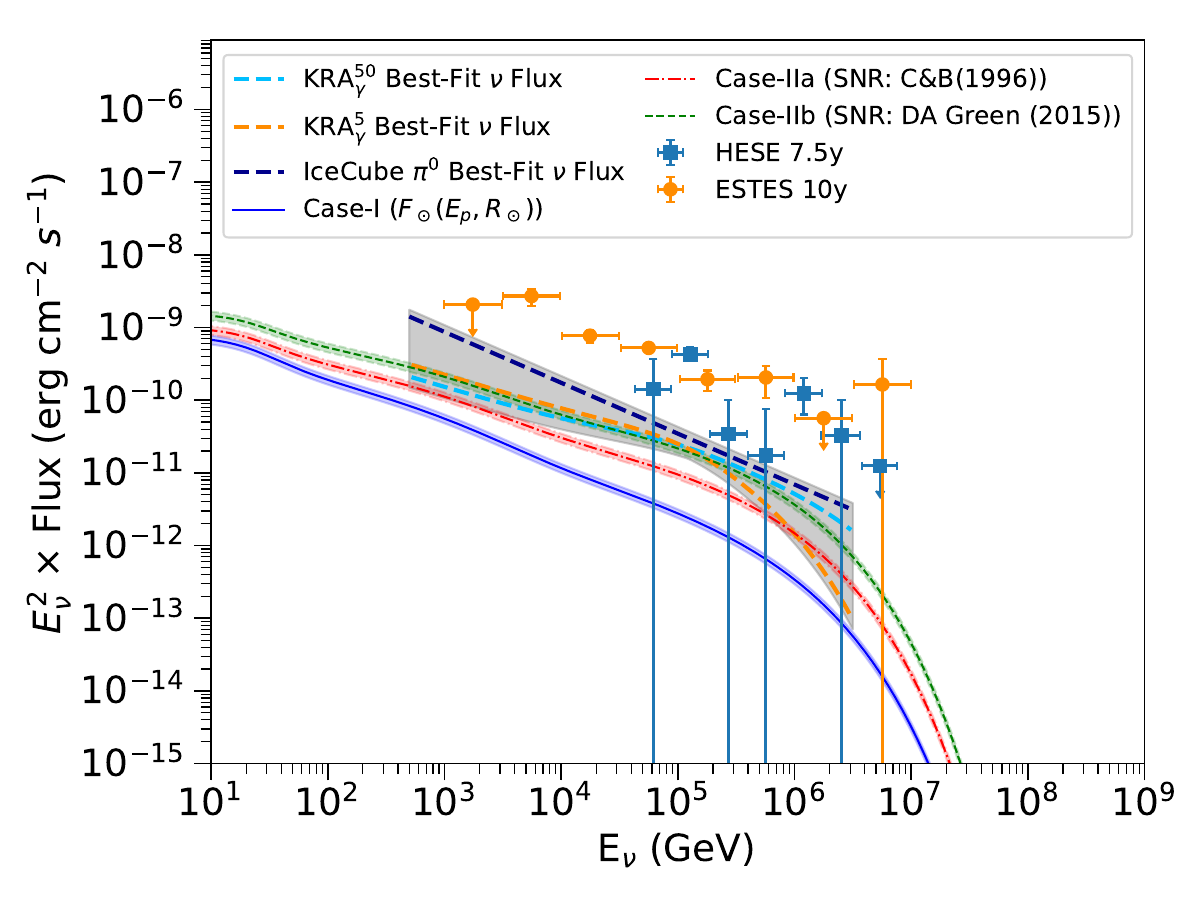}}
\caption{{\it Top panel}: The diffuse gamma-ray flux observed by the Fermi-LAT within the Galactic latitude b $< 10^{\circ}$ \cite{neronov2018multimessenger} along with the flux calculated by combining the contributions from all the individual GMCs located in that region is depicted here. The data points were extracted from Fig. 1 in \cite{neronov2018multimessenger}. {\it Bottom panel}: The diffuse neutrino flux ($\nu + \bar{\nu}$: per flavor) from all the GMCs stacked together in the Galactic plane along with the best-fitted flux of $\pi^0$ emission template and KRA$_{\gamma}$ models is shown here. The shaded region indicates the superimposition between the best-fitted flux of $\pi^0$ emission template, KRA$_{\gamma}^{5}$ and KRA$_{\gamma}^{50}$ models along with their $1\sigma$ uncertainty. The total all-sky astrophysical neutrino flux for starting track-like events \cite{silva2023measurement, abbasi2021icecube} (interacting within a fiducial volume of the detector) is also plotted here for comparison. The data points were extracted from Fig. 5 in \cite{silva2023measurement}.}
\label{fig:DiffFluxStacked}
\end{figure}

Here, we discuss the total gamma-ray and neutrino contributions from the GMC population and compare them with the IceCube observations. The stacked diffuse gamma-ray and neutrino fluxes from the Galactic plane are shown in Fig. \ref{fig:DiffFluxStacked}. On the top panel, the gamma-ray flux observed by Fermi-LAT from the Galactic plane at Galactic latitude $b < 10^\circ$ \cite{neronov2018multimessenger} is shown. We have also included Tibet-AS$\gamma$, ARGO-YBJ, CASA-MIA, and LHAASO gamma-ray observations at different Galactic longitudes and latitudes. These curves are shown in the Appendix of Fig. \ref{fig:TibetARGO} and \ref{fig:LHAASOFlux}, compared with the predictions of our model. The stacked gamma-ray flux from all the clouds located in this region is also displayed for the different cases considered here. 
The bottom panel, instead, portrayed best-fitting fluxes of $\pi^0$ and KRA$_\gamma$ models for neutrino emission observed by the IceCube collaboration \cite{2023Sci380338I}. The $\pi^0$ emission template is based on the Galactic diffuse gamma-ray emission model of Fermi-LAT \cite{ackermann2012fermi}, calculated using GALPROP code \cite{strong1998propagation}. The KRA$_\gamma$ model \cite{gaggero2015gamma} employs a radially dependent diffusion coefficient in the CR transportation code DRAGON \cite{evoli2008cosmic}. 

The blue, red and green curves in Fig. \ref{fig:DiffFluxStacked} are our model predictions for Case-I, Case-IIa and Case-IIb, respectively, and the shaded region indicates 14\% uncertainty in estimating the $\mathcal{A}$ parameter (due to the relatively small uncertainty and large Y-axis range, it is not visible in Fig. \ref{fig:DiffFluxStacked}, but can be seen in Fig. \ref{fig:TibetARGO} and \ref{fig:LHAASOFlux}, as discussed in the Appendix). The contribution of stacked gamma-ray and neutrino fluxes for Case-I is the least while Case-IIb contributes the most. The predictions result in a significant contribution to the observation without oversaturating it. In Fig. \ref{fig:DiffFluxStacked}, we plotted the per-flavor neutrino flux ($\nu + \bar{\nu}$) (assuming the flavor ratio at the Earth is roughly $\nu_e : \nu_{\mu} : \nu_{\tau} = 1 : 1 : 1$ due to neutrino oscillations) calculated using this model that is consistent with the best-fitted KRA$_\gamma^{5}$ model. This suggests that neutrinos observed by the IceCube detector might have a substantial contribution from GMCs. However, the fluxes are almost a factor $\sim$ 2 lower than both Fermi-LAT observed gamma-ray flux \cite{neronov2018multimessenger} and KRA$_\gamma$ model predicted neutrino flux level \cite{2023Sci380338I} (see Fig. 5 of \cite{2023Sci380338I}). These excesses could be due to GCR interaction with diffuse gas in the ISM mostly composed of atomic hydrogen, or to some unresolved sources \cite{fang2023milky, ackermann2012fermi,2023ApJ...957L...6F}, or again to any exotic physics such as dark matter decay~\cite{Fiorillo:2023clw,zuriaga2023multi} which are not taken into account here. Additionally, the low energy gamma-ray flux below 30 GeV could also be due to electromagnetic emission mechanisms, such as bremsstrahlung or inverse Compton interactions \cite{gaggero2015gamma, ackermann2012fermi} (see Sec. \ref{sec:appa}, for comparison with VHE gamma-ray observations). The total astrophysical neutrino flux averaged over the whole sky for starting track-like events interacting within the fiducial region of the detector multiplied by $4\pi$ (coming from per steradian unit conversion) is shown in the \textit{bottom panel} of Fig. \ref{fig:DiffFluxStacked}. In this plot, we have also compared the contribution of the Galactic GMCs towards the total astrophysical neutrino flux. \cite{2023Sci380338I} estimated that the MW Galactic plane contributions may vary from 6 to 13\% of the total astrophysical neutrino flux at 30 TeV, depending on the emission model considered.



\begin{figure}[!ht]
\centering
\subfigure{\includegraphics[width=0.7\linewidth]{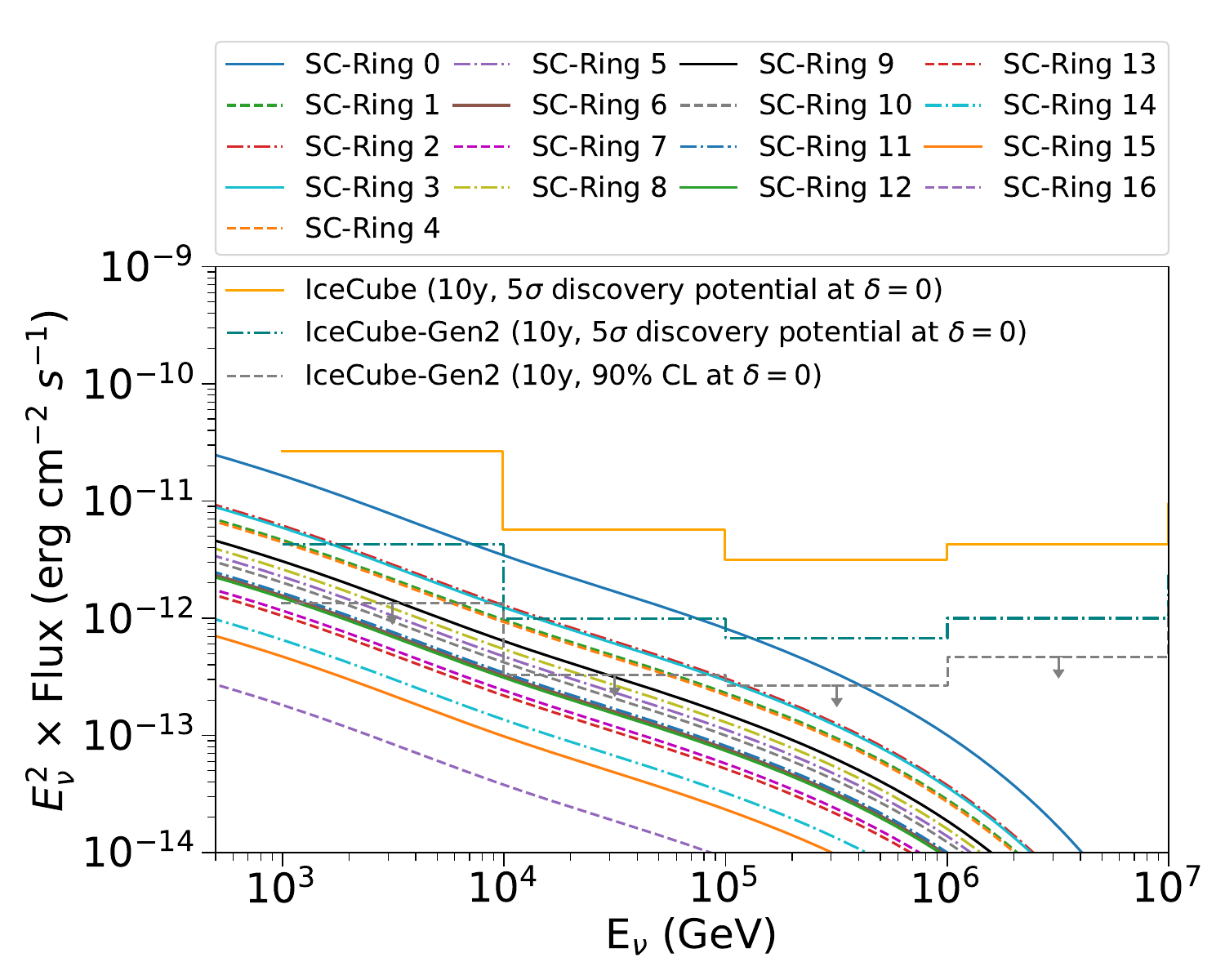}}
\subfigure{\includegraphics[width=0.7\linewidth]{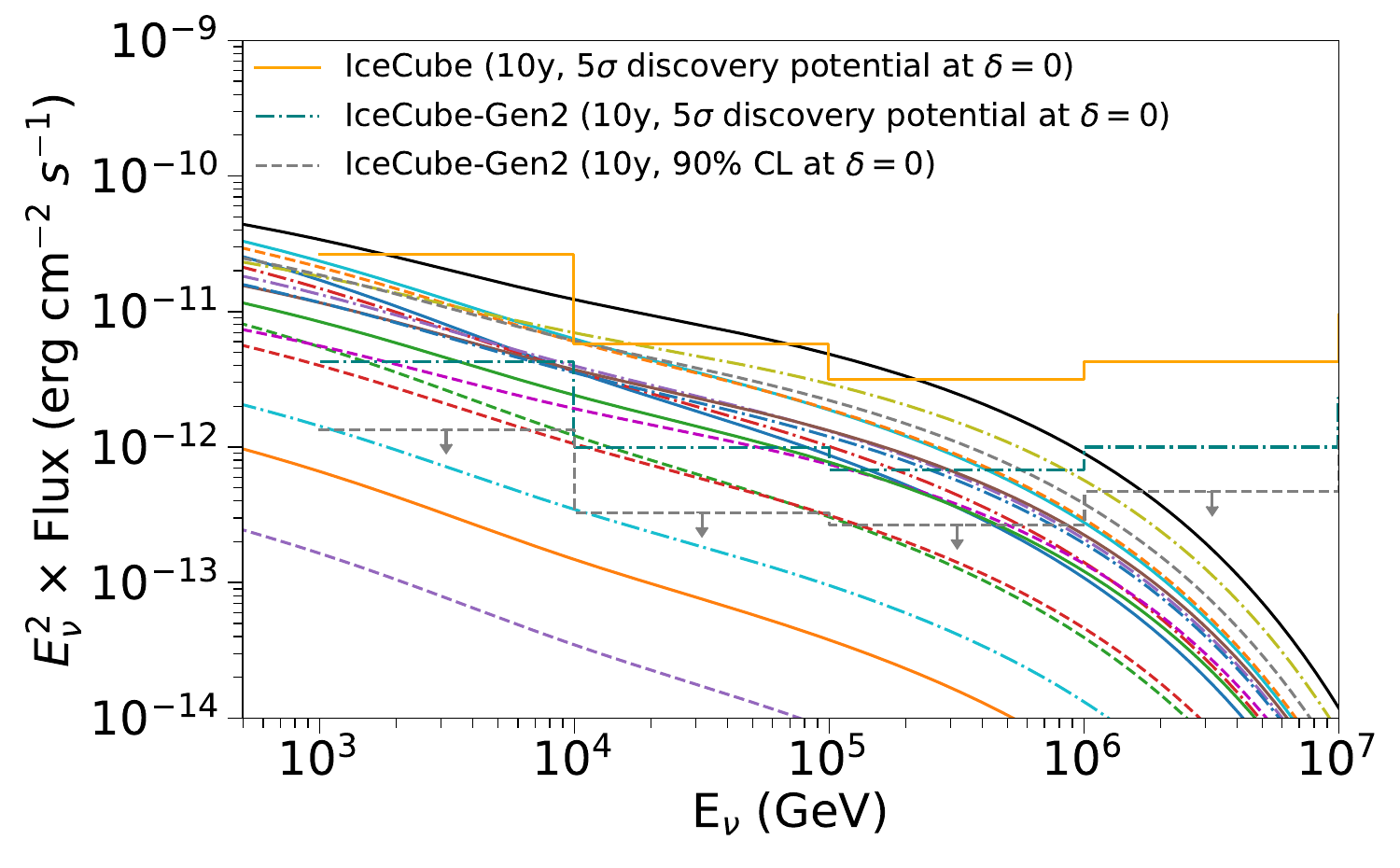}}
\caption{{\it Top panel}: Considering the Sun as the origin ($R_\odot$, 0, 0), the neutrino flux is calculated from GMCs located in concentric circular disks of 1 kpc thickness (SolarCentric SC-Rings) for constant GCR flux scenario (Case-I). {\it Bottom panel}: The galactocentric variation of GCR flux with \cite{green2015constraints} type source distribution is used in this case (Case-IIb) to calculate the flux in SC-Rings. The differential sensitivity limits of the IceCube detector are also shown here for comparison \cite{aartsen2021icecube}. It should be noted that the ring number that contributes the most depends not only on the GCR model assumptions but also on the completeness of the catalog used.}
\label{fig:VariationInNu}
\end{figure}

The variation in the neutrino flux at different distances from the Sun for Case-I and Case-IIb is shown in Fig. \ref{fig:VariationInNu} along with the IceCube sensitivity limits. The stacked neutrino flux from GMCs located around the Sun in concentric circular disks of 1 kpc thickness is shown here (labelled as Ring, and the integers indicate the corresponding ring number). Comparing the top and bottom panels, we can see significant discrepancies in the two cases: the maximum flux for Case-I (top panel) is observed at ring $0$ (i.e., within 1 kpc distance of the Sun), instead, for Case-IIb (bottom panel), the highest flux is concentrated at ring $9$ (i.e., the circular ring closest to the Galactic centre). We can also have an idea about the rings that significantly contribute to the observable diffused neutrino signals by IceCube. Consequently, the ring number that makes the most contribution not only depends on the GCR model assumptions but also on the completeness of the catalogue used.

\begin{figure}[!ht]
    \centering
    \includegraphics[width=\linewidth]{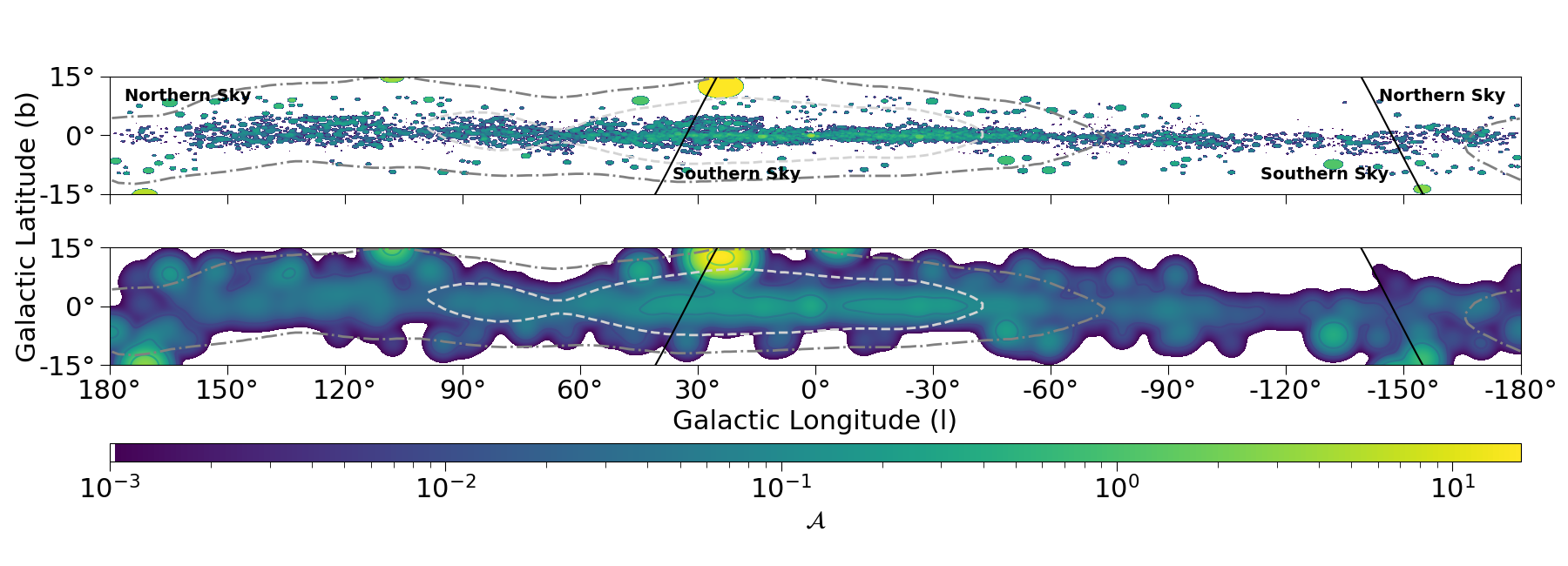}
    \caption{{\it Top panel}: Density distribution of the $\mathcal{A}$ parameter for all GMCs in the Galactic plane (representing Case-I type GCR source distribution). The dotted gray contour lines (dashed and dashed-dot) indicate the 20\% and 50\% of the expected neutrino signal derived from the $\pi^0$ model that matches Fermi-LAT observation. {\it Bottom panel}: A Gaussian smearing of $2^{\circ}$ is introduced to add angular uncertainties in the GMC position. The brightest yellow region in the plot is Aquila Rift GMC location.}
    \label{fig:ADistPlane}
\end{figure}

\section{Spatial Correlation of GMCs and TeVCat Sources Compared to IceCube Significance Correlation Map}
\label{sec:spatialCorrelation}

The density distribution of $\mathcal{A}$ parameter for all GMCs in the galactic plane is represented in Fig. \ref{fig:ADistPlane}. The GMC shape is assumed to be spherical for simplicity, but, actually, it may be different. The circle sizes in the top panel correspond to the angular size of the GMCs in degree square. We introduce a $2^\circ$ angular uncertainty in their sizes, as shown in the bottom panel, using the Gaussian smearing technique. The dark and light grey (dashed-dot and dashed) contour lines represent the region including 50 and 20\% of predicted neutrino signal \cite{2023Sci380338I} and the inclined black lines represent the celestial equator, separating the two hemispheres of the MW sky. The coloured bar at the bottom of the figure represents constant $\mathcal{A}$ values, ranging from $10^{-2}$ to 10 with 10-unit intervals. Aquila Rift GMC (the brightest yellow region in the plot) appears to be the most promising candidate as a potential neutrino source under the condition of Case-I. However, other hot-spot regions can also be seen, and also these regions could be potential neutrino source locations.


\begin{figure}[!ht]
    \centering
    \includegraphics[width=\linewidth]{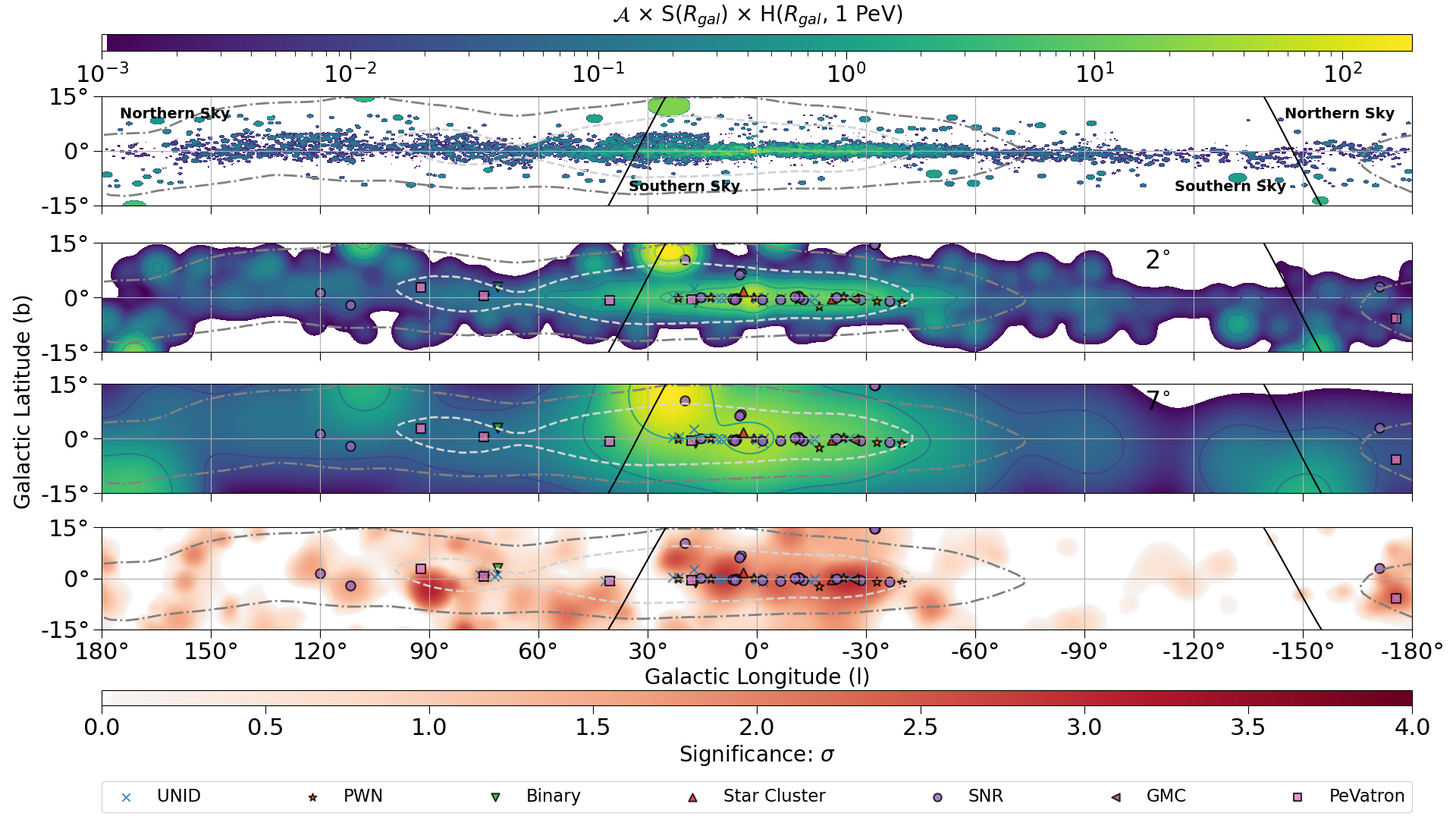}
    \caption{\textit{Top panel:} The density distribution of $\mathcal{A} \times S(R_{gal}) \times H(R_{gal}$, 1\;PeV) in the Galactic plane for \cite{green2015constraints} type GCR source distribution (i.e., Case-IIb). \textit{Middle panels:} The angular size of the GMCs are smeared by 2 and 7$^\circ$ to include an angular uncertainty of their positions. \textit{Bottom panel:} The neutrino significance map of the Galactic plane as recently observed by the IceCube detector \cite{2023Sci380338I}. The TeVCat \& LHAASO PeVatrons gamma-ray sources, which are spatially correlated with the observed neutrino significance, are also plotted here.}
    \label{fig:TeVSourceLocalisation}
\end{figure}

To showcase the effects of diffused source contribution in the observed diffuse neutrino signal from the Galactic plane, we plotted the density distribution of $\mathcal{A} \times S(R_{gal}) \times H(R_{gal}, 1\;PeV)$ in Fig. \ref{fig:TeVSourceLocalisation}. The normalised SNR distribution from \cite{green2015constraints} is used for this case. The top panel displays the density distribution of GMCs, taking into account their actual angular sizes. The middle panels are the same distribution with a Gaussian smearing of $2^\circ$ and $7^\circ$ where the contour lines here correspond to $10^{-2}$, $10^{-1}$, $10$, $10^{1}$, and $10^{2}$. For this case, a clear enhancement near the Galactic centre can be seen as compared to Fig. \ref{fig:ADistPlane}. This density distribution map can be compared with the significance signal observed by the IceCube detector (bottom panel), in order to find their possible sources. Some correlations between observed signals and density distribution can be seen in the figure. As the Gaussian smearing size increases, small-scale structures disappear, but large-scale structures persist. This indicates that GMCs are passive sources of neutrino produced by diffuse GCRs interacting with gas molecules inside them. However, we cannot exclude that the apparent signal missing regions may be due to the sensitivity limits of detectors. 

Finally, we also investigated whether the observed signal was correlated with any astrophysical neutrino sources. The middle and bottom panels show all Galactic gamma-ray sources from the TeVCat catalog \cite{2008ICRC....3.1341W} and LHAASO PeVatrons \cite{cao2021ultrahigh}, corresponding to a 1$\sigma$ significance of the signal detected by IceCube. They could be the possible neutrino sources, but further studies are needed to confirm this. We can observe a significant signal near the Cygnus region, and the contributions from GMCs are not substantial in this region. Therefore, we should investigate this region for potential neutrino sources. The operation of the IceCube-Gen2 observatory will be very useful for the detection of these individual GMCs in the future.

\section{Conclusions}
\label{sec:conclusions}

GCR interactions inside GMCs provide us insights on the non-thermal multimessenger emissions from the Galactic plane. Their neutrino emission is an important asset for revealing Galactic sources of CRs. In addition to GCR flux inside GMCs, CR particles might also be injected by young stellar objects embedded inside them. Detection of neutrino flux from GMCs will be crucial to reveal the properties of primary CR sources inside GMCs. Since the intensity of non-thermal multimessenger emissions is directly proportional to the mass and inversely to the square of the distance of the GMCs, it can put constraints on its mass and distance. 

In this work, we have calculated gamma-ray and neutrino emissions from a total of 8444 individual GMCs and their stacked contributions, taking into account three different GCR distributions (Case-I, Case-IIa and Case-IIb). The neutrino flux calculation using Case-I is a factor of $\sim 12.5/9.0/8.9$ below from the $\pi^0$/KRA$^5_\gamma$/KRA$^{50}_\gamma$ best-fitted model of IceCube observation at $10^5$ GeV; however, for Case-IIb the model can explain $\sim 62.6/86.5/88.3 \%$ neutrino flux of the corresponding models at that energy. Many of these GMCs can be detected by current and future-generation detectors with a large field of view. Till now, only a few GMCs have been detected by the Fermi-LAT \citep{bagh20}, HAWC and HESS telescopes \citep{Albert2021ApJ6A,Sinha_2022icrc90S}. Also, future detectors such as IceCube-Gen2, KM3NeT, SWGO, CTAO, etc., will be able to detect these GMCs. For Case-IIb, the flux of stacked per flavor neutrino from GMCs in the Galactic plane is comparable to the best-fitted KRA$_\gamma^5$ model. This suggests that, under this condition, most of the observed neutrinos in the MW Galactic plane could be associated with GMCs. 

We have also observed a correlation between the neutrino signal significance map and some TeVCat and LHAASO PeVatron sources. Indeed, there could be non-negligible contributions from several other sources that may account for the remaining neutrino flux. For example, diffuse gas interaction (mostly $H_I$) and astrophysical sources (such as SNR, PWN, Binary system, etc.), unresolved GMCs or dark gas clouds, and  CR injection by star-forming regions. Moreover, although there is a lack of evidence, one cannot rule out the possibility of a neutrino flux component originating from the Galactic dark matter halo. In the future, we will study these isolated sources in detail.

\section*{Acknowledgements}
A. Roy is thankful to P. Majumdar and R. Laha for helpful discussions and their inputs. This research has made use of the TeVCat online source catalog (\url{http://tevcat.uchicago.edu}) and Cosmic-Ray Data Base (CRDB) (\url{https://lpsc.in2p3.fr/crdb}).  S.C. acknowledges the support of the Max Planck India Mobility Grant from the Max-Planck Society, supporting the visit and stay at MPP during the project. S.C has also received funding from DST/SERB projects CRG/2021/002961 and MTR/2021/000540. The authors are grateful to the anonymous reviewer for providing insightful comments that helped improve the manuscript.

\bibliography{ref_prop}

\providecommand{\href}[2]{#2}\begingroup\raggedright\begin{thebibliography}{100}

\bibitem{2023Sci380338I}
{Icecube Collaboration}, R.~{Abbasi}, M.~{Ackermann}, J.~{Adams}, J.~A.
  {Aguilar}, M.~{Ahlers} et~al., \emph{{Observation of high-energy neutrinos
  from the Galactic plane}},
  \href{http://dx.doi.org/10.1126/science.adc9818}{\emph{Science} {\bf 380}
  (June, 2023) 1338--1343}, [\href{http://arxiv.org/abs/2307.04427}{{\tt
  2307.04427}}].

\bibitem{Aartsen2019ApJ12A}
M.~G. {Aartsen}, M.~{Ackermann}, J.~{Adams}, J.~A. {Aguilar}, M.~{Ahlers},
  M.~{Ahrens} et~al., \emph{{Search for Sources of Astrophysical Neutrinos
  Using Seven Years of IceCube Cascade Events}},
  \href{http://dx.doi.org/10.3847/1538-4357/ab4ae2}{\emph{\apj} {\bf 886}
  (Nov., 2019) 12}, [\href{http://arxiv.org/abs/1907.06714}{{\tt 1907.06714}}].

\bibitem{nerenov2023PhRvD3044N}
A.~{Neronov}, D.~{Semikoz}, J.~{Aublin}, M.~{Lamoureux} and A.~{Kouchner},
  \emph{{Hadronic nature of high-energy emission from the Galactic ridge}},
  \href{http://dx.doi.org/10.1103/PhysRevD.108.103044}{\emph{\prd} {\bf 108}
  (Nov., 2023) 103044}, [\href{http://arxiv.org/abs/2307.07978}{{\tt
  2307.07978}}].

\bibitem{stecker1979ApJ_19S}
F.~W. {Stecker}, \emph{{Diffuse fluxes of cosmic high-energy neutrinos.}},
  \href{http://dx.doi.org/10.1086/156919}{\emph{\apj} {\bf 228} (Mar., 1979)
  919--927}.

\bibitem{Gupta2013AP75G}
N.~{Gupta}, \emph{{Galactic PeV neutrinos}},
  \href{http://dx.doi.org/10.1016/j.astropartphys.2013.07.003}{\emph{Astroparticle
  Physics} {\bf 48} (Aug., 2013) 75--77},
  [\href{http://arxiv.org/abs/1305.4123}{{\tt 1305.4123}}].

\bibitem{zhang2023ApJ3Z}
R.~{Zhang}, X.~{Huang}, Z.-H. {Xu}, S.~{Zhao} and Q.~{Yuan}, \emph{{Galactic
  Diffuse {\ensuremath{\gamma}}-Ray Emission from GeV to PeV Energies in Light
  of Up-to-date Cosmic-Ray Measurements}},
  \href{http://dx.doi.org/10.3847/1538-4357/acf842}{\emph{\apj} {\bf 957}
  (Nov., 2023) 43}, [\href{http://arxiv.org/abs/2305.06948}{{\tt 2305.06948}}].

\bibitem{2023arXiv230707451V}
V.~{Vecchiotti}, F.~L. {Villante} and G.~{Pagliaroli}, \emph{{Unveiling the
  Nature of Galactic TeV Sources with IceCube Results}},
  \href{http://dx.doi.org/10.3847/2041-8213/acff60}{\emph{\apjl} {\bf 956}
  (Oct., 2023) L44}, [\href{http://arxiv.org/abs/2307.07451}{{\tt
  2307.07451}}].

\bibitem{Gabici2007ApJ6651G}
S.~{Gabici} and F.~A. {Aharonian}, \emph{{Searching for Galactic Cosmic-Ray
  Pevatrons with Multi-TeV Gamma Rays and Neutrinos}},
  \href{http://dx.doi.org/10.1086/521047}{\emph{\apjl} {\bf 665} (Aug., 2007)
  L131--L134}, [\href{http://arxiv.org/abs/0705.3011}{{\tt 0705.3011}}].

\bibitem{kappes_2007ApJ70K}
A.~{Kappes}, J.~{Hinton}, C.~{Stegmann} and F.~A. {Aharonian}, \emph{{Potential
  Neutrino Signals from Galactic {\ensuremath{\gamma}}-Ray Sources}},
  \href{http://dx.doi.org/10.1086/508936}{\emph{\apj} {\bf 656} (Feb., 2007)
  870--878}, [\href{http://arxiv.org/abs/astro-ph/0607286}{{\tt
  astro-ph/0607286}}].

\bibitem{Gonza2009APh437G}
M.~C. {Gonzalez-Garcia}, F.~{Halzen} and S.~{Mohapatra}, \emph{{Identifying
  Galactic PeVatrons with neutrinos}},
  \href{http://dx.doi.org/10.1016/j.astropartphys.2009.05.002}{\emph{Astroparticle
  Physics} {\bf 31} (July, 2009) 437--444},
  [\href{http://arxiv.org/abs/0902.1176}{{\tt 0902.1176}}].

\bibitem{kelner2006energy}
S.~R. Kelner, F.~A. Aharonian and V.~V. Bugayov, \emph{Energy spectra of gamma
  rays, electrons, and neutrinos produced at proton-proton interactions in the
  very high energy regime}, {\emph{Physical Review D} {\bf 74} (2006) 034018}.

\bibitem{ackermann2012fermi}
M.~Ackermann, M.~Ajello, W.~Atwood, L.~Baldini, J.~Ballet, G.~Barbiellini
  et~al., \emph{{Fermi-LAT observations of the diffuse $\gamma$-ray emission:
  implications for cosmic rays and the interstellar medium}}, {\emph{The
  Astrophysical Journal} {\bf 750} (2012) 3}.

\bibitem{abdo2008ApJ_078A}
A.~A. {Abdo}, B.~{Allen}, T.~{Aune}, D.~{Berley}, E.~{Blaufuss}, S.~{Casanova}
  et~al., \emph{{A Measurement of the Spatial Distribution of Diffuse TeV
  Gamma-Ray Emission from the Galactic Plane with Milagro}},
  \href{http://dx.doi.org/10.1086/592213}{\emph{\apj} {\bf 688} (Dec., 2008)
  1078--1083}, [\href{http://arxiv.org/abs/0805.0417}{{\tt 0805.0417}}].

\bibitem{gaggero2015gamma}
D.~Gaggero, A.~Urbano, M.~Valli and P.~Ullio, \emph{Gamma-ray sky points to
  radial gradients in cosmic-ray transport}, {\emph{Physical Review D} {\bf 91}
  (2015) 083012}.

\bibitem{bartoli2015Ap20B}
B.~{Bartoli}, P.~{Bernardini}, X.~J. {Bi}, P.~{Branchini}, A.~{Budano},
  P.~{Camarri} et~al., \emph{{Study of the Diffuse Gamma-Ray Emission from the
  Galactic Plane with ARGO-YBJ}},
  \href{http://dx.doi.org/10.1088/0004-637X/806/1/20}{\emph{\apj} {\bf 806}
  (June, 2015) 20}, [\href{http://arxiv.org/abs/1507.06758}{{\tt 1507.06758}}].

\bibitem{hess2018AnA..1H}
{H.~E.~S.~S. Collaboration}, H.~{Abdalla}, A.~{Abramowski}, F.~{Aharonian},
  F.~{Ait Benkhali}, E.~O. {Ang{\"u}ner} et~al., \emph{{The H.E.S.S. Galactic
  plane survey}},
  \href{http://dx.doi.org/10.1051/0004-6361/201732098}{\emph{\aap} {\bf 612}
  (Apr., 2018) A1}, [\href{http://arxiv.org/abs/1804.02432}{{\tt 1804.02432}}].

\bibitem{abdalla2021ApJ_hesshawc}
H.~{Abdalla}, F.~{Aharonian}, F.~{Ait Benkhali}, E.~O. {Ang{\"u}ner},
  C.~{Arcaro}, C.~{Armand} et~al., \emph{{TeV Emission of Galactic Plane
  Sources with HAWC and H.E.S.S.}},
  \href{http://dx.doi.org/10.3847/1538-4357/abf64b}{\emph{\apj} {\bf 917}
  (Aug., 2021) 6}, [\href{http://arxiv.org/abs/2107.01425}{{\tt 2107.01425}}].

\bibitem{amenomori2021first}
M.~{Amenomori}, Y.~W. {Bao}, X.~J. {Bi}, D.~{Chen}, T.~L. {Chen}, W.~Y. {Chen}
  et~al., \emph{{First Detection of sub-PeV Diffuse Gamma Rays from the
  Galactic Disk: Evidence for Ubiquitous Galactic Cosmic Rays beyond PeV
  Energies}},
  \href{http://dx.doi.org/10.1103/PhysRevLett.126.141101}{\emph{\prl} {\bf 126}
  (Apr., 2021) 141101}, [\href{http://arxiv.org/abs/2104.05181}{{\tt
  2104.05181}}].

\bibitem{2023PhRvL.131o1001C}
Z.~{Cao}, F.~{Aharonian}, Q.~{An}, Y.~X. {Axikegu}, Bai, Y.~W. {Bao},
  D.~{Bastieri} et~al., \emph{{Measurement of Ultra-High-Energy Diffuse
  Gamma-Ray Emission of the Galactic Plane from 10 TeV to 1 PeV with
  LHAASO-KM2A}},
  \href{http://dx.doi.org/10.1103/PhysRevLett.131.151001}{\emph{\prl} {\bf 131}
  (Oct., 2023) 151001}, [\href{http://arxiv.org/abs/2305.05372}{{\tt
  2305.05372}}].

\bibitem{gabici2008_180G}
S.~{Gabici}, A.~M. {Taylor}, R.~J. {White}, S.~{Casanova} and F.~A.
  {Aharonian}, \emph{{The diffuse neutrino flux from the inner Galaxy:
  Constraints from very high energy gamma-ray observations}},
  \href{http://dx.doi.org/10.1016/j.astropartphys.2008.08.005}{\emph{Astroparticle
  Physics} {\bf 30} (Nov., 2008) 180--185},
  [\href{http://arxiv.org/abs/0806.2459}{{\tt 0806.2459}}].

\bibitem{Lunar2015PhRv_1301L}
C.~{Lunardini}, S.~{Razzaque} and L.~{Yang}, \emph{{Multimessenger study of the
  Fermi bubbles: Very high energy gamma rays and neutrinos}},
  \href{http://dx.doi.org/10.1103/PhysRevD.92.021301}{\emph{\prd} {\bf 92}
  (July, 2015) 021301}, [\href{http://arxiv.org/abs/1504.07033}{{\tt
  1504.07033}}].

\bibitem{2023PhRvD.108f1305S}
C.~{Shao}, S.~{Lin} and L.~{Yang}, \emph{{Multimessenger study of the Galactic
  diffuse emission with LHAASO and IceCube observations}},
  \href{http://dx.doi.org/10.1103/PhysRevD.108.L061305}{\emph{\prd} {\bf 108}
  (Sept., 2023) L061305}, [\href{http://arxiv.org/abs/2307.01038}{{\tt
  2307.01038}}].

\bibitem{evoli2007JCAP003E}
C.~{Evoli}, D.~{Grasso} and L.~{Maccione}, \emph{{Diffuse neutrino and
  gamma-ray emissions of the galaxy above the TeV}},
  \href{http://dx.doi.org/10.1088/1475-7516/2007/06/003}{\emph{\jcap} {\bf
  2007} (June, 2007) 003}, [\href{http://arxiv.org/abs/astro-ph/0701856}{{\tt
  astro-ph/0701856}}].

\bibitem{2014PhRvD..89j3002N}
A.~{Neronov}, D.~{Semikoz} and C.~{Tchernin}, \emph{{PeV neutrinos from
  interactions of cosmic rays with the interstellar medium in the Galaxy}},
  \href{http://dx.doi.org/10.1103/PhysRevD.89.103002}{\emph{\prd} {\bf 89}
  (May, 2014) 103002}, [\href{http://arxiv.org/abs/1307.2158}{{\tt
  1307.2158}}].

\bibitem{gagg2015ApJ25G}
D.~{Gaggero}, D.~{Grasso}, A.~{Marinelli}, A.~{Urbano} and M.~{Valli},
  \emph{{The Gamma-Ray and Neutrino Sky: A Consistent Picture of Fermi-LAT,
  Milagro, and IceCube Results}},
  \href{http://dx.doi.org/10.1088/2041-8205/815/2/L25}{\emph{\apjl} {\bf 815}
  (Dec., 2015) L25}, [\href{http://arxiv.org/abs/1504.00227}{{\tt
  1504.00227}}].

\bibitem{ahar2019NatA61A}
F.~{Aharonian}, R.~{Yang} and E.~{de O{\~n}a Wilhelmi}, \emph{{Massive stars as
  major factories of Galactic cosmic rays}},
  \href{http://dx.doi.org/10.1038/s41550-019-0724-0}{\emph{Nature Astronomy}
  {\bf 3} (Mar., 2019) 561--567}, [\href{http://arxiv.org/abs/1804.02331}{{\tt
  1804.02331}}].

\bibitem{Abe2020PhRv02A}
A.~U. {Abeysekara}, A.~{Albert}, R.~{Alfaro}, J.~R. {Angeles Camacho}, J.~C.
  {Arteaga-Vel{\'a}zquez}, K.~P. {Arunbabu} et~al., \emph{{Multiple Galactic
  Sources with Emission Above 56 TeV Detected by HAWC}},
  \href{http://dx.doi.org/10.1103/PhysRevLett.124.021102}{\emph{\prl} {\bf 124}
  (Jan., 2020) 021102}, [\href{http://arxiv.org/abs/1909.08609}{{\tt
  1909.08609}}].

\bibitem{cao2021ultrahigh}
Z.~Cao, F.~Aharonian, Q.~An, Axikegu, L.~Bai, Y.~Bai et~al.,
  \emph{{Ultrahigh-energy photons up to 1.4 petaelectronvolts from 12
  $\gamma$-ray Galactic sources}}, {\emph{Nature} {\bf 594} (2021) 33--36}.

\bibitem{bao2021ApJ2B}
Y.~{Bao} and Y.~{Chen}, \emph{{On the Hard Gamma-Ray Spectrum of the Potential
  PeVatron Supernova Remnant G106.3 + 2.7}},
  \href{http://dx.doi.org/10.3847/1538-4357/ac1581}{\emph{\apj} {\bf 919}
  (Sept., 2021) 32}, [\href{http://arxiv.org/abs/2103.01814}{{\tt
  2103.01814}}].

\bibitem{cristofari2021hunt}
P.~Cristofari, \emph{{The hunt for PeVatrons: The case of supernova remnants}},
  {\emph{Universe} {\bf 7} (2021) 324}.

\bibitem{cardillo2023lhaaso}
M.~Cardillo and A.~Giuliani, \emph{{The LHAASO PeVatron bright sky: what we
  learned}}, {\emph{Applied Sciences} {\bf 13} (2023) 6433}.

\bibitem{olmi2023pulsar}
B.~Olmi, \emph{{The Pulsar Wind Nebulae contribution to gamma-rays}},
  {\emph{arXiv e-prints} (2023) arXiv--2303}.

\bibitem{vieu2023massive}
T.~Vieu and B.~Reville, \emph{Massive star cluster origin for the galactic
  cosmic ray population at very-high energies}, {\emph{Monthly Notices of the
  Royal Astronomical Society} {\bf 519} (2023) 136--147}.

\bibitem{gabici2023cosmic}
S.~Gabici, \emph{Cosmic rays from star clusters}, {\emph{arXiv preprint
  arXiv:2307.01596} (2023) }.

\bibitem{cao2023first}
Z.~Cao, F.~Aharonian, Q.~An, Y.~Bai, Y.~Bao, D.~Bastieri et~al., \emph{{The
  First LHAASO Catalog of Gamma-Ray Sources}}, {\emph{arXiv preprint
  arXiv:2305.17030} (2023) }.

\bibitem{2023arXiv230617285A}
A.~{Ambrosone}, K.~{M{\o}rch Groth}, E.~{Peretti} and M.~{Ahlers},
  \emph{{Galactic Diffuse Neutrino Emission from Sources beyond the Discovery
  Horizon}}, \href{http://dx.doi.org/10.48550/arXiv.2306.17285}{\emph{arXiv
  e-prints} (June, 2023) arXiv:2306.17285},
  [\href{http://arxiv.org/abs/2306.17285}{{\tt 2306.17285}}].

\bibitem{2014PhRvD90h3002K}
M.~{Kachelrie{\ss}} and S.~{Ostapchenko}, \emph{{Neutrino yield from Galactic
  cosmic rays}},
  \href{http://dx.doi.org/10.1103/PhysRevD.90.083002}{\emph{\prd} {\bf 90}
  (Oct., 2014) 083002}, [\href{http://arxiv.org/abs/1405.3797}{{\tt
  1405.3797}}].

\bibitem{peron2022A57P}
G.~{Peron} and F.~{Aharonian}, \emph{{Probing the galactic cosmic-ray density
  with current and future {\ensuremath{\gamma}}-ray instruments}},
  \href{http://dx.doi.org/10.1051/0004-6361/202142416}{\emph{\aap} {\bf 659}
  (Mar., 2022) A57}, [\href{http://arxiv.org/abs/2110.08778}{{\tt
  2110.08778}}].

\bibitem{2014AnA66A142Y}
R.-z. {Yang}, E.~{de O{\~n}a Wilhelmi} and F.~{Aharonian}, \emph{{Probing
  cosmic rays in nearby giant molecular clouds with the Fermi Large Area
  Telescope}}, \href{http://dx.doi.org/10.1051/0004-6361/201321044}{\emph{\aap}
  {\bf 566} (June, 2014) A142}, [\href{http://arxiv.org/abs/1303.7323}{{\tt
  1303.7323}}].

\bibitem{2012ApJ7564A}
M.~{Ackermann}, M.~{Ajello}, A.~{Allafort}, E.~{Antolini}, L.~{Baldini},
  J.~{Ballet} et~al., \emph{{Gamma-Ray Observations of the Orion Molecular
  Clouds with the Fermi Large Area Telescope}},
  \href{http://dx.doi.org/10.1088/0004-637X/756/1/4}{\emph{\apj} {\bf 756}
  (Sept., 2012) 4}, [\href{http://arxiv.org/abs/1207.0616}{{\tt 1207.0616}}].

\bibitem{Yang2015AnA580A..90Y}
R.-z. {Yang}, D.~I. {Jones} and F.~{Aharonian}, \emph{{Fermi-LAT observations
  of the Sagittarius B complex}},
  \href{http://dx.doi.org/10.1051/0004-6361/201425233}{\emph{\aap} {\bf 580}
  (Aug., 2015) A90}, [\href{http://arxiv.org/abs/1410.7639}{{\tt 1410.7639}}].

\bibitem{bagh20}
V.~Baghmanyan, G.~Peron, S.~Casanova, F.~Aharonian and R.~Zanin, \emph{Evidence
  of cosmic-ray excess from local giant molecular clouds}, {\emph{The
  Astrophysical Journal Letters} {\bf 901} (2020) L4}.

\bibitem{1973ApJ185L7B}
J.~H. {Black} and G.~G. {Fazio}, \emph{{Production of Gamma Radiation in Dense
  Interstellar Clouds by Cosmic-Ray Interactions}},
  \href{http://dx.doi.org/10.1086/181310}{\emph{\apjl} {\bf 185} (Oct., 1973)
  L7}.

\bibitem{1983AdSpR23W}
A.~W. {Wolfendale}, \emph{{Gamma rays from giant-molecular clouds}},
  \href{http://dx.doi.org/10.1016/0273-1177(83)90004-2}{\emph{Advances in Space
  Research} {\bf 3} (Jan., 1983) 23--29}.

\bibitem{2001SSRv187A}
F.~A. {Aharonian}, \emph{{Gamma Rays From Molecular Clouds}},
  \href{http://dx.doi.org/10.1023/A:1013845015364}{\emph{\ssr} {\bf 99} (Oct.,
  2001) 187--196}, [\href{http://arxiv.org/abs/astro-ph/0012290}{{\tt
  astro-ph/0012290}}].

\bibitem{2007ApnSS.309..365G}
S.~{Gabici}, F.~A. {Aharonian} and P.~{Blasi}, \emph{{Gamma rays from molecular
  clouds}}, \href{http://dx.doi.org/10.1007/s10509-007-9427-6}{\emph{\apss}
  {\bf 309} (June, 2007) 365--371},
  [\href{http://arxiv.org/abs/astro-ph/0610032}{{\tt astro-ph/0610032}}].

\bibitem{2023JCAP08047R}
A.~{Roy}, J.~C. {Joshi}, M.~{Cardillo} and R.~{Sarkar}, \emph{{Interpreting the
  GeV-TeV gamma-ray spectra of local giant molecular clouds using GEANT4
  simulation}},
  \href{http://dx.doi.org/10.1088/1475-7516/2023/08/047}{\emph{\jcap} {\bf
  2023} (Aug., 2023) 047}, [\href{http://arxiv.org/abs/2305.06693}{{\tt
  2305.06693}}].

\bibitem{ahlers2014PhRvDb3010A}
M.~{Ahlers} and K.~{Murase}, \emph{{Probing the Galactic origin of the IceCube
  excess with gamma rays}},
  \href{http://dx.doi.org/10.1103/PhysRevD.90.023010}{\emph{\prd} {\bf 90}
  (July, 2014) 023010}, [\href{http://arxiv.org/abs/1309.4077}{{\tt
  1309.4077}}].

\bibitem{2014MNRAS.439.3414J}
J.~C. {Joshi}, W.~{Winter} and N.~{Gupta}, \emph{{How many of the observed
  neutrino events can be described by cosmic ray interactions in the Milky
  Way?}}, \href{http://dx.doi.org/10.1093/mnras/stu189}{\emph{\mnras} {\bf 439}
  (Apr., 2014) 3414--3419}, [\href{http://arxiv.org/abs/1310.5123}{{\tt
  1310.5123}}].

\bibitem{Sarmah2023MNRAS144S}
P.~{Sarmah}, S.~{Chakraborty} and J.~C. {Joshi}, \emph{{Probing LHAASO galactic
  PeVatrons through gamma-ray and neutrino correspondence}},
  \href{http://dx.doi.org/10.1093/mnras/stad609}{\emph{\mnras} {\bf 521} (May,
  2023) 1144--1151}, [\href{http://arxiv.org/abs/2301.04161}{{\tt
  2301.04161}}].

\bibitem{myers1978compilation}
P.~C. Myers, \emph{A compilation of interstellar gas properties},
  {\emph{Astrophysical Journal, Part 1, vol. 225, Oct. 15, 1978, p. 380-389.}
  {\bf 225} (1978) 380--389}.

\bibitem{mccray1987soapM}
R.~A. {McCray}, \emph{{Coronal interstellar gas and supernova remnants}},  in
  \emph{Spectroscopy of Astrophysical Plasmas} (A.~{Dalgarno} and D.~{Layzer},
  eds.), pp.~255--278, Jan., 1987.

\bibitem{Jen197845J}
E.~B. {Jenkins}, \emph{{Coronal gas in the Galaxy. I. A new survey of
  interstellar O VI.}}, \href{http://dx.doi.org/10.1086/155846}{\emph{\apj}
  {\bf 219} (Feb., 1978) 845--860}.

\bibitem{mivi17}
M.-A. Miville-Desch{\^e}nes, N.~Murray and E.~J. Lee, \emph{{Physical
  properties of molecular clouds for the entire Milky Way disk}}, {\emph{The
  Astrophysical Journal} {\bf 834} (2017) 57}.

\bibitem{chen20}
B.~Chen, G.~Li, H.~Yuan, Y.~Huang, Z.~Tian, H.~Wang et~al., \emph{{A large
  catalogue of molecular clouds with accurate distances within 4 kpc of the
  Galactic disc}}, {\emph{Monthly Notices of the Royal Astronomical Society}
  {\bf 493} (2020) 351--361}.

\bibitem{2016ApJ...822...52R}
T.~S. {Rice}, A.~A. {Goodman}, E.~A. {Bergin}, C.~{Beaumont} and T.~M. {Dame},
  \emph{{A Uniform Catalog of Molecular Clouds in the Milky Way}},
  \href{http://dx.doi.org/10.3847/0004-637X/822/1/52}{\emph{\apj} {\bf 822}
  (May, 2016) 52}, [\href{http://arxiv.org/abs/1602.02791}{{\tt 1602.02791}}].

\bibitem{hou14}
L.~Hou and J.~Han, \emph{{The observed spiral structure of the Milky Way}},
  {\emph{Astronomy \& Astrophysics} {\bf 569} (2014) A125}.

\bibitem{2001ApJ...547..792D}
T.~M. {Dame}, D.~{Hartmann} and P.~{Thaddeus}, \emph{{The Milky Way in
  Molecular Clouds: A New Complete CO Survey}},
  \href{http://dx.doi.org/10.1086/318388}{\emph{\apj} {\bf 547} (Feb., 2001)
  792--813}, [\href{http://arxiv.org/abs/astro-ph/0009217}{{\tt
  astro-ph/0009217}}].

\bibitem{draine2010physics}
B.~T. Draine, \emph{Physics of the interstellar and intergalactic medium},
  vol.~19.
\newblock Princeton University Press, 2010.

\bibitem{heyer2015molecular}
M.~Heyer and T.~Dame, \emph{{Molecular clouds in the Milky Way}}, {\emph{Annual
  Review of Astronomy and Astrophysics} {\bf 53} (2015) 583--629}.

\bibitem{nakanishi2006three}
H.~Nakanishi and Y.~Sofue, \emph{{Three-dimensional distribution of the ISM in
  the Milky Way galaxy: II. The molecular gas disk}}, {\emph{Publications of
  the Astronomical Society of Japan} {\bf 58} (2006) 847--860}.

\bibitem{bronfman1988co}
L.~Bronfman, R.~Cohen, H.~Alvarez, J.~May and P.~Thaddeus, \emph{{A CO survey
  of the southern Milky Way-The mean radial distribution of molecular clouds
  within the solar circle}}, {\emph{Astrophysical Journal, Part 1 (ISSN
  0004-637X), vol. 324, Jan. 1, 1988, p. 248-266. Research supported by NSF and
  Fondo Nacional de Educacion y Ciencias of Chile.} {\bf 324} (1988) 248--266}.

\bibitem{grabelsky1987molecular}
D.~Grabelsky, R.~Cohen, L.~Bronfman, P.~Thaddeus and J.~May, \emph{{Molecular
  clouds in the Carina arm-Large-scale properties of molecular gas and
  comparison with HI}}, {\emph{The Astrophysical Journal} {\bf 315} (1987)
  122--141}.

\bibitem{digel91}
{Digel S. W}, ``Phd thesis.'' Harvard University, Massachusetts, 1991.

\bibitem{pohl2008three}
M.~Pohl, P.~Englmaier and N.~Bissantz, \emph{{Three-dimensional distribution of
  molecular gas in the barred Milky Way}}, {\emph{The Astrophysical Journal}
  {\bf 677} (2008) 283}.

\bibitem{nakanishi2016three}
H.~Nakanishi and Y.~Sofue, \emph{{Three-dimensional distribution of the ISM in
  the Milky Way galaxy. III. The total neutral gas disk}}, {\emph{Publications
  of the Astronomical Society of Japan} {\bf 68} (2016) 5}.

\bibitem{2014PhRvD89j3003T}
A.~M. {Taylor}, S.~{Gabici} and F.~{Aharonian}, \emph{{Galactic halo origin of
  the neutrinos detected by IceCube}},
  \href{http://dx.doi.org/10.1103/PhysRevD.89.103003}{\emph{\prd} {\bf 89}
  (May, 2014) 103003}, [\href{http://arxiv.org/abs/1403.3206}{{\tt
  1403.3206}}].

\bibitem{schwefer2023ApJ6S}
G.~{Schwefer}, P.~{Mertsch} and C.~{Wiebusch}, \emph{{Diffuse Emission of
  Galactic High-energy Neutrinos from a Global Fit of Cosmic Rays}},
  \href{http://dx.doi.org/10.3847/1538-4357/acc1e2}{\emph{\apj} {\bf 949} (May,
  2023) 16}, [\href{http://arxiv.org/abs/2211.15607}{{\tt 2211.15607}}].

\bibitem{fang2023milky}
K.~{Fang}, J.~S. {Gallagher} and F.~{Halzen}, \emph{{The Milky Way revealed to
  be a neutrino desert by the IceCube Galactic plane observation}},
  \href{http://dx.doi.org/10.1038/s41550-023-02128-0}{\emph{Nature Astronomy}
  (Nov., 2023) }, [\href{http://arxiv.org/abs/2306.17275}{{\tt 2306.17275}}].

\bibitem{menchiari2023probing}
S.~Menchiari, \emph{Probing star clusters as cosmic ray factories},
  {\emph{arXiv preprint arXiv:2307.03477} (2023) }.

\bibitem{strong2007ARNPS_285S}
A.~W. {Strong}, I.~V. {Moskalenko} and V.~S. {Ptuskin}, \emph{{Cosmic-Ray
  Propagation and Interactions in the Galaxy}},
  \href{http://dx.doi.org/10.1146/annurev.nucl.57.090506.123011}{\emph{Annual
  Review of Nuclear and Particle Science} {\bf 57} (Nov., 2007) 285--327},
  [\href{http://arxiv.org/abs/astro-ph/0701517}{{\tt astro-ph/0701517}}].

\bibitem{vos2015new}
E.~E. {Vos} and M.~S. {Potgieter}, \emph{{New Modeling of Galactic Proton
  Modulation during the Minimum of Solar Cycle 23/24}},
  \href{http://dx.doi.org/10.1088/0004-637X/815/2/119}{\emph{\apj} {\bf 815}
  (Dec., 2015) 119}.

\bibitem{2024PhRvL.132e1002V}
F.~{Varsi}, S.~{Ahmad}, M.~{Chakraborty}, A.~{Chandra}, S.~R. {Dugad}, U.~D.
  {Goswami} et~al., \emph{{Evidence of a Hardening in the Cosmic Ray Proton
  Spectrum at around 166 TeV Observed by the GRAPES-3 Experiment}},
  \href{http://dx.doi.org/10.1103/PhysRevLett.132.051002}{\emph{\prl} {\bf 132}
  (Jan., 2024) 051002}.

\bibitem{PhysRevD.102.063002}
F.~Riehn, R.~Engel, A.~Fedynitch, T.~K. Gaisser and T.~Stanev, \emph{{Hadronic
  interaction model SIBYLL 2.3d and extensive air showers}},
  \href{http://dx.doi.org/10.1103/PhysRevD.102.063002}{\emph{Phys. Rev. D} {\bf
  102} (Sep, 2020) 063002}.

\bibitem{2021PhR...894....1A}
M.~{Aguilar}, L.~{Ali Cavasonza}, G.~{Ambrosi}, L.~{Arruda}, N.~{Attig},
  F.~{Barao} et~al., \emph{{The Alpha Magnetic Spectrometer (AMS) on the
  international space station: Part II - Results from the first seven years}},
  \href{http://dx.doi.org/10.1016/j.physrep.2020.09.003}{\emph{\\physrep} {\bf
  894} (Feb., 2021) 1--116}.

\bibitem{2009BRASP..73..564P}
A.~D. {Panov}, J.~H. {Adams}, H.~S. {Ahn}, G.~L. {Bashinzhagyan}, J.~W.
  {Watts}, J.~P. {Wefel} et~al., \emph{{Energy spectra of abundant nuclei of
  primary cosmic rays from the data of ATIC-2 experiment: Final results}},
  \href{http://dx.doi.org/10.3103/S1062873809050098}{\emph{Bulletin of the
  Russian Academy of Science, Phys.} {\bf 73} (June, 2009) 564--567},
  [\href{http://arxiv.org/abs/1101.3246}{{\tt 1101.3246}}].

\bibitem{2022PhRvL.129j1102A}
O.~{Adriani}, Y.~{Akaike}, K.~{Asano}, Y.~{Asaoka}, E.~{Berti}, G.~{Bigongiari}
  et~al., \emph{{Observation of Spectral Structures in the Flux of Cosmic-Ray
  Protons from 50 GeV to 60 TeV with the Calorimetric Electron Telescope on the
  International Space Station}},
  \href{http://dx.doi.org/10.1103/PhysRevLett.129.101102}{\emph{\\prl} {\bf
  129} (Sept., 2022) 101102}, [\href{http://arxiv.org/abs/2209.01302}{{\tt
  2209.01302}}].

\bibitem{2017ApJ...839....5Y}
Y.~S. {Yoon}, T.~{Anderson}, A.~{Barrau}, N.~B. {Conklin}, S.~{Coutu},
  L.~{Derome} et~al., \emph{{Proton and Helium Spectra from the CREAM-III
  Flight}}, \href{http://dx.doi.org/10.3847/1538-4357/aa68e4}{\emph{\apj} {\bf
  839} (Apr., 2017) 5}, [\href{http://arxiv.org/abs/1704.02512}{{\tt
  1704.02512}}].

\bibitem{2019SciA....5.3793A}
Q.~{An}, R.~{Asfandiyarov}, P.~{Azzarello}, P.~{Bernardini}, X.~J. {Bi}, M.~S.
  {Cai} et~al., \emph{{Measurement of the cosmic ray proton spectrum from 40
  GeV to 100 TeV with the DAMPE satellite}},
  \href{http://dx.doi.org/10.1126/sciadv.aax3793}{\emph{Science Advances} {\bf
  5} (Sept., 2019) eaax3793}, [\href{http://arxiv.org/abs/1909.12860}{{\tt
  1909.12860}}].

\bibitem{2022ApJ...940..107C}
G.~H. {Choi}, E.~S. {Seo}, S.~{Aggarwal}, Y.~{Amare}, D.~{Angelaszek}, D.~P.
  {Bowman} et~al., \emph{{Measurement of High-energy Cosmic-Ray Proton Spectrum
  from the ISS-CREAM Experiment}},
  \href{http://dx.doi.org/10.3847/1538-4357/ac9d2c}{\emph{\\apj} {\bf 940}
  (Dec., 2022) 107}.

\bibitem{2019PhRvD.100h2002A}
M.~G. {Aartsen}, M.~{Ackermann}, J.~{Adams}, J.~A. {Aguilar}, M.~{Ahlers},
  M.~{Ahrens} et~al., \emph{{Cosmic ray spectrum and composition from PeV to
  EeV using 3 years of data from IceTop and IceCube}},
  \href{http://dx.doi.org/10.1103/PhysRevD.100.082002}{\emph{\\prd} {\bf 100}
  (Oct., 2019) 082002}, [\href{http://arxiv.org/abs/1906.04317}{{\tt
  1906.04317}}].

\bibitem{2005APh....24....1A}
T.~{Antoni}, W.~D. {Apel}, A.~F. {Badea}, K.~{Bekk}, A.~{Bercuci},
  J.~{Bl{\\\"u}mer} et~al., \emph{{KASCADE measurements of energy spectra for
  elemental groups of cosmic rays: Results and open problems}},
  \href{http://dx.doi.org/10.1016/j.astropartphys.2005.04.001}{\emph{Astroparticle
  Physics} {\bf 24} (Sept., 2005) 1--25},
  [\href{http://arxiv.org/abs/astro-ph/0505413}{{\tt astro-ph/0505413}}].

\bibitem{2017ICRC...35..316A}
C.~J. {Arteaga-Vel{\'a}zquez}, D.~{Rivera-Rangel}, W.~D. {Apel}, K.~{Bekk},
  M.~{Bertaina}, J.~{Bl{\\\"u}mer} et~al., \emph{{Measurements of the muon
  content of EAS in KASCADE-Grande compared with SIBYLL 2.3 predictions}},  in
  \emph{35th International Cosmic Ray Conference (ICRC2017)}, vol.~301 of
  \emph{International Cosmic Ray Conference}, p.~316, July, 2017.
\newblock \href{http://dx.doi.org/10.22323/1.301.0316}{DOI}.

\bibitem{2019AdSpR..64.2546G}
V.~{Grebenyuk}, D.~{Karmanov}, I.~{Kovalev}, I.~{Kudryashov}, A.~{Kurganov},
  A.~{Panov} et~al., \emph{{Energy spectra of abundant cosmic-ray nuclei in the
  NUCLEON experiment}},
  \href{http://dx.doi.org/10.1016/j.asr.2019.10.004}{\emph{Advances in Space
  Research} {\bf 64} (Dec., 2019) 2546--2558}.

\bibitem{2013ApJ...765...91A}
O.~{Adriani}, G.~C. {Barbarino}, G.~A. {Bazilevskaya}, R.~{Bellotti},
  M.~{Boezio}, E.~A. {Bogomolov} et~al., \emph{{Time Dependence of the Proton
  Flux Measured by PAMELA during the 2006 July-2009 December Solar Minimum}},
  \href{http://dx.doi.org/10.1088/0004-637X/765/2/91}{\emph{\apj} {\bf 765}
  (Mar., 2013) 91}, [\href{http://arxiv.org/abs/1301.4108}{{\tt 1301.4108}}].

\bibitem{2005ApJ...628L..41D}
V.~A. {Derbina}, V.~I. {Galkin}, M.~{Hareyama}, Y.~{Hirakawa}, Y.~{Horiuchi},
  M.~{Ichimura} et~al., \emph{{Cosmic-Ray Spectra and Composition in the Energy
  Range of 10-1000 TeV per Particle Obtained by the RUNJOB Experiment}},
  \href{http://dx.doi.org/10.1086/432715}{\emph{\apjl} {\bf 628} (July, 2005)
  L41--L44}.

\bibitem{2013Sci...341..150S}
E.~C. {Stone}, A.~C. {Cummings}, F.~B. {McDonald}, B.~C. {Heikkila}, N.~{Lal}
  and W.~R. {Webber}, \emph{{Voyager 1 Observes Low-Energy Galactic Cosmic Rays
  in a Region Depleted of Heliospheric Ions}},
  \href{http://dx.doi.org/10.1126/science.1236408}{\emph{Science} {\bf 341}
  (July, 2013) 150--153}.

\bibitem{2019NatAs...3.1013S}
E.~C. {Stone}, A.~C. {Cummings}, B.~C. {Heikkila} and N.~{Lal}, \emph{{Cosmic
  ray measurements from Voyager 2 as it crossed into interstellar space}},
  \href{http://dx.doi.org/10.1038/s41550-019-0928-3}{\emph{Nature Astronomy}
  {\bf 3} (Nov., 2019) 1013--1018}.

\bibitem{2013APh42.70E}
A.~D. {Erlykin} and A.~W. {Wolfendale}, \emph{{Cosmic rays in the inner galaxy
  and the diffusion properties of the interstellar medium}},
  \href{http://dx.doi.org/10.1016/j.astropartphys.2012.12.003}{\emph{Astroparticle
  Physics} {\bf 42} (Feb., 2013) 70--75},
  [\href{http://arxiv.org/abs/1212.2760}{{\tt 1212.2760}}].

\bibitem{2016ApJS..223...26A}
F.~{Acero}, M.~{Ackermann}, M.~{Ajello}, A.~{Albert}, L.~{Baldini}, J.~{Ballet}
  et~al., \emph{{Development of the Model of Galactic Interstellar Emission for
  Standard Point-source Analysis of Fermi Large Area Telescope Data}},
  \href{http://dx.doi.org/10.3847/0067-0049/223/2/26}{\emph{\apjs} {\bf 223}
  (Apr., 2016) 26}, [\href{http://arxiv.org/abs/1602.07246}{{\tt 1602.07246}}].

\bibitem{2018PhRvDd3003L}
P.~{Lipari} and S.~{Vernetto}, \emph{{Diffuse Galactic gamma-ray flux at very
  high energy}},
  \href{http://dx.doi.org/10.1103/PhysRevD.98.043003}{\emph{\prd} {\bf 98}
  (Aug., 2018) 043003}, [\href{http://arxiv.org/abs/1804.10116}{{\tt
  1804.10116}}].

\bibitem{pagliaroli2016expectations}
G.~Pagliaroli, C.~Evoli and F.~L. Villante, \emph{Expectations for high energy
  diffuse galactic neutrinos for different cosmic ray distributions},
  {\emph{Journal of Cosmology and Astroparticle Physics} {\bf 2016} (2016)
  004}.

\bibitem{cataldo2019probing}
M.~Cataldo, G.~Pagliaroli, V.~Vecchiotti and F.~Villante, \emph{{Probing
  galactic cosmic ray distribution with TeV gamma-ray sky}}, {\emph{Journal of
  Cosmology and Astroparticle Physics} {\bf 2019} (2019) 050}.

\bibitem{evoli2017cosmic}
C.~Evoli, D.~Gaggero, A.~Vittino, G.~Di~Bernardo, M.~Di~Mauro, A.~Ligorini
  et~al., \emph{{Cosmic-ray propagation with DRAGON2: I. numerical solver and
  astrophysical ingredients}}, {\emph{Journal of Cosmology and Astroparticle
  Physics} {\bf 2017} (2017) 015}.

\bibitem{case1998new}
G.~L. Case and D.~Bhattacharya, \emph{{A new $\Sigma$-D relation and its
  application to the galactic supernova remnant distribution}}, {\emph{The
  Astrophysical Journal} {\bf 504} (1998) 761}.

\bibitem{green2015constraints}
D.~Green, \emph{{Constraints on the distribution of supernova remnants with
  Galactocentric radius}}, {\emph{Monthly Notices of the Royal Astronomical
  Society} {\bf 454} (2015) 1517--1524}.

\bibitem{recchia2016radial}
S.~Recchia, P.~Blasi and G.~Morlino, \emph{{On the radial distribution of
  Galactic cosmic rays}}, {\emph{Monthly Notices of the Royal Astronomical
  Society: Letters} {\bf 462} (2016) L88--L92}.

\bibitem{yang2016radial}
R.~Yang, F.~Aharonian and C.~Evoli, \emph{{Radial distribution of the diffuse
  $\gamma$-ray emissivity in the Galactic disk}}, {\emph{Physical Review D}
  {\bf 93} (2016) 123007}.

\bibitem{kafexhiu2014parametrization}
E.~{Kafexhiu}, F.~{Aharonian}, A.~M. {Taylor} and G.~S. {Vila},
  \emph{{Parametrization of gamma-ray production cross sections for p p
  interactions in a broad proton energy range from the kinematic threshold to
  PeV energies}},
  \href{http://dx.doi.org/10.1103/PhysRevD.90.123014}{\emph{\prd} {\bf 90}
  (Dec., 2014) 123014}, [\href{http://arxiv.org/abs/1406.7369}{{\tt
  1406.7369}}].

\bibitem{engel1992nucleus}
J.~Engel, T.~Gaisser, P.~Lipari and T.~Stanev, \emph{Nucleus-nucleus collisions
  and interpretation of cosmic-ray cascades}, {\emph{Physical Review D} {\bf
  46} (1992) 5013}.

\bibitem{fletcher1994s}
R.~Fletcher, T.~Gaisser, P.~Lipari and T.~Stanev, \emph{{SIBYLL: An event
  generator for simulation of high energy cosmic ray cascades}},
  {\emph{Physical Review D} {\bf 50} (1994) 5710}.

\bibitem{ahn2009cosmic}
E.-J. Ahn, R.~Engel, T.~K. Gaisser, P.~Lipari and T.~Stanev, \emph{{Cosmic ray
  interaction event generator SIBYLL 2.1}}, {\emph{Physical Review D} {\bf 80}
  (2009) 094003}.

\bibitem{kachelriess2014nuclear}
M.~{Kachelriess}, I.~V. {Moskalenko} and S.~S. {Ostapchenko}, \emph{{Nuclear
  Enhancement of the Photon Yield in Cosmic Ray Interactions}},
  \href{http://dx.doi.org/10.1088/0004-637X/789/2/136}{\emph{\apj} {\bf 789}
  (July, 2014) 136}, [\href{http://arxiv.org/abs/1406.0035}{{\tt 1406.0035}}].

\bibitem{mori2009nuclear}
M.~{Mori}, \emph{{Nuclear enhancement factor in calculation of Galactic diffuse
  gamma-rays: A new estimate with DPMJET-3}},
  \href{http://dx.doi.org/10.1016/j.astropartphys.2009.03.004}{\emph{Astroparticle
  Physics} {\bf 31} (June, 2009) 341--343},
  [\href{http://arxiv.org/abs/0903.3260}{{\tt 0903.3260}}].

\bibitem{2011A&A...531A..37D}
T.~{Delahaye}, A.~{Fiasson}, M.~{Pohl} and P.~{Salati}, \emph{{The GeV-TeV
  Galactic gamma-ray diffuse emission. I. Uncertainties in the predictions of
  the hadronic component}},
  \href{http://dx.doi.org/10.1051/0004-6361/201116647}{\emph{\aap} {\bf 531}
  (July, 2011) A37}, [\href{http://arxiv.org/abs/1102.0744}{{\tt 1102.0744}}].

\bibitem{2022A&A...661A..72B}
M.~{Breuhaus}, J.~A. {Hinton}, V.~{Joshi}, B.~{Reville} and H.~{Schoorlemmer},
  \emph{{Galactic gamma-ray and neutrino emission from interacting cosmic-ray
  nuclei}}, \href{http://dx.doi.org/10.1051/0004-6361/202141318}{\emph{\aap}
  {\bf 661} (May, 2022) A72}, [\href{http://arxiv.org/abs/2201.03984}{{\tt
  2201.03984}}].

\bibitem{2014PhRvD..90l2007A}
A.~{Abramowski}, F.~{Aharonian}, F.~{Ait Benkhali}, A.~G. {Akhperjanian}, E.~O.
  {Ang{\"u}ner}, M.~{Backes} et~al., \emph{{Diffuse Galactic gamma-ray emission
  with H.E.S.S.}},
  \href{http://dx.doi.org/10.1103/PhysRevD.90.122007}{\emph{\prd} {\bf 90}
  (Dec., 2014) 122007}, [\href{http://arxiv.org/abs/1411.7568}{{\tt
  1411.7568}}].

\bibitem{aharonian2020probing}
F.~Aharonian, G.~Peron, R.~Yang, S.~Casanova and R.~Zanin, \emph{{Probing the
  sea of galactic cosmic rays with Fermi-LAT}}, {\emph{Physical Review D} {\bf
  101} (2020) 083018}.

\bibitem{ambrogi2018potential}
L.~Ambrogi, S.~Celli and F.~Aharonian, \emph{{On the potential of Cherenkov
  Telescope Arrays and KM3 Neutrino Telescopes for the detection of extended
  sources}}, {\emph{Astroparticle Physics} {\bf 100} (2018) 69--79}.

\bibitem{gammaSens}
{CTAO Performance}, ``Differential sensitivity.''
  \url{https://www.cta-observatory.org/science/ctao-performance}, 2023.

\bibitem{fermiSens}
{Fermi-LAT Performance}, ``Differential sensitivity.''
  \url{https://www.slac.stanford.edu/exp/glast/groups/canda/lat_Performance.htm},
  2023.

\bibitem{aartsen2021icecube}
M.~G. {Aartsen}, R.~{Abbasi}, M.~{Ackermann}, J.~{Adams}, J.~A. {Aguilar},
  M.~{Ahlers} et~al., \emph{{IceCube-Gen2: the window to the extreme
  Universe}}, \href{http://dx.doi.org/10.1088/1361-6471/abbd48}{\emph{Journal
  of Physics G Nuclear Physics} {\bf 48} (June, 2021) 060501},
  [\href{http://arxiv.org/abs/2008.04323}{{\tt 2008.04323}}].

\bibitem{neronov2018multimessenger}
A.~Neronov, M.~Kachelrie{\ss} and D.~Semikoz, \emph{{Multimessenger gamma-ray
  counterpart of the IceCube neutrino signal}}, {\emph{Physical Review D} {\bf
  98} (2018) 023004}.

\bibitem{strong1998propagation}
A.~W. Strong and I.~V. Moskalenko, \emph{Propagation of cosmic-ray nucleons in
  the galaxy}, {\emph{The Astrophysical Journal} {\bf 509} (1998) 212}.

\bibitem{evoli2008cosmic}
C.~Evoli, D.~Gaggero, D.~Grasso and L.~Maccione, \emph{Cosmic ray nuclei,
  antiprotons and gamma rays in the galaxy: a new diffusion model},
  {\emph{Journal of Cosmology and Astroparticle Physics} {\bf 2008} (2008)
  018}.

\bibitem{silva2023measurement}
M.~Silva, S.~Mancina and J.~Osborn, \emph{{Measurement of the Cosmic Neutrino
  Flux from the Southern Sky using 10 years of IceCube Starting Track Events}},
  {\emph{arXiv preprint arXiv:2308.04582} (2023) }.

\bibitem{abbasi2021icecube}
R.~Abbasi, M.~Ackermann, J.~Adams, J.~Aguilar, M.~Ahlers, M.~Ahrens et~al.,
  \emph{{IceCube high-energy starting event sample: Description and flux
  characterization with 7.5 years of data}}, {\emph{Physical Review D} {\bf
  104} (2021) 022002}.

\bibitem{2023ApJ...957L...6F}
K.~{Fang} and K.~{Murase}, \emph{{Decomposing the Origin of TeV-PeV Emission
  from the Galactic Plane: Implications of Multimessenger Observations}},
  \href{http://dx.doi.org/10.3847/2041-8213/ad012f}{\emph{\apjl} {\bf 957}
  (Nov., 2023) L6}, [\href{http://arxiv.org/abs/2307.02905}{{\tt 2307.02905}}].

\bibitem{Fiorillo:2023clw}
D.~F.~G. Fiorillo, V.~B. Valera, M.~Bustamante and W.~Winter, \emph{{Searches
  for dark matter decay with ultrahigh-energy neutrinos endure backgrounds}},
  \href{http://dx.doi.org/10.1103/PhysRevD.108.103012}{\emph{Phys. Rev. D} {\bf
  108} (2023) 103012}, [\href{http://arxiv.org/abs/2307.02538}{{\tt
  2307.02538}}].

\bibitem{zuriaga2023multi}
J.~Zuriaga-Puig, V.~Gammaldi, D.~Gaggero, T.~Lacroix and M.~S{\'a}nchez-Conde,
  \emph{{Multi-TeV dark matter density in the inner Milky Way halo: spectral
  and dynamical constraints}}, {\emph{Journal of Cosmology and Astroparticle
  Physics} {\bf 2023} (2023) 063}.

\bibitem{2008ICRC....3.1341W}
S.~P. {Wakely} and D.~{Horan}, \emph{{TeVCat: An online catalog for Very High
  Energy Gamma-Ray Astronomy}}, {\emph{International Cosmic Ray Conference}
  {\bf 3} (2008) 1341--1344}.

\bibitem{Albert2021ApJ6A}
A.~{Albert}, R.~{Alfaro}, C.~{Alvarez}, J.~R. {Angeles Camacho}, J.~C.
  {Arteaga-Vel{\'a}zquez}, K.~P. {Arunbabu} et~al., \emph{{Probing the Sea of
  Cosmic Rays by Measuring Gamma-Ray Emission from Passive Giant Molecular
  Clouds with HAWC}},
  \href{http://dx.doi.org/10.3847/1538-4357/abfc47}{\emph{\apj} {\bf 914}
  (June, 2021) 106}, [\href{http://arxiv.org/abs/2101.08748}{{\tt
  2101.08748}}].

\bibitem{Sinha_2022icrc90S}
A.~{Sinha}, V.~{Baghmanyan}, G.~{Peron}, Y.~{Gallant}, S.~{Casanova},
  M.~{Holler} et~al., \emph{{Search for enhanced TeV gamma ray emission from
  Giant Molecular Clouds using H.E.S.S.}},  in \emph{37th International Cosmic
  Ray Conference}, p.~790, Mar., 2022.
\newblock \href{http://arxiv.org/abs/2108.01738}{{\tt 2108.01738}}.
\newblock \href{http://dx.doi.org/10.22323/1.395.0790}{DOI}.

\bibitem{luque2022galactic}
P.~D. l.~T. Luque, D.~Gaggero, D.~Grasso, O.~Fornieri, K.~Egberts, C.~Steppa
  et~al., \emph{{Galactic diffuse gamma rays meet the PeV frontier}},
  {\emph{arXiv preprint arXiv:2203.15759} (2022) }.

\bibitem{yan2023origin}
K.~Yan, R.-Y. Liu, R.~Zhang, C.-M. Li, Q.~Yuan and X.-Y. Wang, \emph{{On the
  Origin of Galactic Diffuse TeV-PeV Emission: Insight from LHAASO and
  IceCube}}, {\emph{arXiv preprint arXiv:2307.12363} (2023) }.

\end{thebibliography}\endgroup
\bibliographystyle{JHEP}

\appendix

\counterwithin{figure}{section}
\section{APPENDIX}
\label{sec:appa}

\begin{figure}[!ht]
\centering
\subfigure{\includegraphics[width=0.7\linewidth]{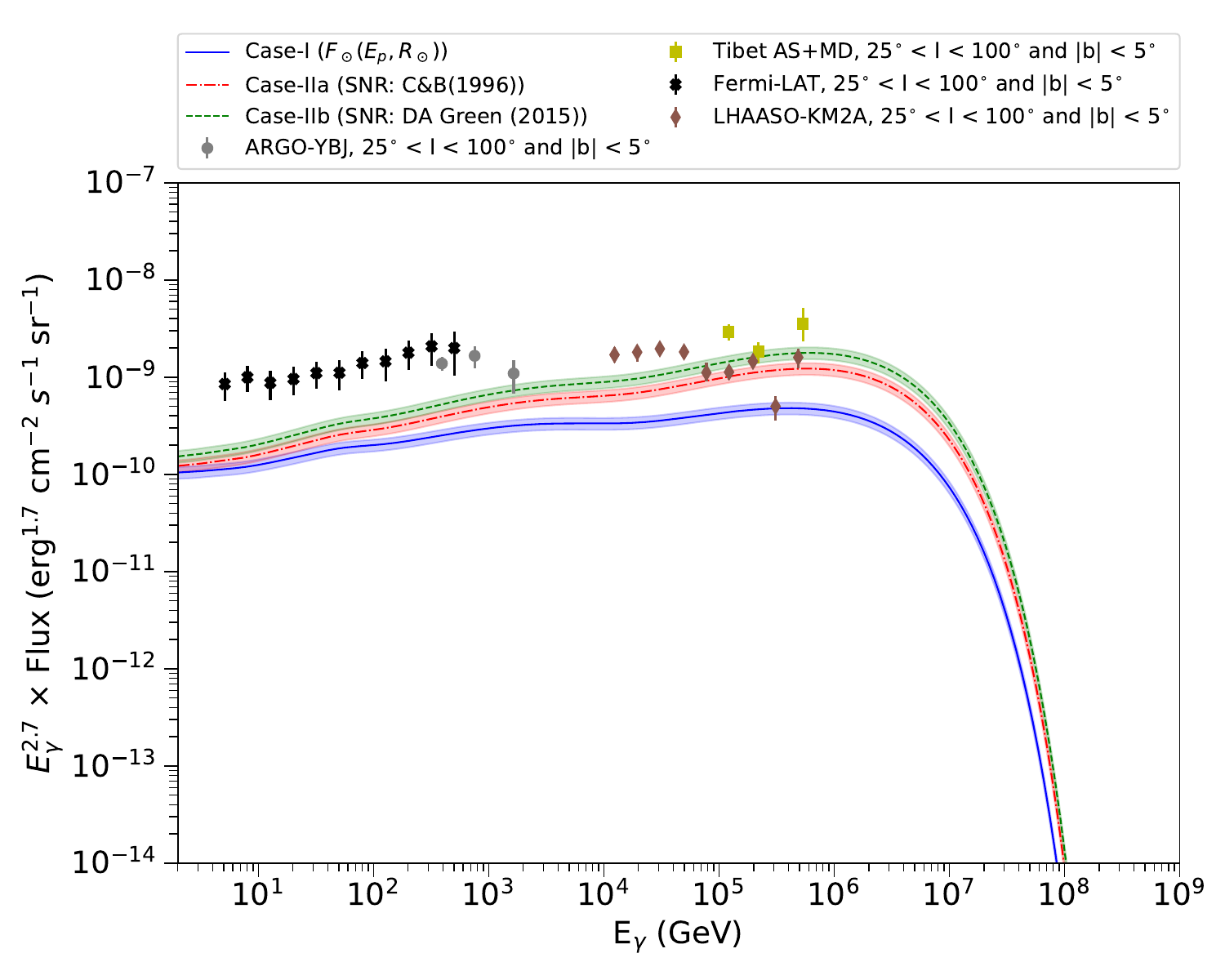}}
\subfigure{\includegraphics[width=0.7\linewidth]{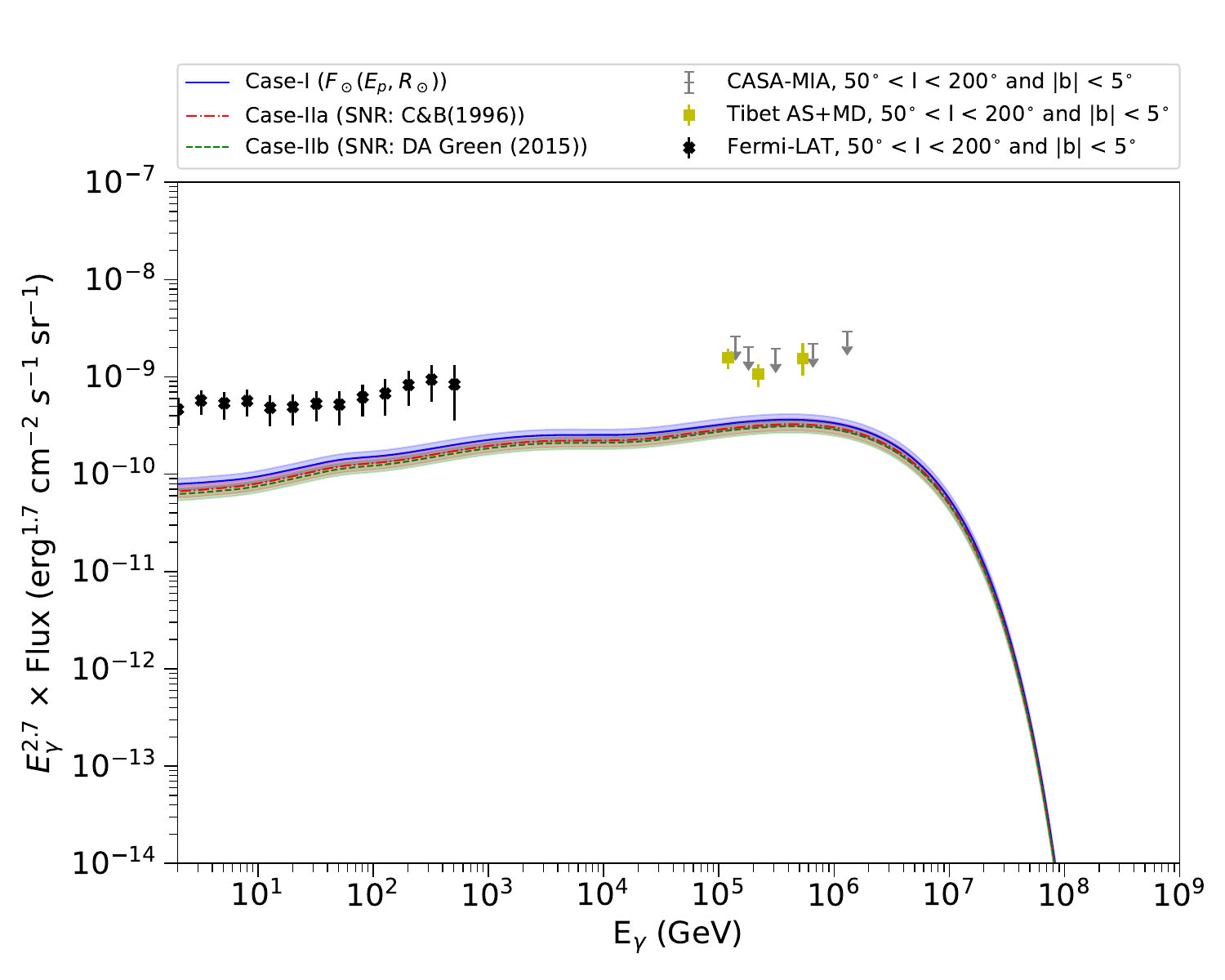}}
\caption{\textit{Top panel:} The diffuse gamma-ray flux as observed by ARGO-YBJ, Tibet AS+MD, Fermi-LAT and LHAASO experiment from the inner part of our Galaxy, along with the expected gamma-ray flux from the GMCs in that region, is depicted here. \textit{Bottom panel:} The comparison between the observed diffuse gamma-ray flux from the outer part of the MW galaxy by Tibet AS+MD, CASA-MIA and Fermi-LAT experiment with the expected diffused signal from the GMCs is presented here. The data points were extracted from Fig. 4 in  \cite{amenomori2021first} and Fig. 4 and 5, in \cite{luque2022galactic}}
\label{fig:TibetARGO}
\end{figure}

For the completeness of the study, we compared the gamma-ray flux observed by Tibet AS$\gamma$, ARGO-YBJ, CASA-MIA, Fermi-LAT and LHAASO experiment in both inner ($25^{\circ} < l < 100^{\circ}$ \& $|b| < 5^{\circ}$) and outer ($50^{\circ} < l < 200^{\circ}$ \& $|b| < 5^{\circ}$) Galactic regions with our calculated flux from the GMCs. The comparison is shown in Fig. \ref{fig:TibetARGO}; it is evident that the calculated flux for Case-IIb is almost two times lower than the observed result at the inner Galactic region. However, as we move towards the outer Galactic region, this factor increases to almost four, and the fluxes for all three cases become nearly identical (see bottom panel of Fig. \ref{fig:TibetARGO}). This growth of the multiplication factor for the outer galaxy is mainly because the number density of atomic hydrogen ($H_I$) starts to increase (see Fig. \ref{fig:surface_dens_Rgal}), and the GCR source contribution decreases.

\begin{figure}[!ht]
\centering
\subfigure{\includegraphics[width=0.7\linewidth]{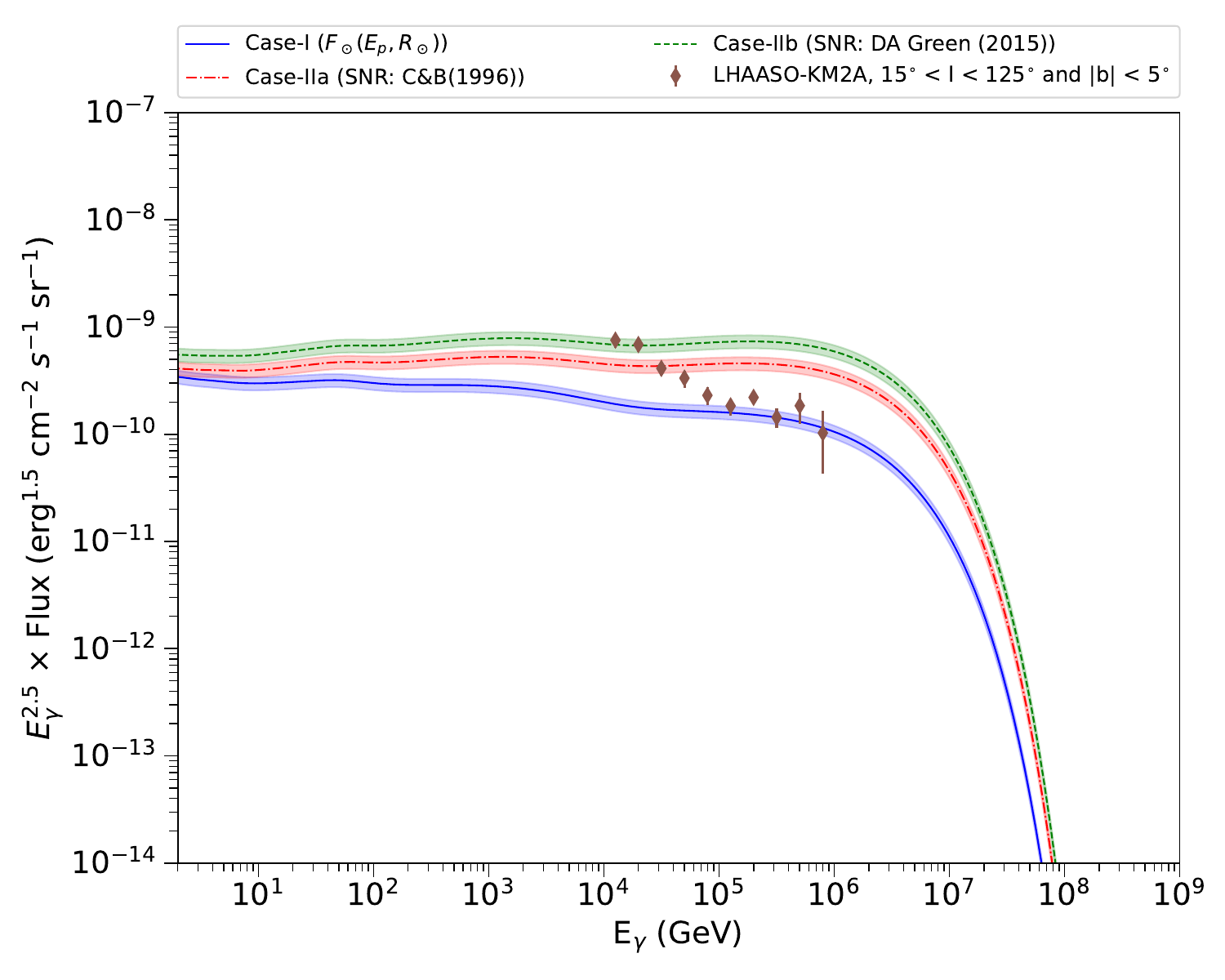}}
\subfigure{\includegraphics[width=0.7\linewidth]{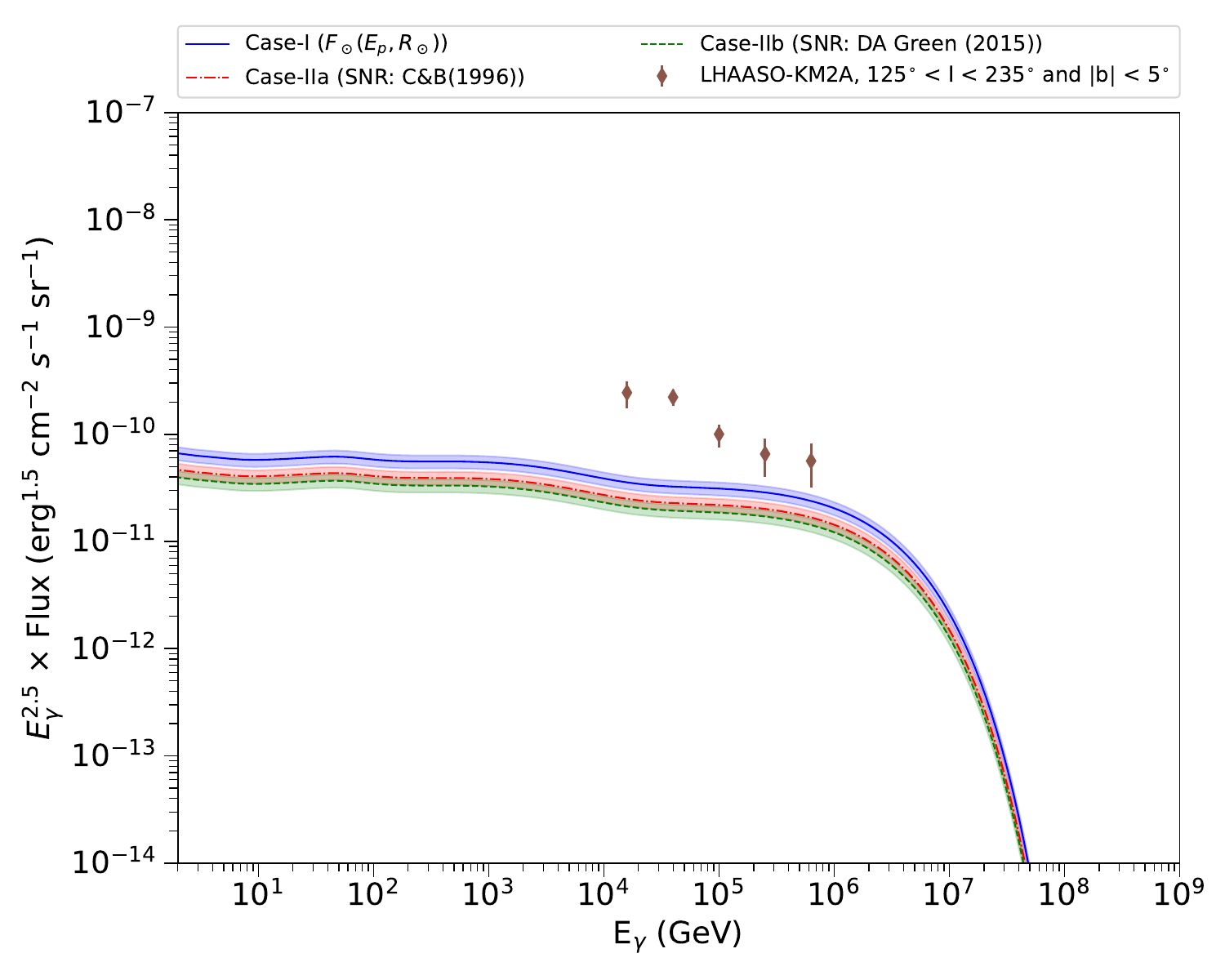}}
\caption{\textit{Top panel:} The truly diffuse gamma-ray signal recently observed by the LHAASO experiment from the inner Galactic plane by masking out the source locations is presented here. The gamma-ray flux from the GMCs in that region calculated considering the three cases of GCR flux are also plotted here without any masking for reference. \textit{Bottom panel:} This plot shows the flux that was observed by the LHAASO experiment with masking and the calculated flux from the GMCs without masking for the outer region of our MW Galaxy. The data points were extracted from Fig. 2 in \cite{2023PhRvL.131o1001C}}
\label{fig:LHAASOFlux}
\end{figure}

The diffuse gamma-ray flux observed by the LHAASO experiment up to PeV energies \cite{2023PhRvL.131o1001C} is shown in Fig. \ref{fig:LHAASOFlux}. However, the comparison between our results and their observations is particularly difficult as they used some masking to exclude source contributions from the inner ($15^{\circ} < l < 125^{\circ}$ \& $|b| < 5^{\circ}$) and outer ($125^{\circ} < l < 235^{\circ}$ \& $|b| < 5^{\circ}$) Galactic regions, which also excludes the GMC contributions from those areas. As a consequence, the observed flux from those regions will be lower than the total expected one. The figure also shows gamma-ray contribution from GMCs in those regions, for reference. However, gamma-ray absorption, which could be significant above 100 TeV, is not taken into consideration here. The observed emission follows a power-law distribution, while the expected diffuse signal shows more or less a flat distribution. According to \cite{yan2023origin}, the gamma-ray flux estimated from interactions between GCR and ISM gas below approximately $10^5$ GeV is lower than the flux observed by the LHAASO for both the inner and outer regions of the Galaxy. Additionally, the gamma-ray flux converted from neutrino flux measurements is consistent with the model predictions \cite{zhang2023ApJ3Z}, thus confirming the existence of the observed gamma-ray excess, whose origin could be leptonic \cite{yan2023origin}.

\begin{figure}[!ht]
\centering
\includegraphics[width=0.8\linewidth]{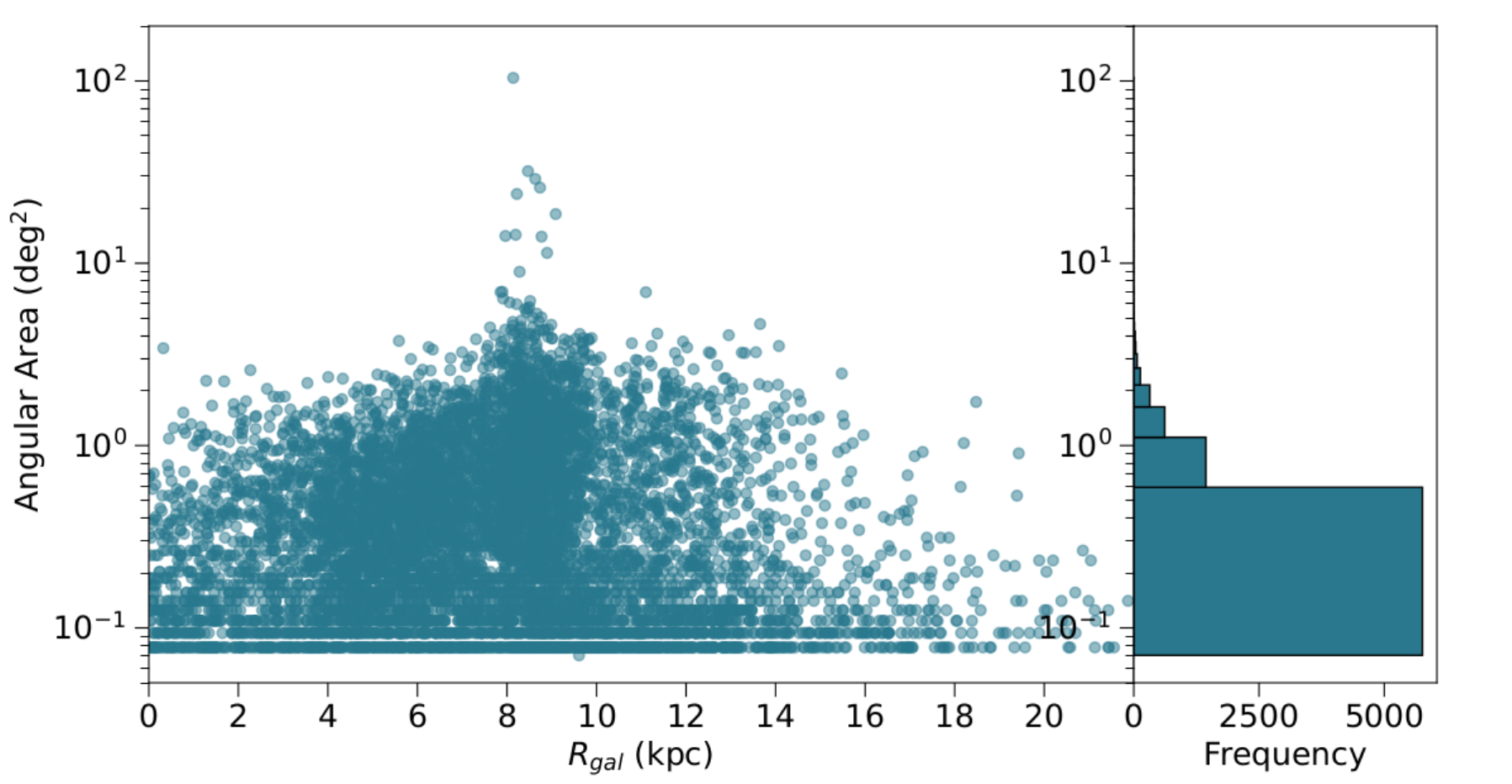}
\caption{The distribution of the angular areas of GMCs (in degree$^2$) along the galactocentric radius is shown here.}
\label{fig:AngArea}
\end{figure}

The Fig. \ref{fig:AngArea} illustrates the angular area distribution of GMCs taken into account in the calculations. Its value ranges from 0.07 degree$^2$ to a maximum value of 104 degree$^2$ with an average value of 0.6 degree$^2$. So, they can be both point-like sources or extended objects.

The TeVCat and LHAASO sources within the $1\sigma$ significance of the observed IceCube signal are listed in Tab.  \ref{Tab:TeVCat1} and \ref{Tab:TeVCat2}. Their nature could be leptonic, hadronic or a combination of both; however, we neglect some source classes here, like pulsars, which are mostly electromagnetic in nature and hence may not contribute to IceCube-detected neutrinos. Here, for simplicity, we categorised the source types into seven groups (i.e., \textit{PWN, Binary, Star Cluster, SNR, GMC, PeVatron and UNID}) in comparison to the actual TeVCat source classification is also shown in Tab. \ref{Tab:TeVCat1}.


\begin{figure}[!ht]
    \centering
    \subfigure{\includegraphics[width=0.48\linewidth]{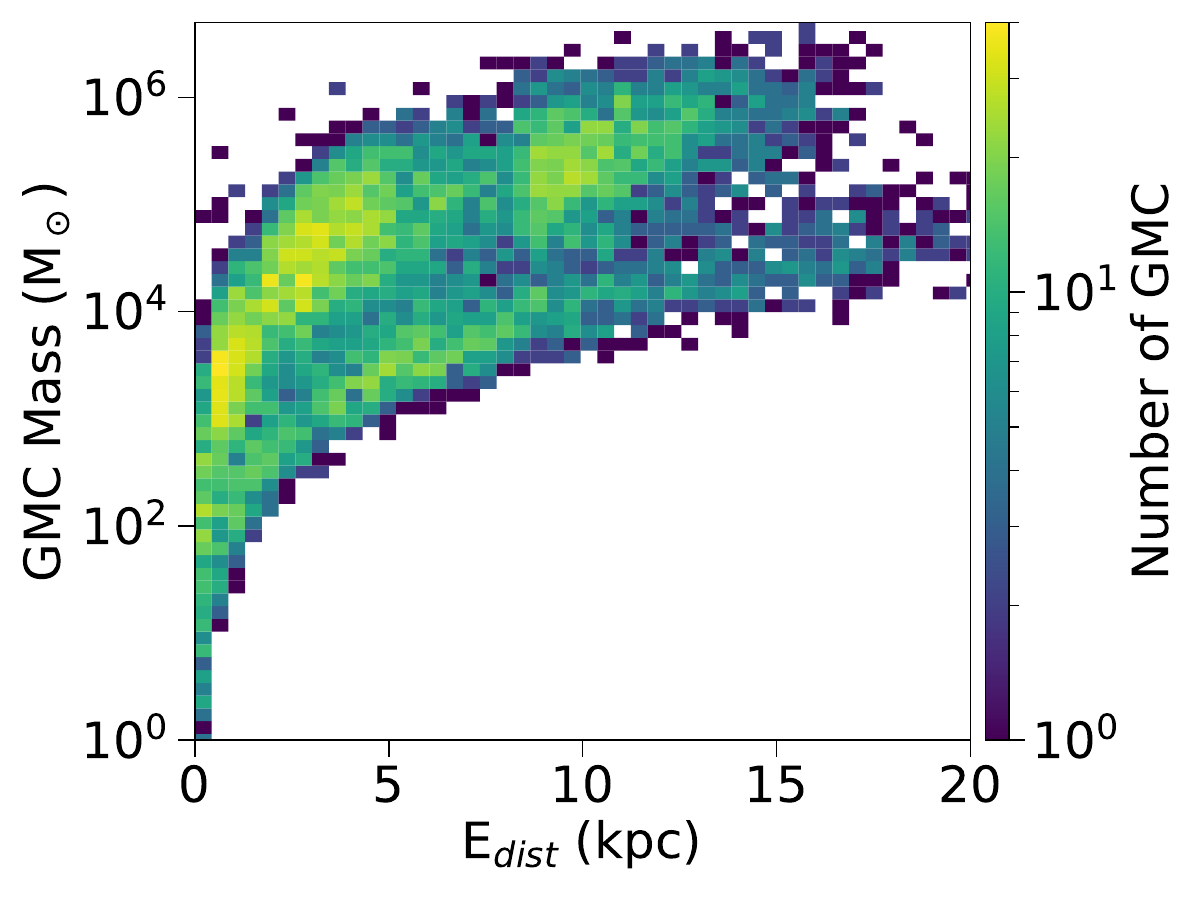}}
    \subfigure{\includegraphics[width=0.48\linewidth]{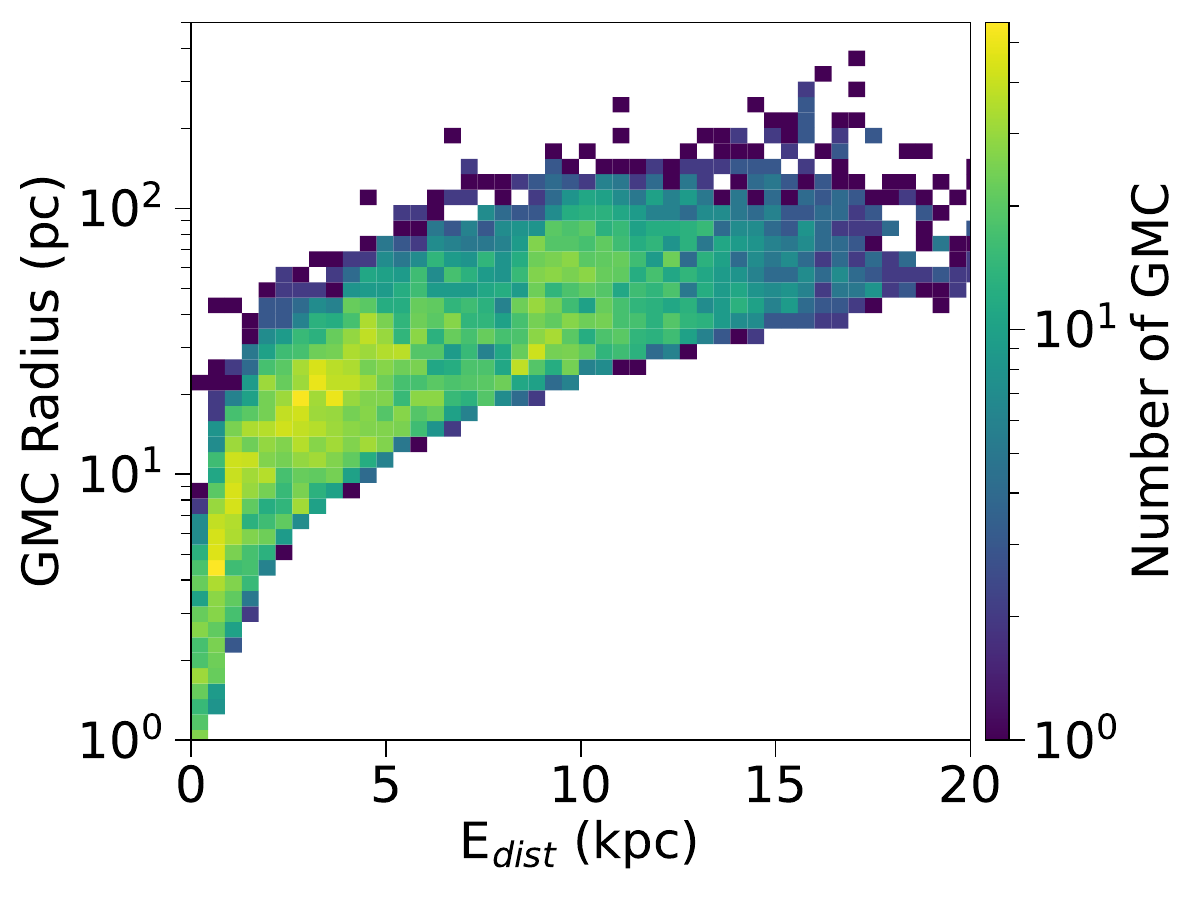}}
    \caption{{\it Left panel}: The 2D-density distribution of GMC Mass vs its distance from the Earth,
 {\it Right panel}: The 2D-density distribution of GMC Radius vs its distance from the Earth.}
    \label{fig:}
\end{figure}

\begin{figure}[!ht]
    \centering
    \subfigure{\includegraphics[width=0.48\linewidth]{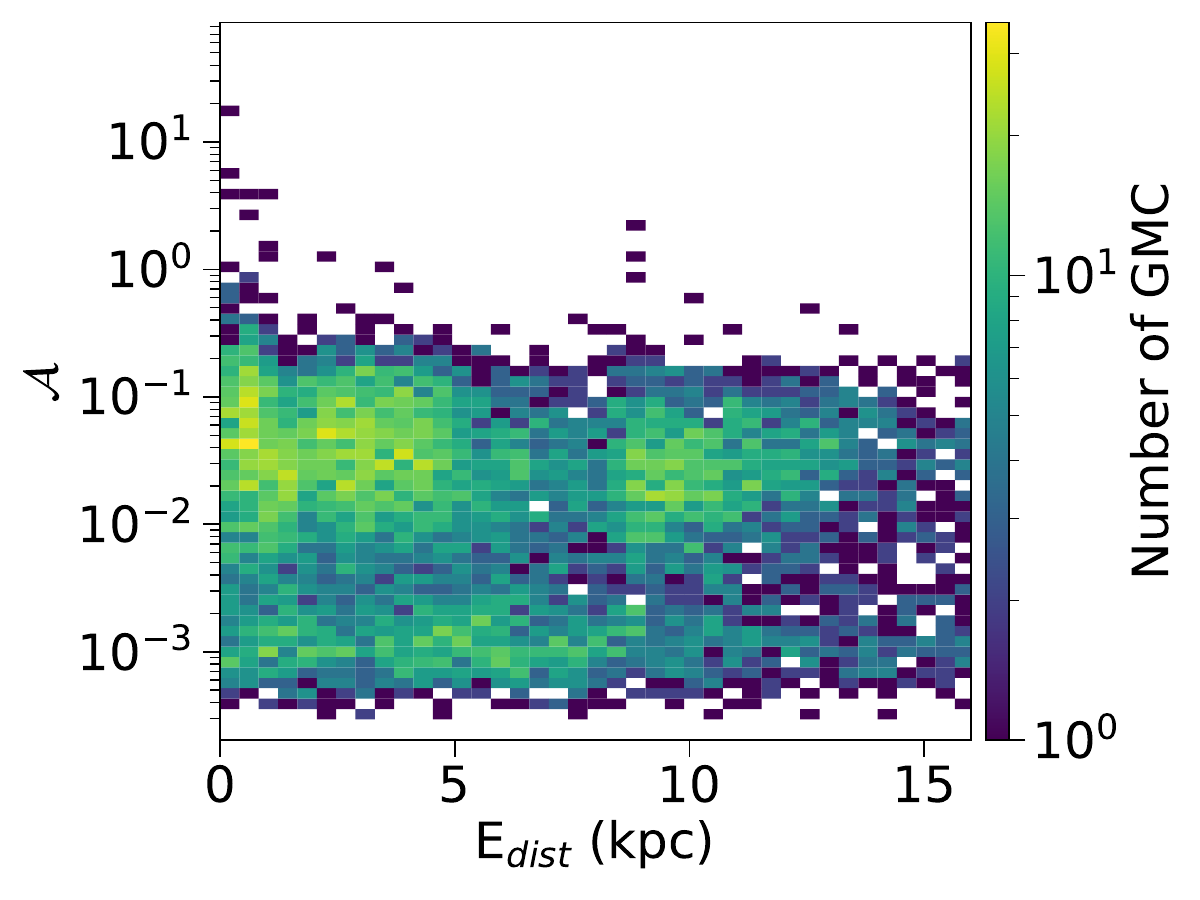}}
    \subfigure{\includegraphics[width=0.48\linewidth]{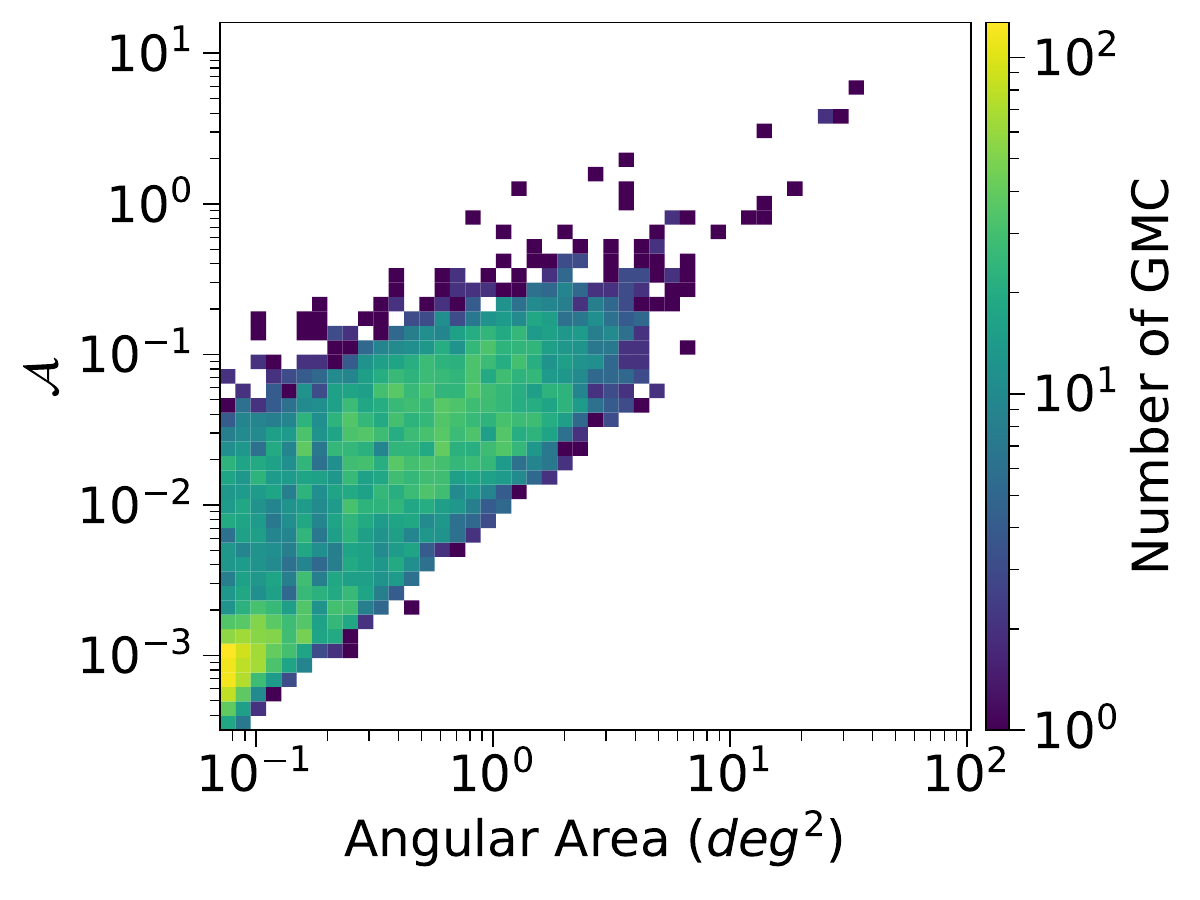}}
    \caption{{\it Left panel}: The 2D-density distribution of $\mathcal{A}$ parameter vs its distance from the Earth {\it Right panel}: The 2D-density distribution of $\mathcal{A}$ parameter vs the angular area of the GMCs.}
    \label{fig:}
\end{figure}




\begin{table}[!ht]
{\footnotesize
\begin{tabular}{lcccc}
\hline\hline
Name & l & b & Types & TeVCat Source Classification \\
 & (deg) & (deg) & & \\
\hline

           Crab & -175.442800 & -5.788800  &  PWN & PWN \\
      MSH 15-52 &  -39.669671 & -1.193044  &  PWN & PWN  \\
SNR G327.1-01.1 &  -32.843900 & -1.078400  &  PWN & PWN  \\
 HESS J1616-508 &  -27.609752 & -0.141240  &  PWN & PWN  \\
 HESS J1632-478 &  -23.615741 &  0.190197  &  PWN & PWN  \\
 HESS J1708-443 &  -16.942474 & -2.375754  &  PWN & PWN  \\
 HESS J1718-385 &  -11.166402 & -0.488102  &  PWN & PWN  \\
SNR G000.9+00.1 &    0.871897 &  0.075976  &  PWN & PWN  \\
 HESS J1813-178 &   12.811507 & -0.026289  &  PWN & PWN  \\
 HESS J1831-098 &   21.850335 & -0.109053  &  PWN & PWN  \\
  MGRO J2019+37 &   74.923416 &  0.514332  &  PWN & PWN  \\

   LS 5039 & 16.902186 & -1.278106 & Binary & Binary \\
Cygnus X-1 & 71.334900 &  3.066600 & Binary & XRB \\

Westerlund 1 & -20.453089 & -0.352882 & Star Cluster & Massive Star Cluster \\
    Terzan 5 &   3.783500 &  1.722900 & Star Cluster & Globular Cluster \\

         IC 443 & -170.927090 &  2.917722 &  SNR  & Shell \\
          Tycho &  120.091500 &  1.402200 &  SNR  & Shell \\
        SN 1006 &  -32.431887 & 14.559082 &  SNR  & Shell \\
     SN 1006 NE &  -32.156506 & 14.564999 &  SNR  & Shell \\
 HESS J1534-571 &  -36.346800 & -0.918500 &  SNR  & Shell \\
 HESS J1614-518 &  -28.480485 & -0.581325 &  SNR  & Shell \\
 HESS J1640-465 &  -21.724900 & -0.036000 &  SNR  & Composite SNR \\
RX J1713.7-3946 &  -12.664481 & -0.472686 &  SNR  & Shell \\
        CTB 37B &  -11.354731 &  0.378331 &  SNR  & Shell \\
        CTB 37A &  -11.584300 &  0.142500 &  SNR  & SNR/Molec. Cloud \\
SNR G349.7+00.2 &  -10.280357 &  0.173978 &  SNR  & SNR/Molec. Cloud \\
   Kepler's SNR &    4.521200 &  6.822400 &  SNR  & SNR \\
 HESS J1731-347 &   -6.457600 & -0.669800 &  SNR  & Shell \\
 SNR G004.8+6.2 &    4.796100 &  6.246200 &  SNR  & SNR \\
 HESS J1745-303 &   -1.290100 & -0.639700 &  SNR  & SNR/Molec. Cloud \\
    RS Ophiuchi &   19.799500 & 10.372100 &  SNR  & Nova \\
HESS J1800-240C &    5.711189 & -0.058772 &  SNR  & SNR/Molec. Cloud \\
HESS J1800-240B &    5.901986 & -0.365181 &  SNR  & SNR/Molec. Cloud \\
           W 28 &    6.656805 & -0.267552 &  SNR  & SNR/Molec. Cloud \\
HESS J1800-240A &    6.141203 & -0.629279 &  SNR  & SNR/Molec. Cloud \\
SNR G015.4+00.1 &   15.409124 &  0.160685 &  SNR  & Composite SNR \\
   Cassiopeia A &  111.711383 & -2.129547 &  SNR  & Shell \\

Cloud 877 & -26.5399  & -0.3101  &  GMC & Giant Molecular Cloud \\~\\

LHAASO J0534+2202 & -175.514138 & -5.830381  & PeVatron & PWN (Crab) \\
LHAASO J1825-1326 &   18.067863 & -0.540475  & PeVatron & UNID \\
LHAASO J1908+0621 &   40.491859 & -0.813784  & PeVatron & UNID \\
LHAASO J2018+3651 &   74.984112 &  0.454934  & PeVatron & UNID \\
LHAASO J2108+5157 &   92.277370 &  2.867798  & PeVatron & DARK \\
       
\hline            
\end{tabular}
}
\caption{TeVCat gamma-ray sources whose position lies within 1$\sigma$ significance map of the IceCube observation of neutrino from Galactic plane.}
\label{Tab:TeVCat1}
\end{table}

\begin{table}[!ht]
{\footnotesize
\begin{tabular}{lcccc}
\hline\hline
Name & l & b & Types \\
 & (deg) & (deg) & \\
\hline
      HESS J1626-490  & -25.229819 &  0.047307 & UNID \\
       HESS J1634-472 & -22.891150 &  0.217191 & UNID \\
       HESS J1641-463 & -21.477200 &  0.090500 & UNID \\
       HESS J1702-420 & -15.696067 & -0.183715 & UNID \\
       HESS J1708-410 & -14.317744 & -0.468802 & UNID \\
       HESS J1729-345 &  -6.556293 & -0.127430 & UNID \\
       HESS J1741-302 &  -1.723900 &  0.050200 & UNID \\
Galactic Centre Ridge &  -0.055138 & -0.043908 & UNID \\
      Galactic Centre &  -0.055138 & -0.043908 & UNID \\
       HESS J1746-308 &  -1.549300 & -1.112500 & UNID \\
        VER J1746-289 &   0.055100 & -0.148100 & UNID \\
       HESS J1746-285 &   0.140100 & -0.114100 & UNID \\
   MAGIC J1746.4-2853 &   0.137100 & -0.120800 & UNID \\
       HESS J1804-216 &   8.400200 & -0.028400 & UNID \\
       HESS J1808-204 &   9.978400 & -0.242700 & UNID \\
       HESS J1809-193 &  11.180020 & -0.087415 & UNID \\
       HESS J1813-126 &  17.310100 &  2.489900 & UNID \\
       2HWC J1814-173 &  13.331700 &  0.126800 & UNID \\
      2HWC J1819-150* &  15.909200 &  0.089700 & UNID \\
       2HWC J1825-134 &  18.116700 & -0.525900 & UNID \\
       HESS J1826-130 &  18.433800 & -0.412600 & UNID \\
       HESS J1828-099 &  21.490100 &  0.379900 & UNID \\
       HESS J1832-085 &  23.205400 &  0.295600 & UNID \\
        MGRO J1908+06 &  40.385271 & -0.785071 & UNID \\
      ARGO J1910+0720 &  41.654007 & -0.881286 & UNID \\
       2HWC J2006+341 &  71.321300 &  1.159000 & UNID \\
       3HWC J2010+345 &  72.140000 &  0.560000 & UNID \\
        VER J2016+371 &  74.938300 &  1.146900 & UNID \\
        VER J2019+368 &  74.993200 &  0.360000 & UNID \\
       MilagroDiffuse &  76.045266 &  0.940755 & UNID \\
\hline            
\end{tabular}
}
\caption{TeVCat unidentified (UNID) gamma-ray sources whose position lies within 1$\sigma$ significance map of the IceCube observation of neutrino flux from Galactic plane.}
\label{Tab:TeVCat2}
\end{table}

\end{document}